\documentclass[plain,biber]{nowfnt}

\usepackage{bm}
\usepackage{mwe}
\usepackage{amsmath}
\usepackage{amssymb}
\usepackage{booktabs}
\usepackage{amsfonts}
\usepackage{multirow}
\usepackage{chemformula}
\usepackage[subrefformat=parens,labelformat=parens]{subfig}
\usepackage[utf8]{inputenc}
\usepackage[show]{chato-notes}

\maintitleauthorlist{
Sebastian Bruch \\
Pinecone, U.S.A. \\
sbruch@acm.org
\and
Claudio Lucchese \\
Ca' Foscari University of Venice, Italy \\
claudio.lucchese@unive.it
\and
Franco Maria Nardini \\
ISTI-CNR, Pisa, Italy \\
francomaria.nardini@isti.cnr.it
}

\usepackage{acronym}
\acrodef{ltr}[LtR]{Learning to Rank}
\acrodef{ltradj}[]{learning-to-rank} 

\newtheorem{tradeoff}{Trade-off}
\newtheorem{challenge}{Efficiency Challenge}

\usepackage{bm}
\newcommand{\queries}{{\mathcal Q}}
\newcommand{\data}{{\mathcal D}}
\newcommand{\doclist}{\bm{x}}
\newcommand{\lablist}{\bm{y}}
\newcommand{\ranker}{R}
\newcommand{\feats}{{\mathcal X}}
\newcommand{\labels}{{\mathcal Y}}

\newcommand{\algo}[1]{{\sc #1}}
\newcommand{\lmart}{\algo{$\lambda$-Mart}}

\newcommand{\cleaver}{\algo{CLEaVER}}
\newcommand{\xcleaver}{\algo{X-CLEaVER}}
\newcommand{\dart}{\algo{Dart}}
\newcommand{\xdart}{\algo{X-Dart}}
\newcommand{\quickscorer}{\algo{QuickScorer}}

\newcommand{\measure}[1]{\textsc{#1}}
\newcommand{\ndcg}{\measure{NDCG}}
\newcommand{\dcg}{\measure{DCG}}
\newcommand{\err}{\measure{ERR}}
\newcommand{\rbp}{\measure{RBP}}

\newcommand{\avgp}{\measure{AP}}
\newcommand{\map}{\measure{MAP}}

\newcommand{\dataset}[1]{{\sc #1}}
\newcommand{\msnsmall}{\dataset{MSLR-WEB10K}}
\newcommand{\msnlarge}{\dataset{MSLR-WEB30K}}
\newcommand{\istellasmall}{\dataset{Istella-S}}

\usepackage{soul}
\usepackage{xcolor}

\title{Efficient and Effective Tree-based and Neural Learning to Rank\footnote{Preprint of article accepted for publication in Foundations and Trends\textregistered{} in Information Retrieval}}

\author[1]{Bruch, Sebastian}
\author[2]{Lucchese, Claudio}
\author[3]{Nardini, Franco Maria}

\affil[1]{Pinecone, U.S.A.; sbruch@acm.org}
\affil[2]{Ca' Foscari University of Venice, Italy; claudio.lucchese@unive.it}
\affil[3]{ISTI-CNR, Pisa, Italy; francomaria.nardini@isti.cnr.it}

\begin{document}

\makeabstracttitle

\begin{abstract}
As information retrieval researchers, we not only develop algorithmic solutions to hard problems, but we also insist on a proper,
multifaceted evaluation of ideas. The literature on the fundamental topic of retrieval and ranking, for instance, has a rich
history of studying the effectiveness of indexes, retrieval algorithms, and complex machine learning rankers, while at the same time
quantifying their computational costs, from creation and training to application and inference. This is evidenced, for example,
by more than a decade of research on efficient training and inference of large decision forest models in \ac{ltr}.
As we move towards even more complex, deep learning models in a wide range of applications, questions on efficiency have once
again resurfaced with renewed urgency. Indeed, efficiency is no longer limited to time and space; instead it has found new,
challenging dimensions that stretch to resource-, sample- and energy-efficiency with ramifications for researchers,
users, and the environment.

This monograph takes a step towards promoting the study of efficiency in the era of neural information retrieval by offering
a comprehensive survey of the literature on efficiency and effectiveness in ranking, and to a limited extent, retrieval.
This monograph was inspired by the parallels that exist between the challenges in neural network-based ranking solutions
and their predecessors, decision forest-based \ac{ltr} models, as well as the connections between the solutions the literature
to date has to offer. We believe that by understanding the fundamentals underpinning these algorithmic and data structure solutions
for containing the contentious relationship between efficiency and effectiveness, one can better identify future directions
and more efficiently determine the merits of ideas. We also present what we believe to be important research directions
in the forefront of efficiency and effectiveness in retrieval and ranking.
\end{abstract}

\chapter{Introduction}
\label{chap:intro}
\acresetall

Search engines are a familiar tool to the reader of this manuscript.
In fact, you have likely arrived at this copy by typing
a few keywords into one and perusing the relevant links and page descriptions
in its results page. Indeed, the abundance of
data on the web makes search engines an integral tool, without
which it would be nearly impossible to discover the right information
and satisfy an information need.

We similarly rely on a suite of other algorithmic tools to get
what is pertinent to us, such as discovering news articles, movies, or songs (recommendation systems),
getting answers to natural language questions (question answering and conversational agents),
finding images depicting a given description (image search),
and many more. What all of these tools have in common is that they
are different manifestations of the \emph{retrieval and ranking} problem,
which seeks to discover a \emph{set of relevant items} from a large collection
and order them according to some \emph{criteria} and with respect to some \emph{context}.

\begin{definition}[The Document Ranking Problem]
  Given a query $q$ (\emph{context}) and a set of documents $D$ (\emph{items}), the goal is to
  order elements of $D$ such that the resulting ranked list
  maximizes a user satisfaction metric $Q$ (\emph{criteria}).
\end{definition}

\begin{figure}[t]
    \centering
    \includegraphics[width=\linewidth]{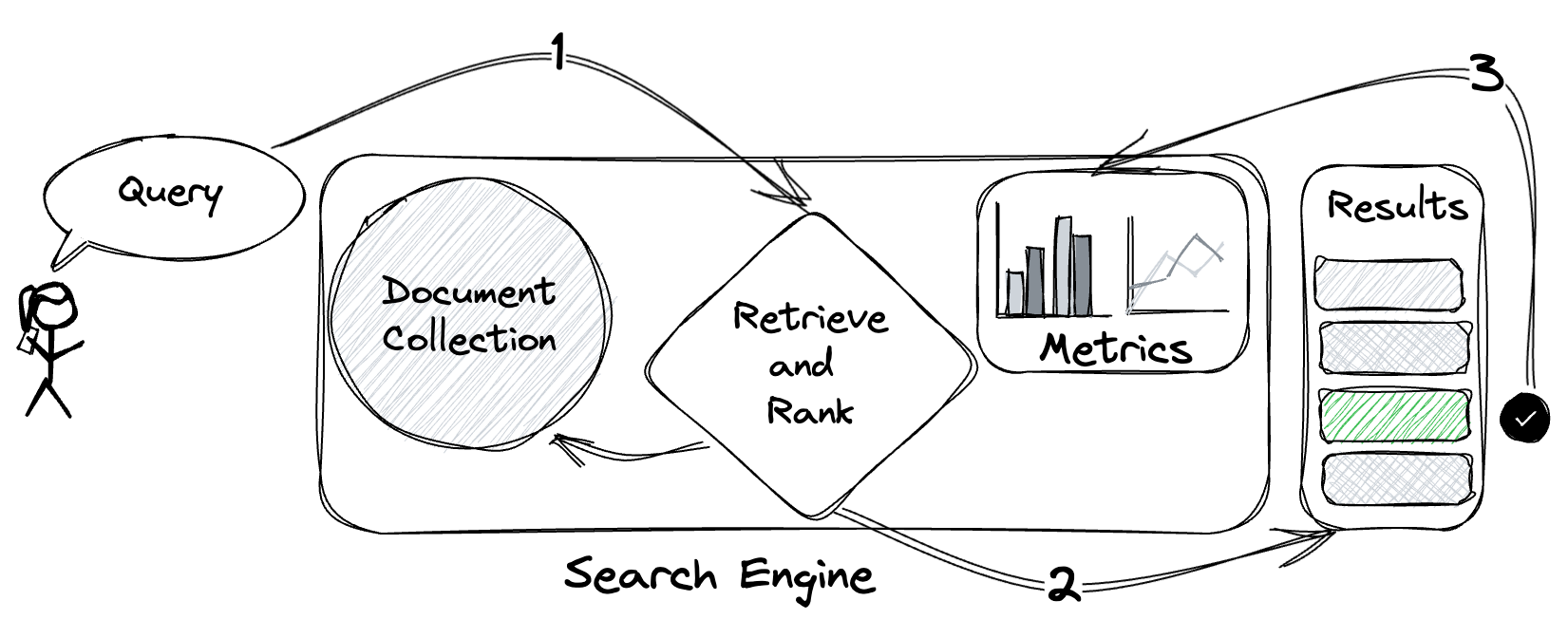}
    \caption{The Document Ranking Problem in the context of web search---our running example.
    The user sends a text query to the search engine (1), which, in turn, \emph{retrieves} the most
    relevant documents from a large collection, and presents them as a \emph{ranked} list (2).
    The user then decides if and to what extent the ranked list satisfies their information
    need, which affects metrics of interest (3).}
    \label{fig:01:web-search}
\end{figure}

We take web search as the theme of this
monograph and delve into the ranking problem in that context. In document ranking,
the query $q$ is an intent expressed (often briefly) as a set of textual keywords or
in natural language, the documents $D$ are (possibly long) texts written in natural language,
and $Q$ is any utility metric that captures the relevance of an ordered list to $q$.
We have illustrated this setup in Figure~\ref{fig:01:web-search}.

Document ranking presents a number of unique questions that are the subject of much research
in the field of information retrieval:
How do we define $Q$ to quantify the perceived quality of a ranked list
and its utility to a user?
How do we capture and interpret implicit, noisy, and sometimes circular user preferences,
which are represented by clicks?
And, more pertinent to this monograph, how do we arrive at a ranked list given a
query, a set of documents, a metric, possibly subject to a set of other constraints?

Over a decade ago, machine learning transformed how we approach the document ranking problem
and answer the questions above. That wave resulted in
a paradigm shift from early statistical methods, heuristics, and hand-crafted rules to determine
the relevance of documents to a query, to what would later be called
\ac{ltr}~\citep{Liu08}, where the relevance of a document to a query
is estimated by a learnt function, hence ``learning'' to rank.
This leap was perhaps best exemplified by
LambdaMART~\citep{export:132652} in the Yahoo! Learning-to-Rank Challenge~\citep{chapelle2011yahoo}.

\begin{figure}[t]
    \centering
    \includegraphics[width=\linewidth]{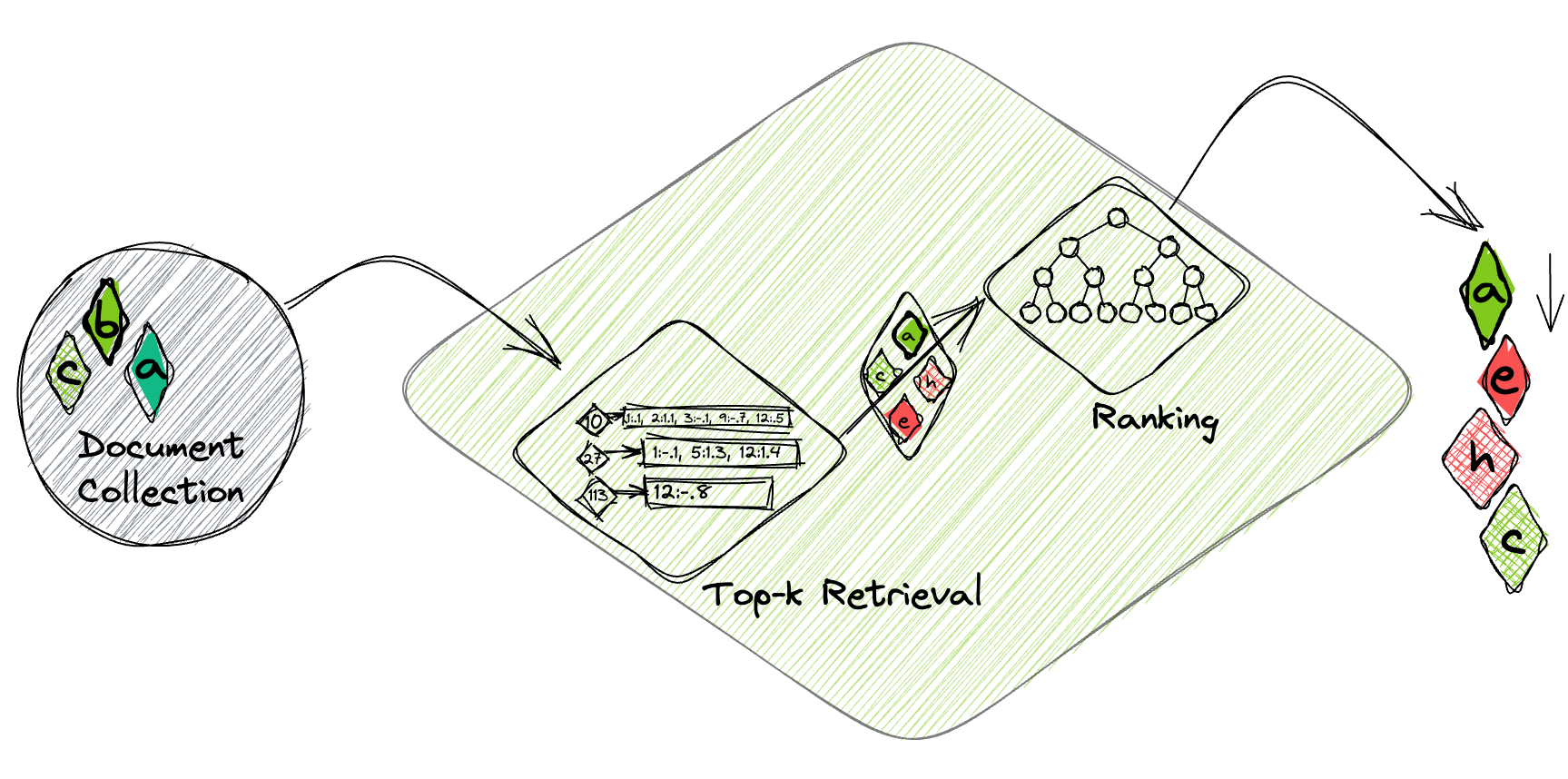}
    \caption{Retrieval and ranking algorithms in a modern search system.
    The \emph{retrieval} algorithm often solves one form of the maximum inner product search (MIPS)
    problem using, for example, an approximate nearest neighbor (ANN) search or an inverted index-based
    top-$k$ retrieval algorithm where closeness is determined by lexical matching scores.
    The \emph{ranking} algorithm may be as simple as an identity function (e.g., in deep learning-based ``dense retrieval'')
    or a complex learnt function such as decision forests or deep learning models.}
    \label{fig:01:retrieval-and-ranking}
\end{figure}

This transformation of the document ranking problem culminated in a framework
that comprises of two distinct algorithms, depicted in Figure~\ref{fig:01:retrieval-and-ranking}:
\emph{top-$k$ retrieval}, which finds a \emph{subset}
of $k$ documents that are more relevant to a query, followed by \emph{ranking} which
orders the documents in the top-$k$ set. In \ac{ltr}, the ranking stage uses an often expensive function
that was trained using supervised or online learning methods, while the retrieval algorithm
solves a form of the maximum inner product search (MIPS) problem. As we will describe later, in ``dense retrieval,''
retrieval is often (but not always) an approximate nearest neighbor search while ranking is the identity function.

\section{The importance of efficiency}
Any solution that addresses the ranking problem, including \ac{ltr},
by definition seeks to maximize a user satisfaction metric, $Q$.
But in many real-world applications achieving the highest \textbf{effectiveness}
is only one of many requirements. We may indeed desire to impose additional
constraints on the ranked list, such as a requirement that ranked
lists fairly represent underrepresented categories; that they guarantee
privacy when the set $D$ consists of documents private to a user;
or that they counter biases and ensure trust.
Each of these additional constraints is an important
objective to optimize in its own right.

An objective that is equally as important as effectiveness in many applications
is the \textbf{efficiency} of the retrieval and ranking systems. For example, it is often imperative
to find the right documents and finalize a ranked list within a small time budget to meet demand and ensure
a timely delivery of information. In fact, a perfectly-ordered ranked list may
be of little value or have a low perceived quality if delivered too late or with substantial
delay.\footnote{
\citet{kohavi2013online}, reporting on an experiment conducted at
Bing, a web search engine, estimated that \emph{``every 100msec improves revenue by 0.6\%.''}}

The question of efficiency gained increasing significance with the rise of \ac{ltr}
whose training and serving require large amounts of computational power.
Indeed, the success of LambdaMART and subsequent decision forest-based
descendents~\citep{ganjisaffar2011bagging,Dato:2016:FRA:3001595.2987380,bruch2021xendcg,lucchese2018selective}
in improving the quality of rankings came at the expense of the efficiency of training and inference.
The training of such models is expensive because we must often (and repeatedly) learn ensembles of hundreds
to thousands of deep decision trees sequentially with gradient boosting~\citep{Friedman01},
with each node in every tree requiring a search in the feature space~\citep{CART}.
To become accurate, these large models need to be trained on vast amounts of data,
often represented as complex features that are in turn costly to compute.
Inference, too, is computationally intensive
because estimating the relevance of a single document to a query requires the traversal of paths,
from roots to leaves, of every decision tree in the model.

\section{Efficiency considerations beyond latency}

A decade later, deep neural networks, and in particular, Transformer-based~\citep{vaswani2017attention}
pre-trained language models advanced the state-of-the-art in ranking
dramatically~\citep{lin2021pretrained,nogueira2020passage,nogueira2019multi,nogueira2020monot5}.
Learnt representations of queries and documents by deep networks, too, offer
a range of opportunities including the development of a new generation
of ``dense'' retrieval methods~\citep{karpukhin-etal-2020-dense,xiong2021approximate},
document expansion techniques~\citep{nogueira2019document}, and others.
These recent developments mark the beginning of a new era known as
Neural Information Retrieval (NIR).

NIR is a leap forward, reaching new highs in quality.
Whatever the reason behind its success may be, NIR achieves a greater
effectiveness than the previous wave of machine learning models like
decision forests on many information retrieval tasks, but with orders of magnitude more
learnable parameters and much greater amounts of data. The new scale
drastically increases the computational and economic costs of model
training and inference. GPT-3~\citep{brown2020language}, for example,
required $285$,$000$ CPU cores and $10$,$000$ GPUs to train, with an
estimated economic cost of $\$4$.$6$M.\footnote{\url{https://lambdalabs.com/blog/demystifying-gpt-3/}}
Although it may be argued that the high cost of training deep models is amortized
because large language models can, through a process known as ``fine-tuning,''
be recycled and reused for a variety of applications with a substantially smaller effort,
it is still a significant price to pay upfront. Furthermore, not all
large neural models can be easily recycled---in fact, that is one of the 
properties~\citet{scells2022sigir-green-ir} call out in their article.
What is more, once trained, the use of such large models in production similarly requires a
nontrivial amount of tensor multiplications and other complex operations.

Due to their alarming computational requirements, NIR models
underline several dimensions of efficiency that have thus far been less
obvious. Crucially, ``efficiency'' is no longer characterized by
low latency, but is instead a concept that amalgamates space-,
sample-, and energy-efficiency, among other emerging factors, as summarized in
Table~\ref{tab:01:efficiency-taxonomy}.

In other words, the inefficiency of
an algorithm cannot and should not be understood solely in terms of negative user experience
due to greater latencies, but instead, we must acknowledge that ineffciency has adverse implications
for resource-constrained researchers and practitioners, and more importantly,
for the environment (in the form of emissions and carbon
footprint)~\citep{scells2022sigir-green-ir,strubell-etal-2019-energy,xu2021green-dl}.
We must therefore acknowledge that, due to environmental factors,
attempting to address the efficiency problem by relying on advances in
hardware systems or by utilizing more resources is not a sustainable long-term solution.
Instead, combating this multi-faceted issue of efficiency necessitates
a careful study and design of efficient algorithms and data structures,
as highlighted by deliberations at recent academic workshops (e.g., the
Workshop on Reaching Efficiency in Neural Information Retrieval~\citep{bruch2022reneuir,bruch2022reneuir-report}).

\begin{table}[t]
    \caption{Taxonomy of a multi-faceted view of ranking efficiency and the stages in which they manifest.}
    \label{tab:01:efficiency-taxonomy}
    \centering
    \begin{tabular}{lp{6.4cm}p{1.6cm}}
    \toprule
    \textsc{Dimension} & \textsc{Definition} & \textsc{Scope} \\
    \midrule
    \textsc{Query} & Time elapsed between the arrival of a query and the presentation of ranked list of documents & Inference \\
    \textsc{Sample} & Number of training examples required to learn a ranking function & Training \\
    \textsc{Space} & Total storage used to serve a ranking model & Training; Inference\\
    \textsc{Training} & Time required to train a ranking model & Training \\
    \textsc{Energy} & Amount of energy required to train a model or evaluate a learnt model on a query-document pair & Training; Inference\\
    \bottomrule
    \end{tabular}
\end{table}

\section{Efficient and effective ranking}

Accuracy by way of ever-increasing complexity presents a challenge:
how do we then optimize for both effectiveness and efficiency? Must we lose
accuracy to find a more efficient solution, inevitably trading off effectiveness for
efficiency and vice versa? These and other similar questions give rise
to a research topic that extends the document ranking problem as follows:

\begin{definition}[The Efficient Document Ranking Problem]
  Given a query $q$ and a set of documents $D$, the goal is to
  order elements of $D$ \emph{efficiently} such that the resulting
  ranked list maximizes a user satisfaction metric $Q$.
\end{definition}

The problem above spawned a line of research in the information retrieval
community to systematically investigate questions of efficiency and explore
the trade-offs between efficiency and effectiveness in ranking models,
leading to several innovations. The community widely adopted multi-stage, \emph{cascade} rankers,
separating light-weight ranking on large sets of documents from costly re-ranking
of top candidates to speed up inference at the expense of
quality~\citep{Wang:2011:CRM:2009916.2009934,asadi2013efficiency,ecir13,culpepper2016dynamic,mackenzie2018query,liu2017cascade,asadi2013phd}.
From probabilistic data structures~\citep{asadi2012bloom,asadi2013bloom},
to cost-aware training and \emph{post hoc} pruning of decision
forests~\citep{asadi2013training,Lucchese:2017:XBD:3077136.3080725,Lucchese:2016:POT:2911451.2914763,Dato:2016:FRA:3001595.2987380},
to early-exit strategies and fast inference algorithms~\citep{cambazoglu10early,LinTKDE,Lucchese:2016:ECS:2911451.2914758,SIGIR2015},
the information retrieval community thoroughly considered the practicality and
scalability of complex ranking algorithms.

In addition to volumes of publications,
the output of this research effort included standardized algorithms and reusable
software packages~\citep{lightgbm,SIGIR2015}. Perhaps more crucially,
the community developed an understanding that quality is not the be-all and
end-all of information retrieval research and that model complexity must be managed (through
more efficient training and inference) and justified (e.g., by contextualizing
quality gains in terms of the amount of computational resources required).

As complex neural network-based models come to dominate the research on document ranking,
it is unsurprising that there is renewed interest in the question above,
not just in the information retrieval community but also in related branches such as natural
language processing. Interestingly, many of the proposals put forward to date
to contain efficiency are reincarnations of past ideas, such as stage-wise ranking
with BERT-based models~\citep{nogueira2019multi,matsubara2020multistage},
early-exit strategies in Transformers~\citep{Soldaini2020TheCT,xin-etal-2020-deebert,xin-etal-2021-berxit},
neural connection pruning~\citep{gordon-etal-2020-compressing,mccarley2021structured,lin-etal-2020-pruning,liu-etal-2021-ebert}, precomputation of representations \citep{10.1145/3397271.3401093}, and enhancing indexes~\citep{zhuang2022reneuir,nogueira2019document,mallia2022sigir,lassance2022sigir}.
Other novel but general ideas such as knowledge
distillation~\citep{jiao-etal-2020-tinybert,sanh2020distilbert,gao2020distillation}
have also proved effective in reducing the size of deep models. Yet other innovative
ideas developed specifically for ranking include efforts to reinvent Transformers
from the ground-up~\citep{mitra2021conformer,Hofstaetter2020_sigir}.

\section{About this monograph}

Given the resurgence of the question of efficiency and the trade-offs between
efficiency and effectiveness in ranking, and the apparent overlap between
the neural and pre-neural ideas to address this question, we believe it is necessary
to present a comprehensive review of this literature with a particular focus on
the document ranking problem. We have thus prepared this monograph in four parts in
the hope that it serves as one such resource.

The first part introduces the document ranking problem and reviews a machine learning
formulation of it in the context of web search in depth. We also describe the architecture of
a modern search engine to illustrate an application of ranking that is of primary interest to this work.
As we explain the ingredients of a search engine and all that is involved
in the training and serving of a ranking model within this framework,
we highlight the costs to efficiency and call out the levers that trade off effectiveness for efficiency.

While the first part of this monograph concerns an abstract, general setup,
the two subsequent parts get more specific and examine two popular
families of ranking algorithms through the lens of efficiency.
One presents a treatment of a branch of \ac{ltr} that is based
on forests of decision trees, while another turns to neural networks and deep learning
methods for retrieval and ranking. Each family presents its own unique challenges and requires its own
set of solutions to explore the Pareto front on the efficiency-effectiveness optimization
landscape.

As the reader will notice, the approaches developed for the two families of ranking algorithms
appear to be---and in many ways, are---independent. But the ideas behind them overlap too.
We attempt, in the last part of the monograph, to identify the common threads that can help translate
ideas from one space to another. We also discuss emerging research directions,
made urgent by the rise of deep neural networks in information retrieval,
and explore open challenges within this space.

\chapter[Learning to Rank: A Machine Learning Formulation of Ranking]{Learning to Rank:\\A Machine Learning Formulation of Ranking}
\label{chap:supervised}

A ranking algorithm, at its core, is a function of a set of documents and a query.
It is no surprise then that its simplicity or complexity is determined by how we represent documents and queries.

Let us, for example, strip away grammar and sentence structure from a text query or document.
That leaves us with a bag-of-words perspective of the text: a query or document is simply a
multiset of terms from a fixed vocabulary. Modeled this way, we can represent
documents and queries na\"ively as vectors in a space that
has as many dimensions as there are terms in our vocabulary, and where each dimension
records the frequency of the corresponding term in that document or query.
Perhaps we would further weight each dimension to reflect its
``importance''~\citep{sparck1972statistical}---an article like ``the'' that
occurs frequently in a large subset of documents but that carries little information
would have a lower weight.

\begin{figure}[t]
    \centering
    \includegraphics[width=\linewidth]{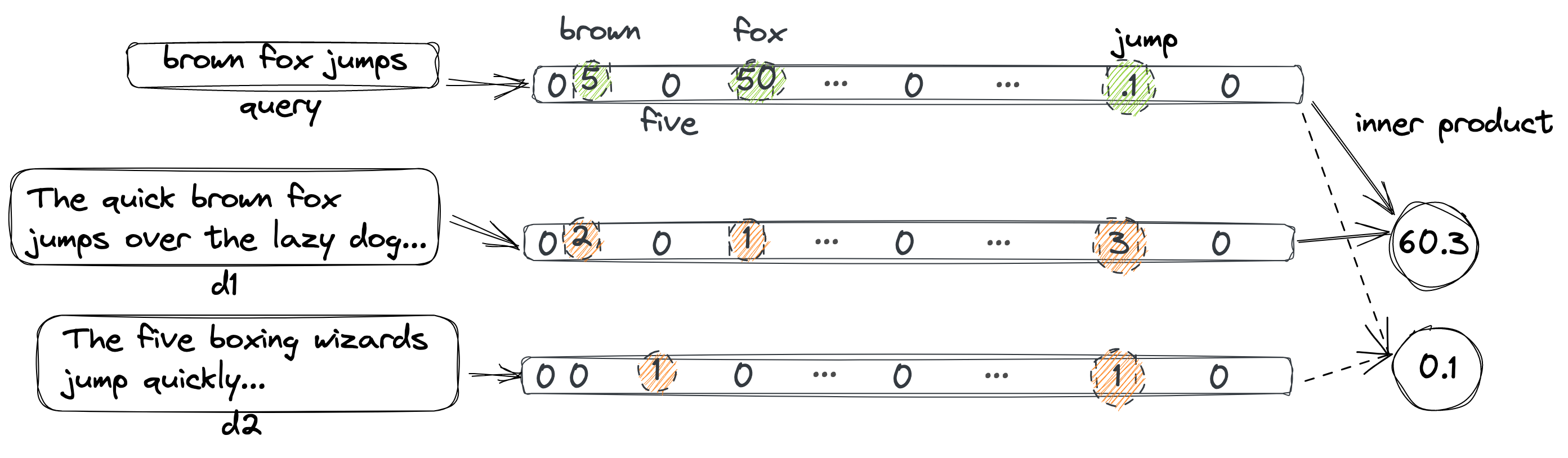}
    \caption{Vector representation of query and documents in the bag-of-words model.
    The relevance of a document to a query may be estimated using the inner product of their vectors.}
    \label{fig:02:bow-inner-product}
\end{figure}

In the vector space construction above we can measure the relevance of a document to a query
using a similarity measure between query and document vectors~\citep{salton1988term},
such as cosine similarity or inner product, as illustrated in Figure~\ref{fig:02:bow-inner-product}.\footnote{
If this reminds the reader who is familiar
with neural information retrieval of neural rankers, it is because in both models queries
and documents are represented as vectors. While similar in principle, in the bag-of-words model, these vectors are sparse
vectors of basic statistics such as (weighted) term frequencies, while a neural ranking function \emph{learns} a dense or sparse
vector representation of its input from the raw data.}
Alternatively, because each vector is a distribution over
a vocabulary, we may use a probabilistic approach
based on language models~\citep{ponte1998language} to measure query-document similarity (e.g.,
the likelihood of observing a query given a document).
In either case, the ranking function is simple: it computes the similarity scores for pairs
of query and document vectors and sorts the documents in decreasing order of similarity.

Term frequencies in a bag-of-words model only carry
so much information, and there is indeed far richer signals
available in queries and documents to aid ranking.
Term co-occurrences and increasingly more advanced semantic features provide
greater insight than individual words, for example.
Another source of useful features, especially in the context of
web search, is the structure within documents, where ``fields''
such as titles and sections carry different
weights~\citep{jones2000probabilistic,robertson2004simple}. Continuing with
the web search example, there are numerous other indicators of
relevance in the web graph (such as the anchor text of
incoming links, in-degree and out-degree of a web page,
number of references in social network streams) and in
the user interaction with the search system (e.g., clicks,
user sessions, effect of query reformulation).

The list of statistics used in modern search engines
is indeed long, encompassing many facets of documents and queries beyond term frequencies.
Benchmark ranking datasets, for example, represent queries and documents with hundreds or thousands
of statistics. As examples, there are 136 features for a query-document pair in
the MSLR datasets,\footnote{Available at \url{https://www.microsoft.com/en-us/research/project/mslr/}.}
and 700 in the Yahoo! Learning to Rank Challenge datasets.\footnote{Available at \url{https://webscope.sandbox.yahoo.com/catalog.php}.}

\begin{figure}
    \centering
    \includegraphics[width=\linewidth]{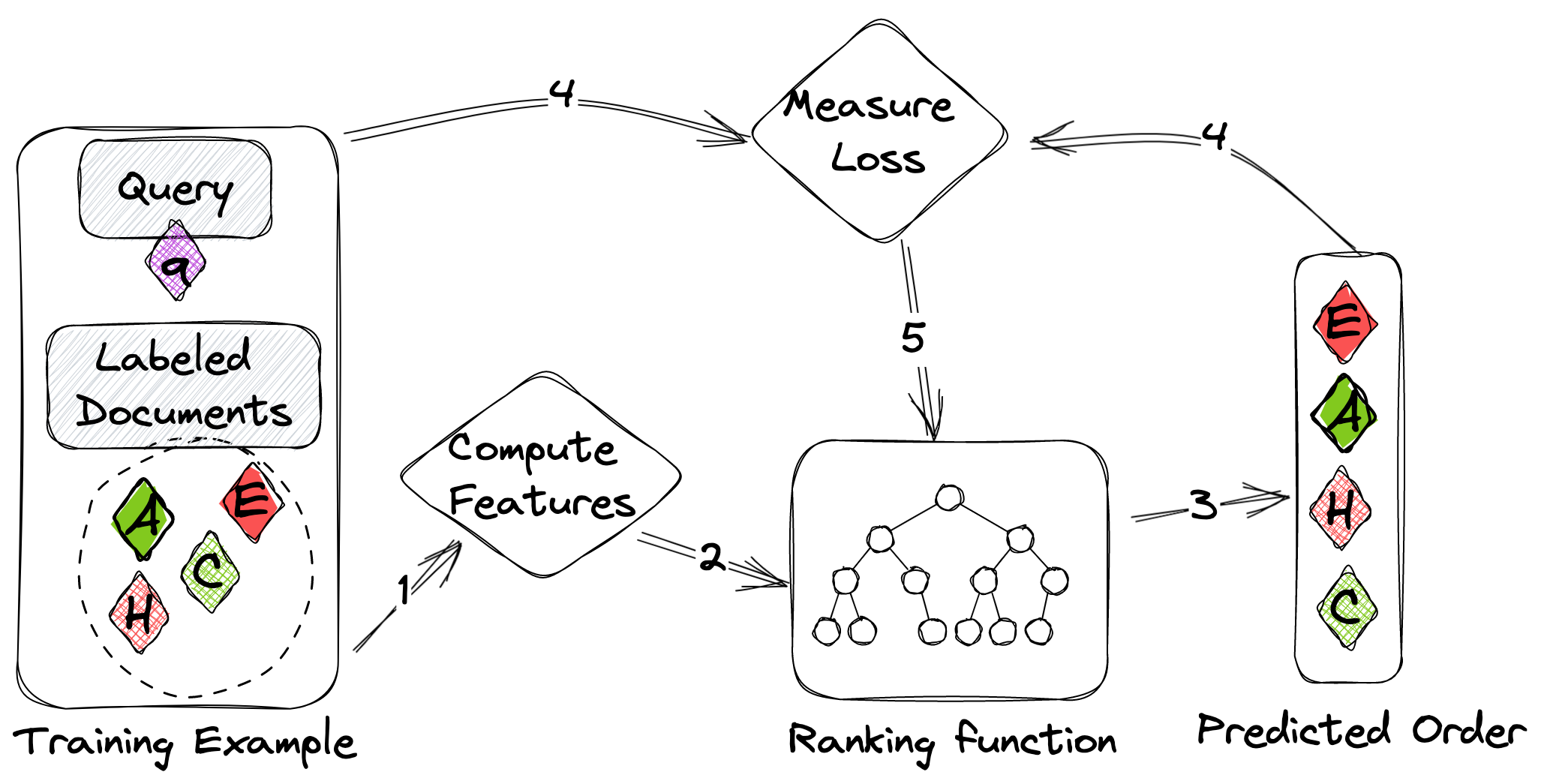}
    \caption{Visualization of a supervised framework for learning a ranking function.
    A training example in ranking consists, at least, of a query and a set of (partially) labeled documents.
    Each example is represented by a vector of features, either learnt using a deep learning model or
    engineered based on statistics. A ranking function is then trained to produce ranked lists so as to
    minimize the difference between the produced ranked list and the ideal one.}
    \label{fig:02:ltr}
\end{figure}

As the size and complexity of the query and document representations grow,
it becomes impractical to hand-craft a ranking function and
ineffective to use basic similarity measures. 
It is rather more practical and effective to view the ranking problem as a supervised learning task---a
task known as \ac{ltr}, visualized in Figure~\ref{fig:02:ltr}.
We should highlight that, other texts often use \ac{ltr} to refer specifically to the first
wave of machine learning-based methods that came before deep learning. In this monograph,
we view neural networks and deep learning as just another hypothesis class in the \ac{ltr}
framework and use \ac{ltr} to discuss both pre- and post-neural research in ranking in a unified framework.

Equipped with a rich representation of queries and documents as vectors of feature values, and
a set of labels indicating the relevance of a document to a query, we have one piece of the puzzle to learn a ranking
function. To complete the picture, we also need a method of evaluation
 to determine the success of the learnt ranking function, and an optimization objective to enable learning.
In this chapter, we unpack each of these components in the context of ranking.

\section{Ranking datasets}
To understand the structure of a ranking dataset, we find it helpful to
first revisit other tasks in the family of supervised learning algorithms
which have a simpler construction.
Take regression and classification as its well-known members.
A typical regression or classification dataset is made up of a number of
examples that, respectively, come with a value and a class label.
An example could be an image to be classified as a cat or a dog, or
a piece of text to be decided as relaying a positive or negative sentiment.
Each example is assumed to have been sampled independently of others,
but from the same underlying data distribution. Examples are either given
as vectors of feature values---hence, tabular or structured---or represented
in their raw, unstructured form such as text, or a mix of both.

Ranking adds a new dimension to the construction above, with an example
now comprising of two parts: A query and a \emph{set} of documents.\footnote{Note that,
while ``query'' and ``document'' may suggest textual content,
these notions, in fact, extend to a variety of multi-modal
domains such as a recommendation engine where a query may be a user
``profile'' and documents are movies.}
The objective of a ranking algorithm is
to sort the set of documents with respect to the query in some
order. In web search, that order is often determined by relevance,
but other factors such as diversity, fairness, freshness, personalization,
or a combination of those may play a role.

The additional dependence of a ranking example on query and document affects how we
represent and label a ranking dataset. The next two sections explain its implications.

\subsection{Representation}

How a query and documents are represented is largely similar
to the regression or classification setting, with tabular features summarizing
examples or unstructured data from which latent features may be learnt. The one
difference in ranking, however, is that in addition to extracting features from
the query, and another set from individual documents, we must also obtain signals that
jointly describe a query-document pair.

Consider the query alone. We may classify the query and record its type
(e.g., news vs.\ health) or categorize a user
session~\citep{bennett2010classification,jiang2016query,lucchese2013discovering}.
These signals allow the ranking model to better determine the user intent and
match the query to a document from the appropriate class, and to adaptively use the available features (e.g.,
recency should be considered as a more discriminative signal if we are searching
for news articles rather than historical facts).

Now consider the document in isolation. There are important signals in the structure of a web document,
such as the number of incoming links, or the PageRank value, its quality or spamminess, or other more advanced
link analysis features~\citep{henzinger2000link}. The MSLR dataset, for example, includes some
rather surprising document features such as the number of slashes in a document's URL, click count,
and dwell time, capturing aspects of user interaction and experience!

Now consider the numerous signals that may describe or capture the relevance of a document to a query.
The MSLR dataset, for example, uses a vector of 136 real-valued features for each query-document pair.
These features include signals from a vector space model (such as the weighted sum of term frequencies
of query terms in that document~\citep{salton1988term}), as well as the expected relevance according to
different language models~\citep{ponte1998language}. Another class of features that can be computed
jointly for a query-document pair is term proximity~\citep{rasolofo2003term}: the terms of a query with
multiple terms are likely to appear within a short span in relevant documents.
It is common to compute such content-based features not just from the body of a document,
but also from its other parts such as its URL, title, and anchors, resulting in a fine-grained
feature set.

Table~\ref{table:representation:features} summarizes some of the most common features used to represent a query-document pair, and categorize by whether they model a document, a query, a user or their combination. We refer the interested reader to an analysis of these features by \citet{macdonald2012usefulness}.

\begin{table}
\caption{Typical features to represent a query-document pair.}
\label{table:representation:features}
\small
\begin{tabular}{p{50pt}p{130pt}p{110pt}}
\toprule
\textsc{Category} & \textsc{Description} &  \textsc{Feature}\\
\midrule
Document    & Properties of body and URL (MSLR datasets) & Number of slashes in URL; length of URL; page length \\
Document    & Properties of the page as a node in the Web graph (MSLR datasets) & Inlink count; outlink count; PageRank \\
Document & Document quality & Spam score \citep{cormack2011efficient}; fraction of stop-words; fraction of visible terms \citep{quality-biased-ranking} \\
\midrule
Query & Query classification \citep{bennett2010classification,jiang2016query} & Query topic; query intent \\
\midrule
Query-Document  &  Properties of query-document match (MSLR datasets) & Boolean model; TF-IDF; BM25; vector space model; language model \\
Query-Document & Proximity-sensitive matching \citep{rasolofo2003term} & Term proximity  \\
\midrule
User-Document & Properties of user interaction with the page (MSLR datasets) & URL click count; dwell time  \\
\midrule
User-Query-Document & Properties of the user interaction with the page in response to a query (MSLR datasets) & Query-URL click count  \\

\bottomrule
\end{tabular}
\end{table}

\subsection{Relevance labels}

Viewed as a simple yes or no question---is this and only this ranked list
correct?---the ranking problem reduces to a classification one,
where a single binary label suffices for each example.
But the ranking question is seldom this coarse and
it is unlikely there is just a single correct ranking.
Instead, for a single query, many different ranked lists may equally make sense.
For instance, it hardly matters how documents that are irrelevant to a query
are ordered relative to each other, so long as they appear below the more relevant ones.

In fact, in most applications of ranking, labels are defined at the
granularity of query-document pairs: Is this document relevant to that query?
If so, what is its degree of relevance? In most practical settings where
document collections are vast and query possibilities endless, labels
may only be defined for a fraction of documents. Documents that are examined
and subsequently labeled too are typically not drawn uniformly randomly from
the collection. In these ways, labels in a ranking dataset are different from
those in classification datasets.

We distinguish between two different methods for labeling document with relevance labels.
The first approach is to manually assign labels. It is well known that major web search engines
have been using thousands of quality raters for this purpose \citep{the_guardian_2017}.
And in this case very accurate guidelines for raters exists~\citep{gomes_2017}
so that the labeling process is accurate, consistent, and reliable.
For instances, it might be preferable to collect preference judgments among document pairs~\citep{carterette2008here}
(i.e., is document $A$ more relevant to the query $q$ than document $B$?).

Manually labeling collections of query-document pairs in a na\"ive manner
clearly does not scale in terms of space (number of documents) or time (new queries and new documents come in every day)
and can therefore be cost-prohibitive.
To scale this effort, researchers have developed other methods of collecting feedback
such as by using active learning to minimize the number of annotations but still maximize the
quality of the training data \citep{long2010active}. Interestingly, \citet{yilmaz2009deep} also
show that, for the purposes of training a ranking model,
it is more advantageous to collect a larger dataset of \emph{shallow}
judgements (i.e., fewer annotated documents per query) as opposed to a smaller dataset of
\emph{deep} judgments (i.e., with a large number of annotations per query).

An alternative to (targeted) manual labeling, one can look to a different source of information
to deduce relevance labels. One important source is the \emph{implicit} feedback generated by
the users of a ranking system. Consider, for instance, the number of clicks received by a
document in response to a query, or even the absence of any click, or the editing and reformulation
of a query by a user after an unsatisfactory search results list, among other
signals~\citep{joachims2005accurately,joachims2002optimizing,radlinski2005query}.
All of those actions can be mined for and translated into relevance labels!
Of course, this data is subject to different biases: users are more likely to click on the top document in the result list;
users are unlikely to traverse multiple result pages; and the system may not be able
to return the most relevant result for a query anyway.
But this noisiness and these biases can be modeled using ``click'' models~\citep{chuklin2015click}
and treated counterfactually~\citep{oosterhuis-www2020-counterfactual,joachims2017unbiased}
to enable unbiased learning and evaluation of ranking models.
We refer the reader to the vast literature on these topics for details.

\subsection{Notation}

Let us introduce some notation to summarize the discussion in this section and formally define a typical \ac{ltr} dataset.
Such a dataset $\data$ comprises of a set of triplets $( q, \doclist, \lablist)$.
The vector\footnote{Throughout this monograph, we denote vectors as lowercase letters in bold.}
$\doclist$ includes the documents that we want to rank in response to the query $q$.
For a given query-document pair $(q, x_i)$, where $x_i$ is a member of $\doclist$, the true relevance
of $x_i$ with respect to $q$ is encoded by $y_i$ in the vector $\lablist$ of relevance labels.

Each $(q, x_i)$ is represented in some feature space $\feats$ that captures properties of the query (e.g., its likelihood, category), of the document (e.g., its incoming links), and of their relationship (e.g., the number of occurrences of the query in the document).
We denote by $\labels$ the set of possible labels. This is typically \emph{binary} indicating relevant vs.\ non-relevant, or \emph{graded} where $\labels$ is restricted to a small set of integers such as $\labels=[4] \triangleq \{0,1,2,3,4\}$, encoding different degrees of relevance with larger grades corresponding with stronger relevance.

A \emph{ranker} $\ranker$ is a function that, given a pair $(q, \doclist) \in \feats^n$, produces a permutation vector $\bm{\pi} \in \mathbb{Z}^n$, $\pi_r \in [n]$, of the $n$ items in $\doclist$. Such a permutation defines the ranking produced by $\ranker$ as follows: the item ranked at position $r$ is the  $\pi_r$-th item in $\doclist$: $x_{\pi_r}$. An \emph{ideal} ranking $\bm{\pi}^*$ is one that sorts documents in decreasing order of their relevance labels, where $y_{\pi^*_r}\geq y_{\pi^*_{r+1}}$. Note that there might be multiple ideal rankings.

\section{Ranking metrics}

Earlier in our discussion, we touched on ways in which
labels in ranking are different from those in classification.
It is not surprising then that evaluation metrics that help us
assess the quality of a ranked list too are different from their
classification or regression counterparts.

Ranking metrics attempt to measure the utility of a ranked list to a user.
It is therefore helpful to consider the important factors in the way
users interact with a ranked list.
First, users expect the relative ordering of documents to be correct.
That is, documents that are placed higher
in the ranked list (i.e., towards the top of the list) should satisfy the information
needs of a user better than documents lower on the list, and that
as the user goes down the list, documents become less relevant to the query.
Second, user attention has a skewed distribution with much of it focused on the
top of the ranked list. In other words, users typically do not examine
all documents at every position with equal probability or care.
As such, higher positions carry more weight.

The information retrieval literature offers a great number of
metrics that are designed specifically on the basis of the factors above.
Most of these have the following additive form:
\begin{equation}
Q@k(\bm{\pi}, (q,\doclist,\lablist)) = 
\frac{1}{Z}
\sum_{1\leq r\leq k} \textsc{Gain}(r)\cdot \textsc{Discount}(r).
\label{equation:supervised:metric}
\end{equation}
At a high level, the additive nature of the formulation above reflects the view that
each document contributes to the overall quality $Q$ of a ranking $\bm{\pi}$ independently of others.
Typically, we only consider the top $k$ high-ranking documents when computing $Q$ and denote it by $Q@k$,
reflecting the assumption that user attention dissipates past position $k$.
Within this framework, a document at rank position $r$ provides a contribution of $\textsc{Gain}(r)$ to the metric,
which is typically a function of its label $y_{\pi_r}$. However, this contribution wanes
as $r$ grows, by a factor of $\textsc{Discount}(r)$, a decreasing function of $r$.
Note that, both $\textsc{Gain}(\cdot)$ and $\textsc{Discount}(\cdot)$
are typically formulated based on a model of user behavior or ``click'' models~\citep{chuklin2015click}.
The constant $Z$ is often used as a normalization factor to ensure that the
metric $Q$ lies within the unit interval. Finally, given a test ranking dataset of
examples $(q, \doclist, \lablist)$ and their corresponding ranked lists $\bm{\pi}_q$,
we compute $Q@k$ for each example and report its mean as the average quality.

Specific instances of Equation~(\ref{equation:supervised:metric}) differ in how they define $\textsc{Gain}$ and $\textsc{Discount}$.
We review a few metrics that are commonly used in the ranking literature in this section,
but encourage the reader to refer to \citep{Liu08} for a more thorough treatment.

As an example, consider Rank-Biased Precision ($\rbp$) by \citet{moffat2008rank}.
The authors define $\textsc{Gain}(r) = y_{\pi_r}$, but to formulate $\textsc{Discount}$,
they make the assumption that, at any given point, a user inspects the next document on the list
with probability $p$ and abandons the ranked list altogether with probability $1-p$. On that basis, they take
the probability that a user reaches rank $r$ as the discount factor: $\textsc{Discount}(r) = p^{r-1}$.
The normalization constant $Z$ is $1-p$, the inverse of the average number of documents
that a user inspects.

As another example for graded relevance, consider one of the most popular metrics known as
Normalized Discounted Cumulative Gain ($\ndcg$) \citep{jarvelin2000ir}.
The gain is computed as $\textsc{Gain}(r) = 2^{y_{\pi_r}} - 1$, leading to a dynamic where a document with label $4$
is about twice as important as a document with label $3$. The discount is computed as $\textsc{Discount}(r)=\frac{1}{\log_2(r+1)}$.
When $Z=1$, the resulting metric is called Discounted Cumulative Gain ($\dcg$).
To compute $\ndcg$, however, we normalize by the \emph{ideal} $\dcg$ by setting 
$Z=\dcg@K(\bm{\pi}^*, (q,\bm{x},\bm{y}))$.

Both $\rbp$ and $\ndcg$ assume that the user examines the next document on the ranked list
independently of the relevance of the documents observed along the way. In a more practical click model, however,
users are likely to stop their inspection of the remainder of a ranked list soon after they find the document
that satisfies their information need. With the goal of capturing such a behavior,
\citet{chapelle2009expected} propose the Expected Reciprocal Rank ($\err$).
The gain there is defined as $\textsc{Gain}(r) = p_r\prod_{i=1}^{r-1}(1-p_i)$,
where $p_i$ is the probability that the user is satisfied with the document at rank $i$ and stops inspecting the rest of the list.
This probability is generally a function of relevance, for example, $p_i = \frac{2^{y_{\pi_i}}-1}{2^{\max \labels}}$.
The gain of the document at rank $r$ is thus the probability that the user finds the first satisfactory document
at that position, having judged as non-relevant all the preceding documents.
The discounting mechanism is simply $\textsc{Discount}(r)=\frac{1}{r}$, and $Z=1$ as no normalization is necessary.
Experiments show that $\err$ correlates better with user satisfaction \citep{chapelle2009expected}.

\begin{table}
\caption{Common evaluation metrics for ranked lists}
\label{table:supervised:metrics}
\small
\begin{tabular}{llc}
\toprule
\textsc{Relevance} & \textsc{Name} & \textsc{Metric as a function of} $\bm{\pi} \textsc{ and } (q,\doclist,\lablist)$ \\
\midrule
Binary
  & \map
  & $\frac{1}{\sum_r y_{\pi_r}} \sum_{\substack{1\leq r \leq n,\\ y_{\pi_r}=1}} \precision@r $, $\precision@r = \frac{1}{r}\sum_{1 \leq i \leq r} y_{\pi_i}$ \\ \\
\midrule
\multirow{7}{*}{Graded}
  & $\rbp$
  & $(1-p) \cdot \sum y_{\pi_r} \cdot p^{r-1}$ \\ \\
  & $\dcg@k$
  & $\sum_{1\leq r\leq k} \frac{2^{y_{\pi_r}} - 1}{\log_2(r+1)}$ \\ \\
  & $\ndcg@k$
  & $\dcg@k(\bm{\pi}, (q,\bm{x},\bm{y})) / \dcg@k(\bm{\pi}^*, (q,\bm{x},\bm{y}))$ \\ \\
  & $\err@k$
  & $\sum_{1\leq r\leq k}  p_r\prod_{i=1}^{r-1}(1-p_i)\cdot \frac{1}{r}$,
  $p_i = \frac{2^{y_{\pi_i}}-1}{2^{\max \labels}}$ \\
\bottomrule
\end{tabular}
\end{table}

The metrics we have reviewed thus far are based on Equation~(\ref{equation:supervised:metric}),
but not all ranking metrics belong to this family. One such example is Mean Average Precision ($\map$) \citep{Buckley:2005}
for binary relevance. Let us parse this metric one term at a time. Let Precision at $k$, denoted by $\precision@K$,
be the fraction of relevant documents among the top $k$ documents. Then define Average Precision ($\avgp$) as follows: 
$$
\avgp(\bm{\pi}, (q,\bm{x},\bm{y})) = \frac{1}{\sum_r y_{\pi_r}} 
\sum_{\substack{1\leq r \leq n,\\ y_{\pi_r}=1}} \precision@r.
$$
$\map$ is the mean of this value computed over all queries in a dataset.

Table \ref{table:supervised:metrics} summarizes the metrics we have reviewed.
We conclude by highlighting that measures such as $\ndcg$ and $\err$ are very difficult to optimize
and that changes that may appear small have a significant impact in practice. According to \citep{chapelle2012large},
the differences between major revisions of Bing, ``involve changes of over half a percentage point, in absolute terms, of $\map$ and $\ndcg$.''

\section{Learning objectives}

We have just seen what factors are good indicators of the quality of a ranked list
and how ranking metrics evolved to take those factors into consideration.
In this section, we review how we learn a ranker that produces high-quality ranked lists.

While we defined a ranker $\ranker$ to be a function that permutes documents $\doclist$
in response to a query $q$, in practice, $\ranker$ instead computes a \emph{relevance score} for
every query-document pair $(q, x_i)$ and subsequently sorts $x_i$'s in decreasing order of relevance
to produce a permutation. This two-step trick greatly simplifies the learning of a ranker $\ranker$,
which is also known as a \emph{scoring} function.

How do we learn such a scoring function given a labeled training dataset?
At a high level, it is natural to take a ranking metric $Q$ and learn an $\ranker$ that maximizes it,
with the intuition that a ranking function trained to maximize $Q$ should produce
high-quality ranked lists as measured by $Q$.

While the instinct to use a ranking metric as the learning objective may be natural,
whether that is sensible depends on the optimization method itself.
Consider, for example, gradient-based optimizers that are commonplace in machine
learning. For an objective to be optimized by such an optimizer, it must
have meaningful gradients. A ranking metric, being a function of discrete ranks,
does not offer gradients that are all that interesting:
small perturbations of relevance scores computed by a ranking function often do not
lead to a change in ranks, and as such, the gradients of a ranking metric with respect to
relevance scores are typically either zero or nonexistent due to discontinuities.

The popularity and effectiveness of gradient-based optimizers and
their unfortunate incompatibility with ranking metrics
bring us to an important research topic that offers a way to reconcile
the two: \emph{surrogate} objectives.
The idea is to devise or derive from ranking metrics an objective that is differentiable
and consistent. It must be differentiable so that its gradients can inform an optimizer
of the correct direction to follow. It must be consistent with a ranking metric so that,
in expectation, optimizing it leads to an optimal metric as well.

The \ac{ltr} literature has long sought and studied surrogate
objectives that are differentiable and, while not necessarily consistent,
exhibit a behavior that is intuitively in keeping with ranking metrics.
To help explain the differences between existing surrogate ranking objectives,
let us place them into one of three buckets based on their behavior:
pointwise, pairwise, and listwise.

\begin{figure}[t]
    \centering
    \centerline{
        \subfloat[Pointwise]{
        \includegraphics[width=0.5\linewidth]{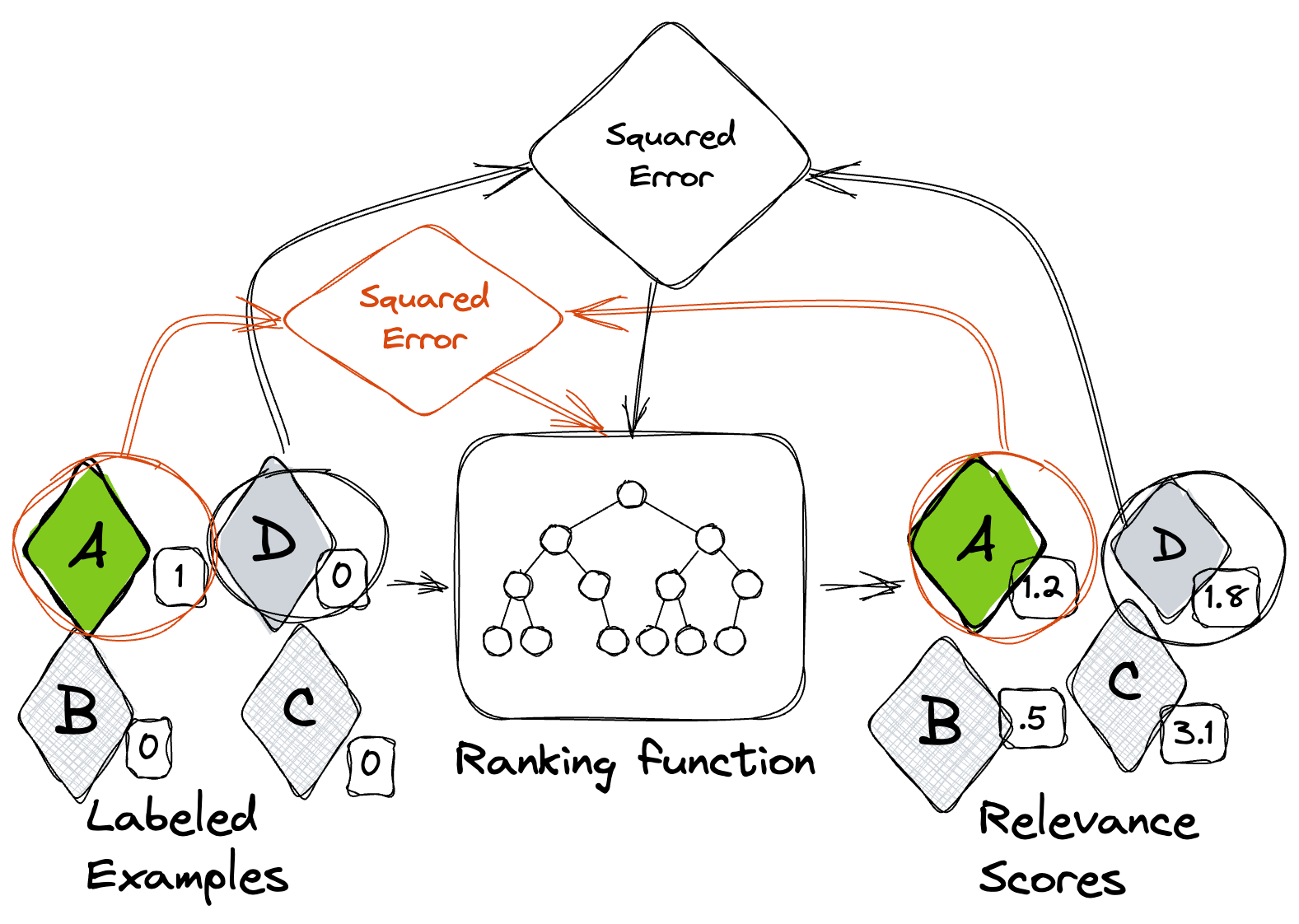}
        \label{fig:02:pointwise-pairwise-losses:pointwise}}
        \subfloat[Pairwise]{
        \includegraphics[width=0.5\linewidth]{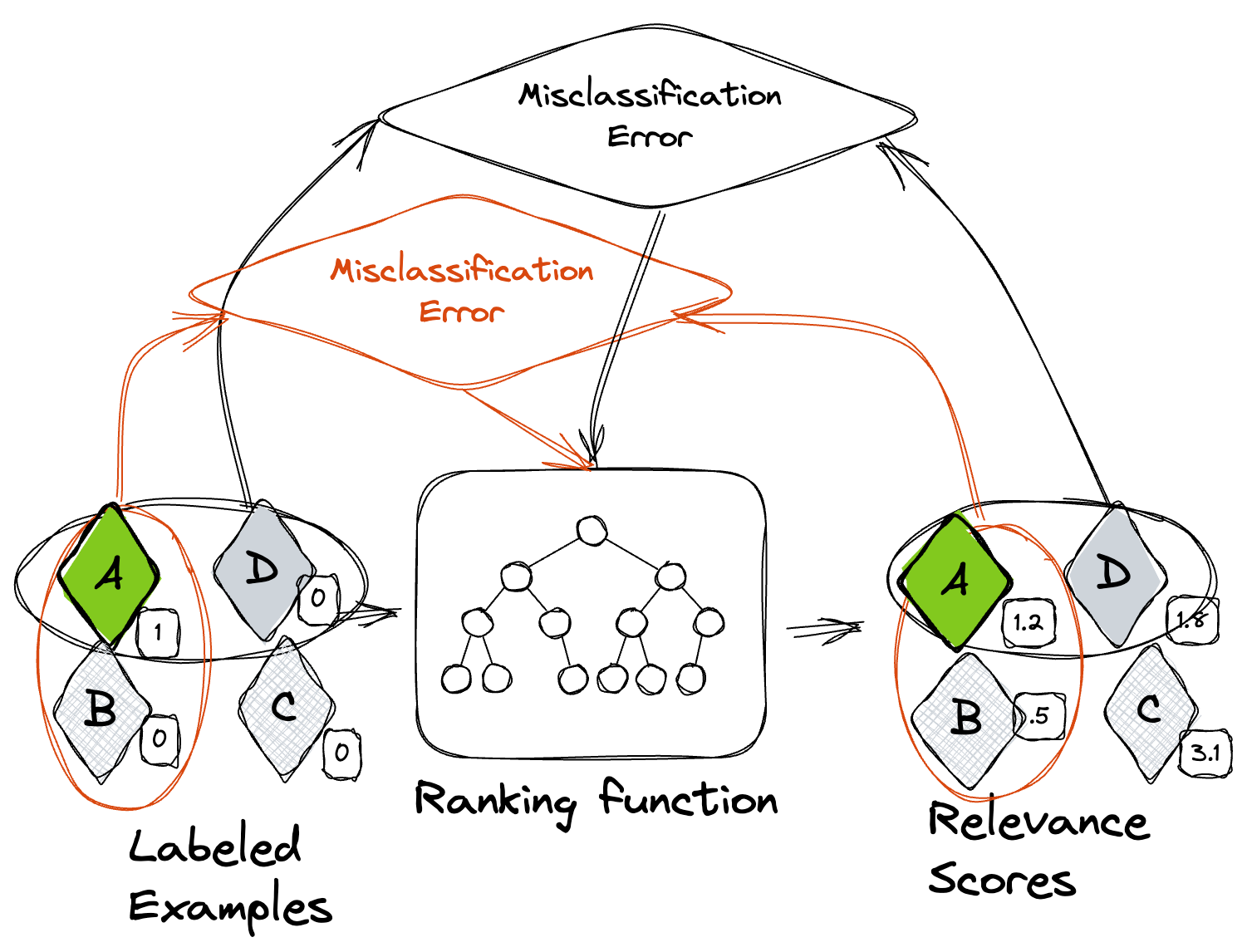}
        \label{fig:02:pointwise-pairwise-losses:pairwise}}
    }
    \caption{Illustration of the machinery of pointwise and pairwise ranking objectives for
    four documents ($\{A, B, C, D\}$) in the context of a single query, with numbers attached to each document
    indicating a relevance label or predicted score.
    In pointwise methods, (a), the predicted relevance score of every document is compared to its label.
    In pairwise methods, (b), the function is evaluated in terms of its accuracy in predicting the correct order
    among pairs of documents.}
    \label{fig:02:pointwise-pairwise-losses}
\end{figure}

The intuition behind pointwise methods is to reduce the ranking problem
to one of regression, multi-class classification, or ordinal regression~\citep{Liu08}.
In regression, for example, we may optimize the squared difference between
the true relevance label and predicted relevance score of a query-document pair
in expectation, known as the mean squared error, as illustrated in Figure~\subref*{fig:02:pointwise-pairwise-losses:pointwise}.
When cast this way, as noted earlier, the question becomes one of
predicting the degree of relevance of each document with respect to a given query
independently of others. Furthermore, this framing of the ranking problem
implicitly requires an absolutist view of relevance: a document is either
relevant or it is not.

We hinted in our earlier discussion that such a view is hardly appropriate in general.
For the vast majority of applications, a stronger view is to consider relevance as a
relative concept: a document is \emph{more} (or \emph{less}) relevant than another.
The next wave of surrogate ranking objectives reflect this paradigm shift.

Pairwise methods are closer to the relative definition of relevance in that
they model error as a function of not a single, isolated document, but of pairs
of documents: When sorted by their relevance scores,
does the resulting order between any pair of documents correctly reflect our preference
between them? This is illustrated in Figure~\subref*{fig:02:pointwise-pairwise-losses:pairwise}.
Presented this way, the question becomes
one of preference learning via binary classification and, as such, any classification objective
serves as a suitable surrogate. RankNet~\citep{burges2005learning},
Ranking-SVM~\citep{joachims2002optimizing}, and RankBoost~\citep{freund2003efficient}
offer examples of this approach.

To make the idea more concrete, consider RankNet, whose surrogate objective
was argued to correlate with NDCG~\citep{cao2007learning}. Given two documents
$x_i$ and $x_j$, it maps the difference between their relevance scores ($o_{ij}$)
to a probability using the logistic function: $P_{ij}=1/(1+e^{-o_{ij}})$.
This probability can be understood as the strength of the predicted order between
the pair. When we have computed these probabilities for every pair,
it is simply a matter of optimizing its cross entropy ($C_{ij}$) with the ground truth $\overline{P}_{ij}$,
which is $1$ if $x_i$ is more relevant than $x_j$ and $0$ otherwise:
$C_{ij}=-\overline{P}_{ij}\log(P_{ij})-(1-\overline{P}_{ij})\log(1-P_{ij})$.

In repeated experiments, pairwise methods have proven successful, particularly when compared
with their pointwise counterparts. The empirical success that ensued the shift above motivated
the research community to extend the idea of preference learning from a pair of documents to
an entire list of documents. In other words, similar to how ranking metrics quantify the quality
of an entire ranked list, we seek to quantify the ranking error in terms of the induced order among
a list of documents, not just between pairs. The surrogate objectives that have emerged
from this research effort are known collectively as the class of listwise methods.

\begin{figure}
    \centering
    \includegraphics[width=\linewidth]{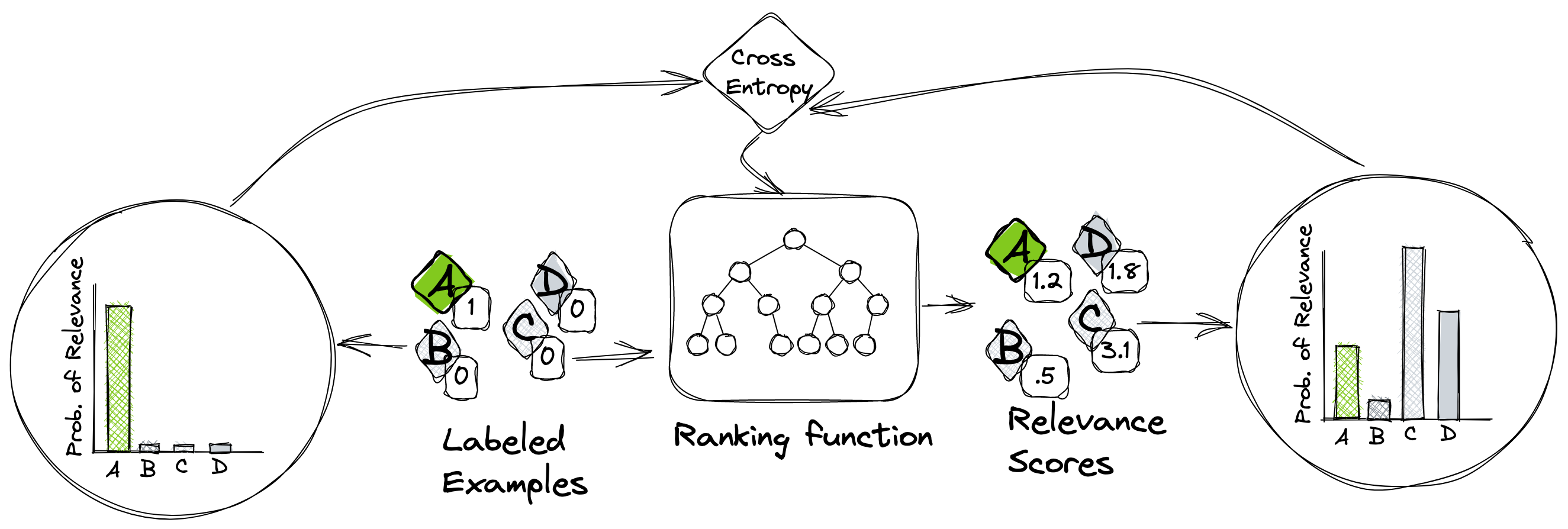}
    \caption{Illustration of the ListNet~\citep{cao2007learning} objective
    for a single query with four documents, with numbers attached to each document
    indicating a relevance label or predicted score. The labels and scores are separately
    projected onto the probability simplex to form a distribution over documents.
    Subsequently, the cross entropy between the two distributions is taken as a measure of how
    far the ranked lists are from each other.
    }
    \label{fig:02:listwise-losses}
\end{figure}

The \ac{ltr} literature contains a great number of listwise methods.
ListNet~\citep{cao2007learning} and ListMLE~\citep{xia2008listmle} take a probabilistic
approach by applying the Plackett-Luce model to estimate the probability of permutations.
This is illustrated in Figure~\ref{fig:02:listwise-losses}.
More interestingly, others like ApproxNDCG~\citep{qin2010general,bruch2019approxndcg}
or SoftRank~\citep{taylor2008softrank} derive smooth approximations to ranking metrics.
LambdaRank and LambdaMART~\citep{export:132652} extend the RankNet objective using a heuristic
where the contribution to the error from a pair of documents is shrunk or amplified by
a multiplicative factor that correlates with the amount of change in $\ndcg$ (or any other metric) if the
two documents traded ranks. A more recent work~\citep{bruch2019xendcg,bruch2021xendcg}
modifies the ListNet objective to improve its consistency.
\citet{oosterhuis2021plrank} proposed an approach to directly optimize the Plackett-Luce
model for ranking.

The listwise methods cited above are but a few representatives of a large class
of algorithms in the machine learning and information retrieval
literature~\citep{NEURIPS2021_b5200c61,NEURIPS2020_ec24a54d,NEURIPS2019_d8c24ca8,10.5555/3524938.3525027,10.1145/3477495.3531849}.
We return to these methods in later chapters and study some of them
in greater detail. But first, to complete the supervised learning formulation,
we must discuss hypothesis classes.

\section{Hypothesis classes}
Equipped with a training dataset, a ranking objective, and an optimizer,
we are ready to learn a ranker $\ranker$. There is, however, one final piece of
the supervised learning puzzle: What relationship do we hypothesize exists between
a relevance score and the features that represent a query-document pair?
In other words, what family of functions do we think $\ranker$ belongs to?

In its simplest form, $\ranker$ may be a linear function, parameterized by
a set of coefficients and a bias term that can be learnt from data so as to optimize
our objective. It is clear that such a function, on its own, does not model any
nonlinear relationship that may exist in the data, and, as such, is generally less
effective than other, more complex families of functions. However, its simplicity
facilitates formal and rigorous analysis and allows us to provide certain
guarantees on performance. That is why linear rankers are favored in the vast
literature on online \ac{ltr} with bandit algorithms~\citep{radlinski2008learning,yue2009interactively,yue2012k,hofmann2013reusing,hofmann2013balancing,kveton2015cascading}.

Another class of algorithms take $\ranker$ to be a \emph{decision forest}:
an ensemble of decision trees, typically with real-valued leaves.
At a high level, a decision tree is a piecewise constant function that is
learnt by recursively partitioning the feature space into disjoint spaces
and assigning a value to each partition. Learning many of these decision trees
and putting them together into a forest in an additive manner yields highly complex functions,
capable especially of modeling \emph{tabular} features.
Indeed, past studies have shown decision forest-based rankers to be highly
effective~\citep{ganjisaffar2011bagging,szummer2011semi,bruch2021xendcg},
among them, LambdaMART~\citep{export:132652} remains the state of the art.

In recent years, the success of neural networks and deep learning in related
areas of research has led to a rise in deep learning methods for ranking,
where $\ranker$ is taken to be an often complex neural network with specialized
modules for processing textual data. The effectiveness of pioneering methods such as
ConvDNN~\citep{SIGIR2015ltr-shorttext}, DSSM~\citep{huang2013learning},
and others~\citep{mitra2016dual,mitra2017learning,DBLP:conf/sigir/DehghaniZSKC17,borisov2016neural}
attests to the potential of neural networks in ranking. In particular,
the ability of deep neural networks in learning an effective representation
for query-document pairs from raw, unstructured data opens a new frontier
in the ranking research.

Decision forests and deep neural networks represent the most common classes of functions in \ac{ltr}.
While these classes are highly effective, their inherent complexity leads to a number of challenges.
To illustrate one such challenge, consider inference.
Computing a relevance score for a query-document pair from a decision forest
involves traversing many decision branches in a large number of decision trees.
Similarly, doing a forward pass through a deep neural network to compute a relevance
score requires a large number of matrix multiplications, each of a considerable size.
Finally, producing a single ranked list for a query involves the computation of relevance scores
for a large number of query-document pairs.
Doing so within a small time budget, therefore, necessitates efficient data structures and
inference algorithms. We explore these specialized tools for decision forests and
neural networks in the remainder of this work.

\chapter{Efficiency Challenges in Learning to Rank}
\label{chap:costs}
The modern web search engine is a complex software with one main objective:
to identify and return the subset of documents that are more relevant to a user query
from a much larger set of all known documents.
In the preceding chapter, we reviewed the ingredients of an \ac{ltr} model and the machinery of its supervised training
without explaining how a trained model is used within a search engine and what challenges we may face in
adopting a complex ranker for the task above. We examine these unexplored questions in this chapter
by describing the anatomy of a ranking pipeline and identifying the costs and efficiency challenges associated with each component
at a high level.

Before we even get to the ranking part of a search engine, we should address a more immediate problem.
It is clear that, due to the sheer size of document collections, it is simply infeasible
to rank all documents known to a search engine in response to a query with a
complex \ac{ltr} model. Instead, we usually first apply a lightweight \emph{retrieval} mechanism
to find a smaller subset of documents that potentially match a query. This may be a dense retrieval method
over representations learnt by a deep neural network where a match is determined by how similar the representation
of a document is to the representation of the query, or it may be a statistical score defined for terms and phrases
from the vocabulary where a document is deemed a potential match if it scores high---the latter is also known as
sparse or lexical retrieval.

We do not delve into the algorithmic details of dense retrieval methods which often (but not always) use approximate
nearest neighbor search algorithms, or lexical retrieval methods which often (but not always) operate over inverted indices.
However, we highlight the importance of efficient index structures and top-$k$ retrieval algorithms over index structures,
and present the following efficiency challenge:

\begin{challenge}
Given a query $q$ and a large collection of documents $\mathcal{D}$,
we seek a \emph{space-efficient} data structure known as an index $\mathcal{I}$ to represent $\mathcal{D}$
and a {time-efficient} algorithm $\mathcal{A}_\textsc{Retrieve}$ that operates on $\mathcal{I}$ and returns the top-$k$ documents
that are most similar to $q$.
\end{challenge}

A great body of Information Retrieval literature and beyond investigate this particular challenge.
We refer the interested reader to these works and citations therein for more
details~\citep{asadi2013efficiency,asadi2012bloom,asadi2013bloom,wang2021graphanns,malkov2016hnsw,petri2019accelerated,mackenzie2021anytime,ding2011bmwand,broder2003wand,mallia2022sigir}.
Throughout the rest of this monograph, we take for granted the existence of an efficient index and retrieval
algorithm as a first step in processing a user query, and focus instead on the \ac{ltr} stage.

While the retrieval step above greatly reduces the problem size, it does not change the overarching goal;
we must still train a ranker and apply it to every retrieved document to compute relevance scores
and return a ranked list.

Consider the training of an \ac{ltr} model. As discussed in Chapter~\ref{chap:supervised}, we need labeled training data
in the form of queries and sets of documents, which we then use to learn the parameters of a ranking function with
the objective of optimizing a ranking loss. It is clear that the efficiency of the training procedure depends on the
size of the data collection (as larger datasets require larger storage capacity and lead to longer training duration)
as well as the complexity of the parameterized function (as a larger set of parameters requires exponentially more tuning).
In addition to memory and time requirements, a training procedure that utilizes more data and requires more parameter updates
is likely to result in higher energy consumption. This last point is particularly acute when the parameterized function
is the class of deep neural networks~\citep{scells2022sigir-green-ir,strubell-etal-2019-energy,xu2021green-dl}.
Together, these factors present the following efficiency challenge:

\begin{challenge}
We seek a \emph{sample-efficient} learning algorithm $\mathcal{A}_\textsc{Train}$---requiring as few training data points
as possible---to learn a parameterized function $f(\cdot, \cdot; \Theta)$ with \emph{minimal complexity} required, in
an \emph{energy-} and \emph{time-efficient} manner, such that $f$ yields a desired quality measure on unseen data.
\label{efficiency-challenge:training}
\end{challenge}

\begin{figure}[h!]
	\centering
	\includegraphics[width=.9\textwidth]{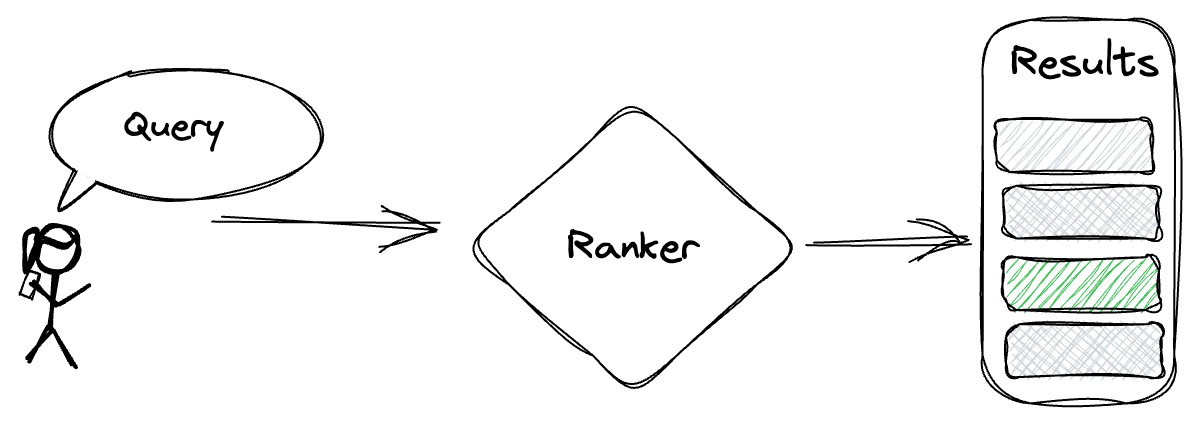}
	\caption{Single-stage ranking pipeline}
	\label{fig:single-stage}
\end{figure}

Once we have trained a model efficiently, we must apply the learnt function $f$ to user queries in production.
A na\"ive design to accomplish this goal would be to use an \ac{ltr}
model in a single stage, as depicted in Figure~\ref{fig:single-stage}.
The resulting ranking architecture is aptly called the \emph{single-stage} pipeline.

It turns out that even with an effective retrieval method, the set of matching documents may yet be too large
for an \ac{ltr} ranker to process \emph{efficiently}. That is because
computing a single relevance score requires the execution of two potentially expensive operations.
First, the set of query-document pairs must be translated into feature vectors---a
phase that is known as ``feature computation'' or ``feature extraction.''
Second, the model must be applied to each feature vector, which as discussed in the previous chapter,
may involve computationally-expensive operations such as tree traversal or matrix multiplication.
As the complexity of features and models increase, the cost incurred by these operations may become prohibitive,
to the point where computing relevance scores for an entire set of retrieved documents may be impractical.
As such, the inference procedure above involves addressing a number of efficiency challenges,
which we summarize as follows:

\begin{challenge}
Given a query $q$ and a retrieved set of documents $S_q$ by algorithm $\mathcal{A}_\textsc{Retrieve}$,
we seek a \emph{time-efficient} algorithm $\mathcal{A}_\textsc{Infer}$ that first represents the set
$\{(q, x_i) \,|\, x_i \in S_q\}$ in a $|S_q| \times d$-dimensional
feature space $\mathcal{X} \subset \mathbb{R}^{|S_q| \times d}$ and subsequently applies the function $f(\cdot, \cdot; \Theta)$
learnt by $\mathcal{A}_\textsc{Train}$ to each query-document pair and orders them in decreasing order of relevance scores.
\label{efficiency-challenge:inference}
\end{challenge}

We have so far described the challenges inherent in retrieval, and training and inference of an \ac{ltr} model.
In the remainder of this chapter, we will describe high-level ideas that help address some of these challenges.
We start, however, with inference and visit training efficiency last.

\section{Efficient inference}
How may we achieve effective but efficient ranking given a trained model or a collection of trained models
to address Challenge~\ref{efficiency-challenge:inference}?
At a high level, the answer is quite intuitive and follows how we split the ranking problem to
one of retrieval-then-rank: the set of retrieved documents
can go through multiple stages, where each stage weeds out less-relevant documents and
passes to the next stage a more promising but much smaller subset, and where each stage uses
a more complex ranking model with increasingly sophisticated features than the stages before it.
This paradigm is known as the \emph{multi-stage} ranking pipeline.
But to understand how we arrived at this solution, we must dissect
the two operations involved (i.e., feature computation and model inference)
and identify the factors that contribute to the overall cost.

\subsection{Feature computation}
Feature computation deals with the computation of query-document features
that are given as input to the model to compute relevance scores.
This task can be computationally expensive for a number of reasons.
First, the number of features used in modern rankers is typically large
with hundreds of features describing a single query-document pair. Second, each feature has its own intrinsic complexity:
it can be a composition of more basic signals, or itself be the output of another machine-learnt model.
This added complexity is justifiable because more sophisticated features
often offer a higher discriminative power than basic, cheap-to-compute features such as term frequency.

On that basis, determining the appropriate set of features involves an implicit trade-off:
\begin{tradeoff}[]
  Using a large number of sophisticated features likely leads to improved ranking quality
  but also increased overall query processing time.
\end{tradeoff}

\begin{figure}[h!]
	\centering
	\includegraphics[width=1\textwidth]{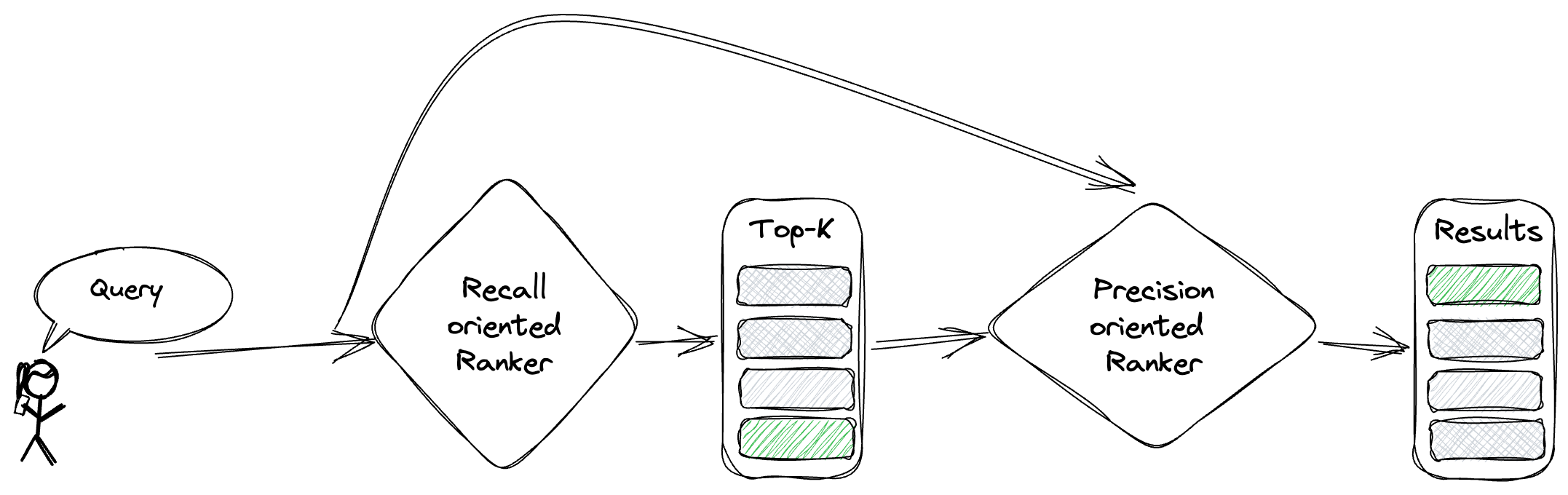}
	\caption{Two-stage ranking pipeline}
	\label{fig:two-stage}
\end{figure}

\subsubsection{The two-stage design}

One idea to rein in the cost of feature computation is to
break up the ranking pipeline into two stages, as we illustrate in Figure~\ref{fig:two-stage}.
The first stage in this design is in charge of executing a recall-oriented ranking of documents.
It is important for this stage to employ simple rankers using cheap-to-compute features, to keep
at bay the total cost of ranking the potentially large set of retrieved documents.
The second stage, which only ever observes the top-$K$ documents (also known as ``candidates'') as ranked by the first stage,
is free to apply a precision-oriented, complex ranker to re-rank the candidates
and produce a final ranked list.

The two-stage design greatly shrinks the set of documents for which we must compute
expensive features, and as a result, reduces the overall cost of feature computation.
It therefore makes it feasible to use complex \ac{ltr} models to produce effective
ranked lists but do so efficiently.
However, materializing this design requires choosing one key parameter:
the cut-off value $K$ that caps the number of documents that the second stage must re-rank.
This choice presents our second trade-off:

\begin{tradeoff}
  A large $K$ leads to a larger set of candidates to re-rank,
  in turn, increasing the cost of the second-stage ranker while potentially facilitating a higher-quality final ranked list.
  A small $K$, on the other hand, enables faster ranking in the second stage by passing a smaller set of candidates to re-rank,
  while potentially hurting quality.
\end{tradeoff}

The trade-off above has been the subject of much research in the past.
\citet{macdonald2013whens} demonstrated empirically that the cut-off value does indeed
affect ranking performance. The authors evaluated the impact of $K$ on two public document collections and gave a detailed analysis of the
performance of two-stage pipelines where the first stage used statistical retrieval models (e.g., BM25 and DPH)
and the second stage applied pointwise, pairwise and listwise \ac{ltr} models.

Approaching this trade-off from a slightly different angle,
\citet{ecir13} investigated the recall bias of the first stage ranker.
They found that the performance of the first stage affects the second-stage ranker in two unsurprising ways:
(1) by influencing the quality of the training data available to learn the ranking model; and,
(2) by controlling the number of relevant documents observed by the learnt model.
The authors then showed that by using a learnt, yet fast model in the first stage,
not only did recall improve in the first stage, but so did the overall performance of the second stage ranker.

This ability to trade off effectiveness for efficiency makes the two-stage design suitable
for real-world applications where quality and speed are both critical to users.
There is indeed evidence in the literature to support this speculation.
\citet{relyahoo}, for example, describe the query processor at Yahoo search engine as a
distributed system deployed on hundreds of machines where each search node retrieves
``hundreds of thousands'' of candidates for a subsequent stage to re-rank.
Another known deployment of this architecture is Alibaba’s
e-commerce search engine~\citep{liu2017cascade}.

\begin{figure}[t]
	\centering
	\includegraphics[width=\linewidth]{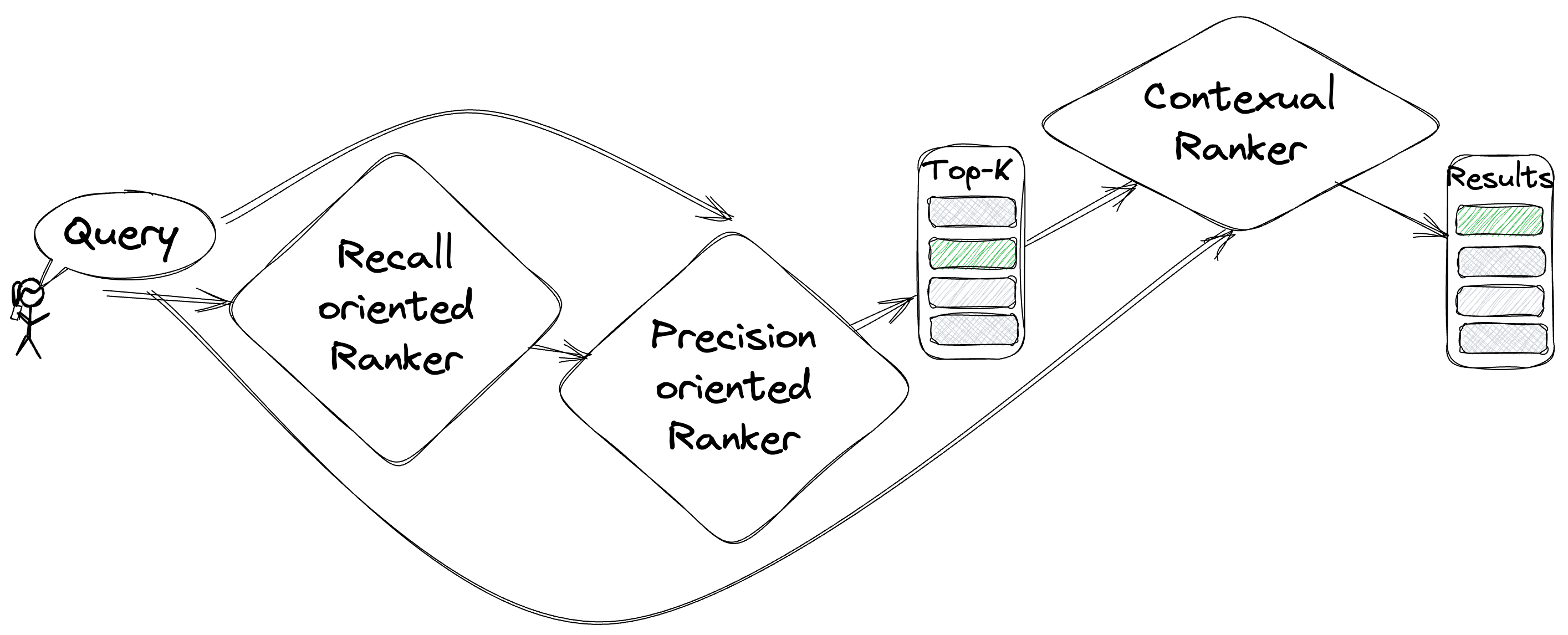}
	\caption{Three-stage ranking pipeline}
	\label{fig:three-stage}
\end{figure}

\subsubsection{The multi-stage design}

With all the benefits the two-stage design has to offer, as observed by study after study,
and all the knobs it provides to choose the right balance between efficiency and effectiveness,
it is natural to wonder if one could simply extend the design to more than two stages.
In the work by~\citet{relyahoo}, for example, the ranking pipeline is actually comprised of
three stages, as shown in Figure~\ref{fig:three-stage}.
As in the two-stage design, the first two stages, which they call ``Core Ranking,''
find top candidates for a query and re-rank those to produce a high-quality ranked list.
The third stage, dubbed ``Contextual Re-ranking,'' extracts features that capture contextual information
about the entire list (e.g., rank, feature mean and variance, normalized features, topicality),
and uses the resulting richer feature set to re-rank the candidates again.
This idea of leveraging contextual, rank-based features showcases the flexibility of
a stage-wise view of ranking, and has been shown in other independent studies to greatly
improve both ranking quality, and, when applied wisely, speed~\citep{lucchese2015speeding}.

\begin{figure}[t]
	\centering
	\includegraphics[width=.9\textwidth]{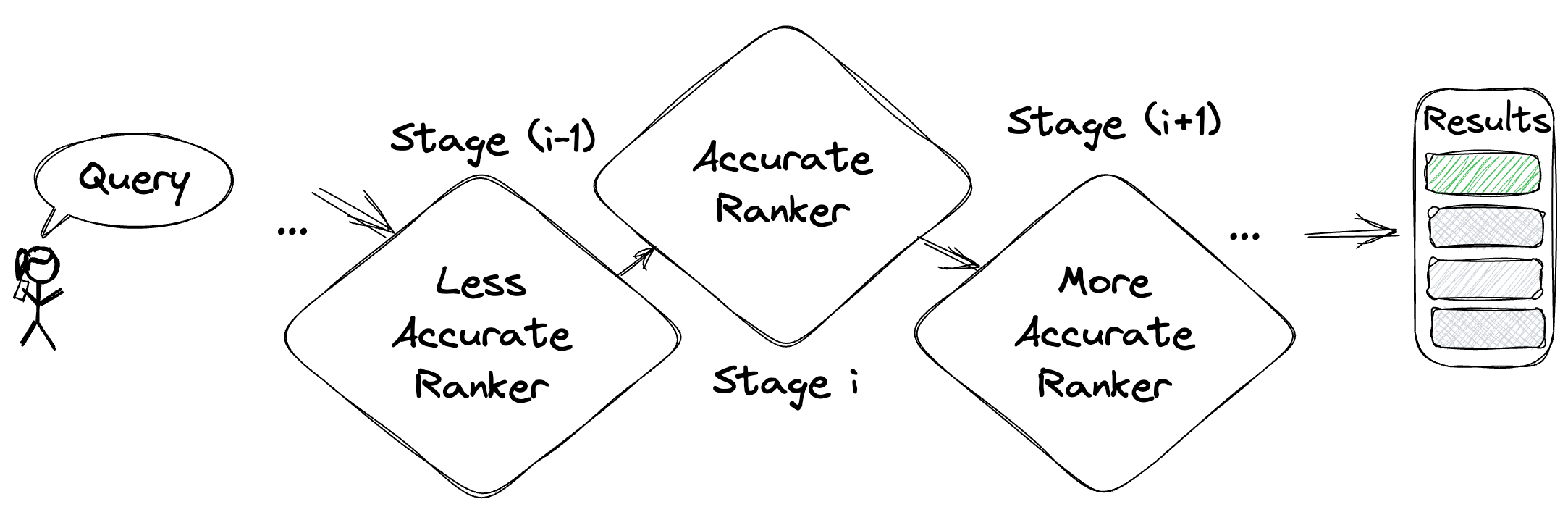}
	\caption{Multi-stage ranking pipeline}
	\label{fig:multi-stage}
\end{figure}

Given the success of a progression from two stages to three,
it is tempting to generalize the design to $N$ stages, as shown in Figure~\ref{fig:multi-stage}.
Although potentially more effective, the \emph{multi-stage} ranking pipeline
is characterized by an increased complexity due to the sequential nature of query processing:
each stage has to wait for the output of its predecessor to begin processing the input candidates.
There are other questions too: Which features and which model should be used in each stage?
How many documents should each stage re-rank?

\citet{Roi_SIGIR17} study a subset of these questions: Suppose we have, in some way, arrived
at a particular number of stages in a multi-stage ranking system. Given this particular scaffolding,
can we select features and the number of candidate documents passed between consecutive stages
so as to maximize effectiveness and efficiency of the overall cascade? \citet{Roi_SIGIR17} formalize
this problem using the concept of \emph{regularization} from machine learning and present an
optimization framework to minimize the ``cost'' of a cascade---defined as the cost of computing
a particular feature and the number of documents for which this feature must be computed---while
maximizing its ranking precision. For example, through $\ell 1$-regularization, one can enforce a
certain degree of sparsity in the set of features used within a single stage of the cascade;
a more aggressive sparsity rate yields a more compact, but potentially less effective stage.

In a follow-up study,~\citet{10.1145/3289600.3290986} identify another gap in the construction of multi-stage ranking
systems: The models employed within individual stages are often learnt independently of one another,
while in reality the decisions and rankings of one stage affects that of subsequent stages.
The authors posit that the ``stage-wise'' ranking loss and the global effectiveness and efficiency objectives
of a cascade can be optimized \emph{jointly} using backpropagation. The key insight that
enables gradient-based optimization of a cascade is that whether a document enters a stage but is dropped
within that stage (i.e., document is \emph{covered} by stage) can be expressed as an indicator function,
which can be smoothed and differentiated.

\subsection{Model inference}

We have just seen how the multi-stage design offered a way to manage the cost
of feature computation by introducing levers that allow us to trade off speed for quality.
That included using simpler features in early stages and computing more complex features
in later stages where we have fewer documents to re-rank.
In our discussion, we also hinted that rankers in early stages are typically ``simpler''
and that we are free to use more ``complex'' models as we get to later stages.
But what makes a ranking model more complex than others?

To answer the question above, it helps to consider the wide array of algorithms
that the \ac{ltr} literature has to offer.
Many of these learn a ranking function with very few parameters,
therefore requiring few operations to compute a relevance score for a given feature vector.
Examples include Coordinate Ascent \citep{tseng1988coordinate}, Ridge regression \citep{hoerl1970ridge}, SVM-Rank \citep{joachims2002optimizing}, and RankBoost \citep{freund2003efficient}. But there are also
models that comprise of deep learning modules or hundreds or thousands of deep decision trees,
resulting in large matrices to be multiplied sequentially or an exponentially large number
of comparisons to evaluate recursively. These include GBRT~\citep{Friedman01},
Initialized GBRT~\citep{mohan2011web}, LambdaMART~\citep{export:132652},
and large language model-based Rankers~\citep{lin2021pretrained}.

Even once we choose an \ac{ltr} algorithm, we are often in control of the complexity
of the model it learns. For example, in tree-based algorithms, we can cap the number of
leaves each tree is allowed to have, or limit the maximum number of trees in the ensemble,
all by adjusting the corresponding hyperparameters in the training algorithm.

Given the diverse set of algorithmic choices before us,
it is not surprising that numerous studies have in the past conducted
a comparative analysis of models based on their
complexity~\citep{tax2015cross,capannini2016quality, liu2017cascade,macdonald2013whens,scells2022sigir-green-ir}.
\citet{capannini2016quality}, for example, show that complex models---in particular,
those based on decision trees---achieve significantly higher quality.
They conclude that choosing the best model depends on the time budget available for query processing,
and propose an objective---the Area under the Quality Cost Space (AuQC)---to
compare different algorithms in terms of their accuracy-latency requirements.
\citet{scells2022sigir-green-ir}, as another example, compare a range of models from decision tree-based to
large language model-based rankers and observe significantly higher energy consumption in more complex models, adding a
new but important dimension to the efficiency of ranking algorithms.

These empirical observations lead to a third trade-off between efficiency and effectiveness
stemming from inherent complexities of \ac{ltr} models:
\begin{tradeoff}
  Models that have fewer parameters and thus require fewer operations for evaluation
  are fast to execute and consume less energy typically at the expense of ranking quality. The flip side is that
  more complex models achieve higher effectiveness but incur a significantly higher computational cost and energy consumption.
\end{tradeoff}

\section{Efficient training}
Unlike inference, there have been relatively few studies in the \ac{ltr} literature that investigate
Challenge~\ref{efficiency-challenge:training}: efficient training. This is because up until the advent of deep learning, training even the most complex
decision forests for \ac{ltr} required a relatively modest number of training data points and the algorithms used to
learn individual decision trees themselves would complete reasonably fast on general-purpose CPUs. As such, most training
procedures were considered sample-, space-, and time-efficient, thereby rendering efficiency in training a non-issue.

That changed with the arrival of deep learning models, whose training needs vast datasets---thereby resulting in larger
sample and space requirements---and involves computationally-intensive operations---in turn, requiring
longer training duration on specialized, energy-hungry hardware.

The march towards ever larger datasets and ever more complex deep learning-based ranking models
led \citet{scells2022sigir-green-ir} to study the training efficiency challenges with a particular focus on
environmental impact. The authors conducted a comparative study of widely-used \ac{ltr} models in terms of
time-efficiency and effectiveness, as well as their power usage, which can be translated into the amount of \ch{CO2}
emissions. Their comprehensive study reveals that the training of deep learning models yields orders of magnitude
larger emissions as compared to decision forests. They conclude their study by offering a framework for Information
Retrieval researchers to alleviate some of the environmental costs of developing deep learning retrieval and ranking
models.

As shown by a few recent works~\citep{scells2022sigir-green-ir} and events~\citep{bruch2022reneuir,bruch2022reneuir-report},
there is increasing
interest in the efficiency challenges during training, driven by the urgency created by the environmental costs of
recent ranking models. But more research is needed to help understand the trade-offs and offer solutions.

The challenges and trade-offs reviewed in this chapter capture the existing research in the efficient
\ac{ltr} literature. In the subsequent chapters, we present a detailed analysis of state-of-the-art solutions that
explore these trade-offs to improve the efficiency of \ac{ltr} models. As we pointed out earlier, because decision
forests and neural networks require different types of intervention, we study them separately.

\chapter{Tree-based Learning to Rank}
\label{chap:gbdt}

Consider a basic supervised learning task where we have a labeled set of data points
$\data = \{(x_i, y_i)\,|\, 1 \leq i \leq |\data|\}$ with $x_i \in \mathbb{R}^d$ being
a $d$-dimensional real-valued vector of features and $y_i \in \mathbb{R}$ a target label.
As usual, we wish to learn a function $F: \mathbb{R}^d \rightarrow \mathbb{R}$ that predicts a label $y$ for an example $x$ by optimizing the empirical loss $\mathcal{L}_\data$:
\begin{equation*}
  \mathcal{L}_\data(F) = \frac{1}{|\data|} \sum_{(x, y) \in \data} \ell(y, F(x)),
\end{equation*}
where $\ell(\cdot)$ is a loss function such as the Squared Error (MSE).

A familiar approach to learning $F$ is to parametrize it with a set of parameters $\Theta$,
$F(\cdot;\,\Theta)$, and find a $\Theta^\ast$ that leads to an optimal $\mathcal{L}_\data$.
In other words, we first choose the family of functions that we believe represents the
relationship between examples and labels---which may be a line, a neural network,
or a decision tree---and then find the right shape by adjusting its parameters.
We typically find $\Theta^\ast$ iteratively by applying gradient descent
where at each iteration we take a step proportional to the negative gradient of $F$ with respect to
the current $\Theta$:
\begin{equation*}
  \Theta^{(t+1)} \leftarrow \Theta^{(t)} -\eta \nabla_{\Theta}F,
\end{equation*}
where $\eta$ is the learning rate.

Contrast the above with a different formulation where we assume that $F$
can be broken up into additive components as follows: $F(x) = \sum_{t} f_t(x)$.
$f_t$'s, which are known as \emph{weak learners}, may be any arbitrary function
including parameterized families of functions, $f_t(\cdot;\,\Theta_t)$.
We learn $F$, again, iteratively, but here at the $m$-th iteration we
learn the weak learner $f_m$ to approximate the \emph{residual} error: the negative gradient of the loss
$\mathcal{L}_\data$ with respect to the current function $F^{(m-1)}=\sum_{t=1}^{m-1} f_t(x)$.
This quantity is also known as the \emph{pseudo-response} and is defined as follows:
\begin{equation*}
  g_m(x_i) = -\frac{\partial{\mathcal{L}_\data(y_i, F)}}{\partial{F}} \Big|_{F=F^{(m-1)}(x_i)}.
\end{equation*}
For example, when $\mathcal{L}_\data$ is the MSE, $g_m$ is simply $y-F^{(m-1)}$.
Said differently, $f_m$ is learnt in a supervised manner on a new copy of the
dataset $\data^{(m)} = \{(x_i,\, g_m(x_i))\,|\, x_i \in \data \}$  where the residuals
are now the labels. Finally, we scale $f_m$ by a regularizing
\emph{shrinkage} factor, $\nu$, and add it to $F^{(m-1)}$ to obtain $F^{(m)}$.

The learning framework just described is known as gradient boosting.
It can be thought of as performing gradient descent in the space of functions,
where instead of adjusting the parameters of a function at each step to reduce
error, we learn an entirely new function that accounts for the leftover error
and add it to the ensemble.

Within this framework, when the weak learners are the class of decision trees,
we refer to the resulting \emph{forest} as gradient boosted decision trees or GBDTs.
Similarly, when the decision trees have real-valued leaves---also known as
regression trees---we use GBRTs as a shorthand.

GBRTs are central to a powerful class of \ac{ltr} algorithms and,
as such, are the topic of the next few chapters. But before we proceed,
we must explain how GBRTs are adapted to the ranking task.

\section{GBRTs and learning to rank}

It is easy to see that the gradient boosting framework described earlier is quite general and flexible,
and that it can extend to specific learning tasks simply through the use of a differentiable loss function
that is appropriate for the desired task.
In theory, then, \ac{ltr} with GBRTs is a matter of
plugging in a ranking loss function as $\mathcal{L}_\data$ that is
applicable to a ranking dataset
$\data = \{ (q_i, \bm{x}_i, \bm{y}_i)\,|\, 1 \leq i \leq |\queries| \}$,
which, as a reminder, is a set of tuples with each tuple comprising of a query $q \in \queries$,
a set of documents belonging to that query $\bm{x}=\{x_1, x_2, \ldots, x_{|\bm{x}|} \}$,
and a set of relevance labels $\bm{y}$ corresponding to those documents.
The only challenge specific to \ac{ltr}, as we explained
in Chapter~\ref{chap:supervised}, is deriving a surrogate smooth loss function
that is more amenable to gradient boosting than ranking metrics.

As we have already seen, the need for surrogate losses is
a central question in \ac{ltr} research whenever the optimization algorithm
requires meaningful gradients. It is in no way unique to gradient boosting or GBRTs.
Naturally then, any of the ranking loss functions reviewed in Chapter~\ref{chap:supervised}
are reasonable candidates and, indeed, some such as the cross-entropy ranking loss~\citep{bruch2021xendcg}
have been implemented with GBRTs.

What makes GBRTs stand out, however, is the observation that all one needs to
conduct a boosting step are the residuals: we need not necessarily have a closed-form
loss function $\mathcal{L}_\data$, so long as its $g_m$'s are known to us.
This simple observation inspired LambdaMART~\citep{export:132652},
an early but influential \ac{ltr} algorithm.

LambdaMART designs the residuals at each iteration heuristically and leaves
the existence of a loss function with those gradients to assumption. Concretely,
the residual of document $x_k \in \bm{x}$ belonging to a query $q$
is a sum of pairwise quantities as follows:
\begin{equation*}
  g_m(x_k) = \sum_{l:\,x_l \in \bm{x}} \lambda_{k,l}.
\end{equation*}
Each $\lambda_{k,l}$ is the multiplication of two factors.
One measures the distance between the scores of documents $x_k$ and $x_l$
using a sigmoid transformation. The other sorts the documents in $\bm{x}$ by their
scores up to the current iteration (i.e., $F^{(m-1)}(x_i)$),
swaps the positions of documents $x_k$ and $x_l$, and measures the
change in the metric we wish to optimize. So, for instance, if our metric of interest
is \ndcg, $\lambda_{k,l}$ would materialize as follows:
\begin{equation*}
  \lambda_{k,l} = \frac{1}{1 + e^{-(s_k - s_l)}} \times \Delta\ndcg_{k, l},
\end{equation*}
where $s_o = F^{(m-1)}(x_o)$ and $\Delta\ndcg_{k,l}$ is the change in $\ndcg$
when $x_k$ and $x_l$ trade positions in the ranked list. Note that
the algorithm uses relevance labels $\bm{y}$ to compute $\ndcg$.

Once the residuals of every document of every query are computed,
we have all that is necessary to finalize one boosting step following
the general recipe of gradient boosting. Repeating this process
results in a forest of GBRTs that can be readily used as a ranker.

\section{The prominance of tree-based learning to rank}

GBRT-based \ac{ltr} algorithms rose to prominence not just in academic research,
but also in real-world applications in industry thanks to their unrivaled effectiveness.
In tasks ranging from Ads Click Prediction at Facebook~\citep{he2014practical}
and Microsoft~\citep{ling2017model}, to product and document ranking at Amazon~\citep{sorokina2016amazon}
and Yahoo!~\citep{relyahoo}, to forecasting and recommendations at Yandex,
GBRT-based rankers have played a major role.
Winning solutions in many machine learning competitions in recent years too were centered around GBRTs.
LambdaMART, for example, was the winner of the Yahoo! \ac{ltr} Challenge~\citep{chapelle2011yahoo}.
In a competition hosted by Kaggle in 2015 too the majority of the winning solutions used GBRTs~\citep{xgboost}.
According to the same source, so did the top 10 teams who qualified in the KDDCup 2015.

The expansion of GBRTs and GBRT-based \ac{ltr} algorithms to a larger and more varied range of
applications has inspired novel implementations of gradient-boosted tree learning algorithms.
Among them  XGBoost~\citep{xgboost}, LightGBM~\citep{lightgbm}, and CatBoost~\citep{catboost} offer
state-of-the-art results with comparatively lightweight training routines. That, in turn, led to a variety of evaluation and analysis frameworks for \ac{ltr}~\citep{LUCCHESE2020100614,10.1007/978-3-030-99739-7_38,macavaney:arxiv2020-abnirml,10.1145/3077136.3084140}.

The instances above highlight the continued importance of GBRTs in machine learning---and, in particular,
in \ac{ltr}---even in the face of the recent successes of deep learning. That is the reason why
we study this family of algorithms in the next few chapters of this monograph.

\chapter{Training Efficient Tree-based Models}
\label{chap:learning}

Is it possible to train a tree-based \ac{ltr} model that is efficient during inference?
In other words, given we have identified and are aware of the factors that challenge the efficiency of an
inference algorithm in Chapter~\ref{chap:costs}, can we use that knowledge to train a model that does not
incur high efficiency costs during its application? This chapter reviews methods that explore the trade-offs
between inference efficiency and effectiveness \emph{while} learning a ranking model.
Broadly, these methods approach the problem in two different ways: (1) by presenting variations of the learning algorithm to address efficiency while training the ranking model and (2) by applying post-hoc optimization to a trained model in a post-processing phase. We review these methods in order.

\section{Optimizing inference efficiency while learning}
There is a large class of methods that aim to reach inference-efficiency while learning a ranking model. This research effort dates back to the work of~\citet{Wang:2010:RUT:1871437.1871452} who first introduced the notion of \emph{temporally-constrained ranked retrieval} as a desirable property of a ranking algorithm. They argued that, equipped with this property, the ranking algorithm can cope with diverse users and information needs, and better manage load and variance in query execution time.

As a way to induce the property above in an \ac{ltr} model, \citet{WangSIGIR10} proposed a unified framework for jointly optimizing effectiveness \emph{and} inference efficiency during learning. The idea is simple: Instead of optimizing an effectiveness metric during the training of a model, the new framework optimizes a hybrid metric that balances efficiency and effectiveness. Dubbed the ``Efficiency-Effectiveness Trade-off'' metric (EET) is a weighted harmonic mean of some measure of efficiency and effectiveness, which can be stated more formally as follows:
\begin{displaymath}
EET(q) = \frac{(1 + \beta^2) \cdot \gamma(q) \cdot \sigma(q)}{\beta^2 \cdot (\gamma(q) + \sigma(q))},
\end{displaymath}
where $\sigma(q)$ and $\gamma(q)$ are two functions $\mathbb{R} \to [0,1]$ mapping efficiency and effectiveness of a model for a given query $Q$ into the unit interval.

Most query effectiveness metrics have the form above, including  precision, recall, average precision, and NDCG. The authors use average precision for $\gamma(\cdot)$ in their work, but any of the other metrics too can be plugged into the hybrid metric. As for the measure of efficiency, $\sigma(\cdot)$, the authors take the query execution time (measured in seconds) as input and map it to a efficiency score in the unit interval such that $0$ and $1$ represent an inefficient and efficient ranking, respectively. For example, consider the following mappings: \emph{constant}, \emph{step}, \emph{exponential}, and \emph{step + exponential}.

\begin{figure}
	\centering
	\includegraphics[width=0.8\linewidth]{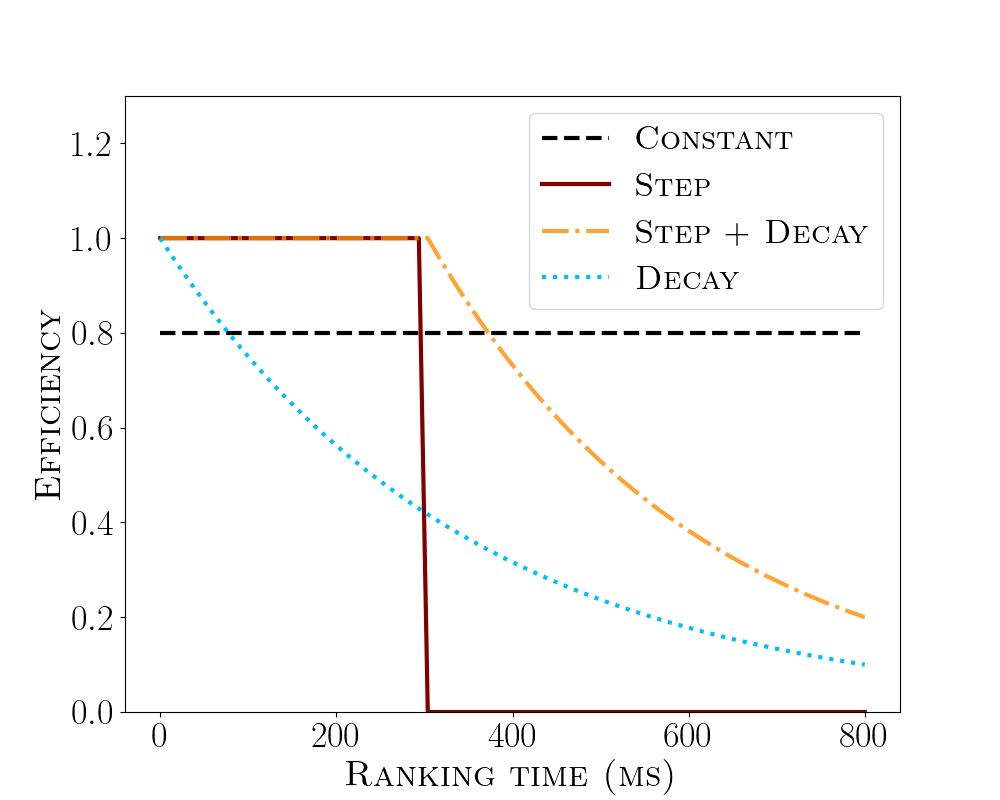}
	\caption{The four different $\sigma(q)$ functions measuring efficiency introduced by Wang \emph{et al.}. The figure is redrawn based on \citep{WangSIGIR10}.\label{efficiencyFuncWang10}}
\end{figure}

 The \emph{constant} function always maps the query execution time to the same value independent of the actual execution time. The \emph{step} function computes a score of $1$ for all queries whose execution time does not exceed a given threshold (e.g., $300$ msec.) and $0$ for all other queries. The \emph{exponential decay} function allows for a softer penalization of increasing query execution times. Finally, the \emph{step + exponential decay} function is a combination of the two previous functions that allow for a soft penalization of query execution times above a given threshold. Figure~\ref{efficiencyFuncWang10} shows the four different $\sigma(q)$ functions introduced to quantify the efficiency behavior of a query $q$.

The EET metric is a weighted harmonic mean. In fact, the contribution of efficiency and effectivenss factors to the final measure is controlled by a hyper-parameter, $\beta$, which in the original work of~\citet{WangSIGIR10} is set to $\beta = 1$. Note that, EET measures the efficiency-effectiveness trade-off on a per-query basis. As such, the training algorithm optimizes the mean EET, called MEET(q) and defined as $\frac{1}{N} \sum^{N}_{i = 1} EET(q_i)$, to learn a ranking model on a set of $N$ training queries.

Suppose we wish to learn a linear function of the following form, as in the work of~\citet{WangSIGIR10}:
\begin{displaymath}
	S(q,d) = \sum \lambda_i f_i(q,d),
\end{displaymath}
where $q$ is a query, $d$ is a document, $S(q,d)$ is the ranking score produced by the function, $f_i(q,d)$ is a feature computation function, and $\lambda_i$ is the weight assigned to feature $i$. How may we make the function more efficient to compute during inference? One idea is to introduce an $L_1$ penalty in the learning objective, thereby encouraging the model to learn sparse weights. That results in a function $S(\cdot, \cdot)$ where a large number of $lambda_i$'s have a value close to $0$. It is now reasonable to disregard feature whose weights are close to $0$ as their contribution to the final score is small, thereby obviating the need to compute those features for a query-document pair during inference, and as a result improving inference efficiency with little impact on effectiveness.

\citet{WangSIGIR10} put that hypothesis to the test and conduct experiments on three TREC web collections: Wt10g, Gov2 and ClueWeb09 (part B). The empirical results show that models learnt by optimizing MEET achieve a good balance between effectiveness and efficiency. A comparison of the query likelihood (QL) model or the sequential dependence model (SD)~\citep{metzler2007linear} shows that mean query evaluation time for MEET-optimized models is greater than that of QL, but less than that of SD. As expected, increasing the decay rate in a exponential decay flavor of MEET reduces the mean query evaluation time, suggesting that MEET is able to take efficiency into consideration during the learning process.

In a later study, \citet{Xu:2013:CTC:3042817.3042834} observe that in real-world web search engines the time available for evaluating and applying a machine learning model is budgeted~\citep{kohavi2013online, SIGIR2015, chapelle2011boosted, zheng2008general}. As discussed before, this time budget is typically spent on computing features and evaluating the model itself. In the models studied by~\citet{Xu:2013:CTC:3042817.3042834}, the inference time is often dominated by the computation required to perform feature values. Reducing the number of features required to confidently rank documents for a given query thus should greatly improve efficiency.

\begin{figure}[tb]
	\centering
	\includegraphics[width=\textwidth]{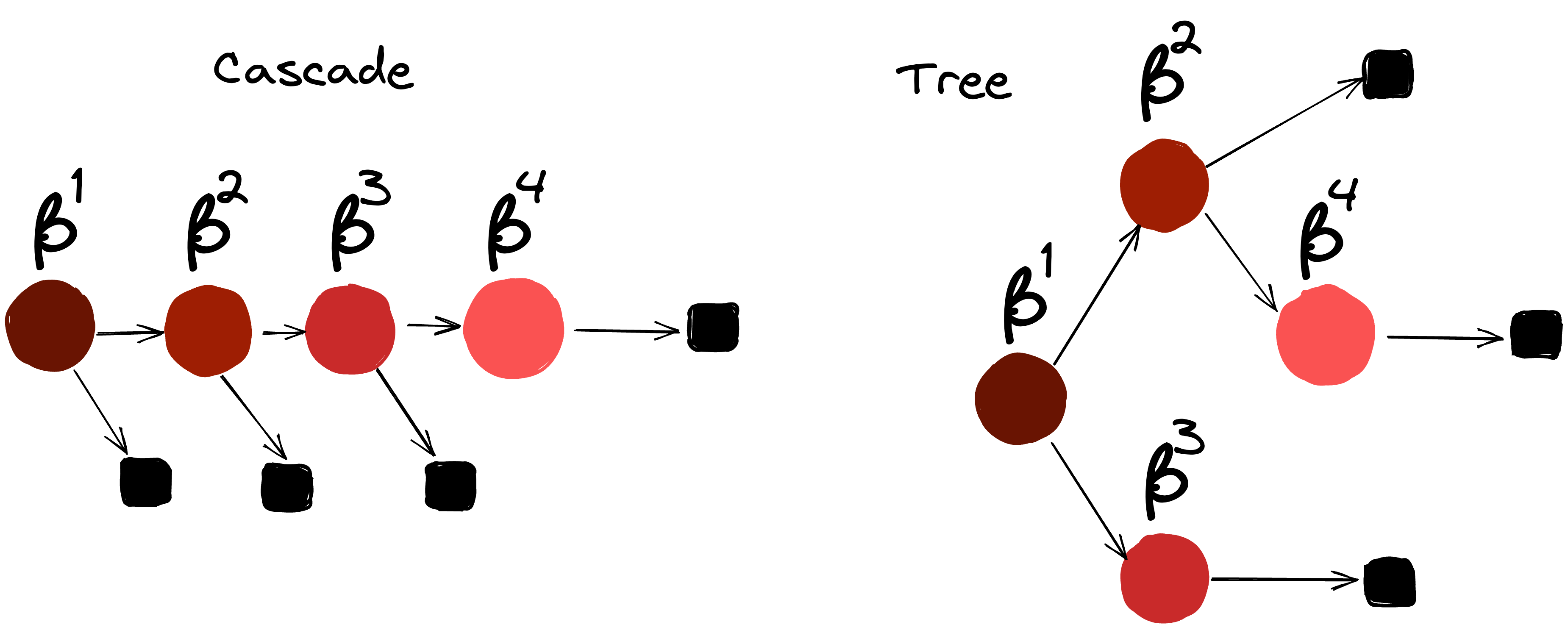}
	\caption{Cascade of classifiers (CSCC) vs Tree of classifiers (CSTC). Circular nodes represent classifiers (with their parameters $\beta$). Squares represent predictions. The color of each classifier indicates the number of inputs passing through it (darker shades indicate that a larger number of queries are evaluated by that classifier). The figure is redrawn based on \citep{Xu:2013:CTC:3042817.3042834}.\label{xu.treeClassifiers}}
\end{figure}

In contrast to~\citet{WangSIGIR10} and similar works~\citep{efron2004least,dredze2007learning} which discard entire features from the feature set and reduce costs equally across all queries, \citet{Xu:2013:CTC:3042817.3042834} take a more dynamic, query-dependent approach. Figure~\ref{xu.treeClassifiers} describes their solution pictorially. The figure on the left shows a model that comprises a cascade of classifiers (CSCC) where each classifier in the cascade terminates the inference for queries that it considers ``easy,'' thereby reducing the total amount of computation necessary to confidently predict a relevance score for a query-document pair.
Contrast that with Figure~\ref{xu.treeClassifiers} (right), which instead illustrates a \emph{tree of classifiers} (CSTC). The tree is used to classify test queries flowing along individual paths from root to leaf nodes. Each path computes different features and is optimized for a specific sub-partition of the input space. This tree structure allows us to decrease the feature computation cost by computing only the features that benefit a given input the most. This is possible because the input space is partitioned by the tree and different features are only computed when they contribute most heavily to the final ranking quality.

Learning such a structure requires an query-dependent feature selection strategy and a dynamic allocation of time budgets for features used in different tree paths (e.g., infrequent paths require a larger share of the inference time budget). As such, we need a new learning objective to optimize. \citet{Xu:2013:CTC:3042817.3042834} introduce the following loss function to learn a CSTC:
\begin{equation*}
\min_{\beta^0,\theta^0,\ldots,\beta^{|V|}\theta^{|V|} } \sum_{v^k \in \mathcal{V}} \underbrace{\big(
\frac{1}{n} \sum_{i=1}^n p_i^k \ell_i^k + \rho \lvert \beta^k \rvert
\big)}_{\text{regularized loss}} +
\lambda \sum_{v^l \in \mathcal{L}} p^l \underbrace{\big[
\sum_\alpha c_\alpha \sqrt{\sum_{v^j \in \pi^l} (\beta_\alpha^j)^2}
\big]}_{\text{test-time cost penalty}}.
\end{equation*}
This loss has two terms: a typical regularized loss and a inference-time cost penalty.
These terms together encourage the learning process to re-use features that are already computed along a specific path, rather than
computing additional features.

\citet{Xu:2013:CTC:3042817.3042834} provide a comprehensive empirical evaluation of CSTC on the Yahoo! Learning to Rank dataset, with a comparison against several state-of-the-art methods such as stage-wise regression~\citep{Friedman01}, early-exit strategies~\citep{cambazoglu10early}, and Cronus optimized~\citep{chen2012classifier} in terms of NDCG@5.

Results confirm that CSTC achieves a higher ranking quality at a small fraction of the computational cost of other methods. The early exit methods achieve limited gains because the inference cost is dominated by feature computation, rather than model evaluation cost. On the other hand, CSTC has the ability to identify features that are most beneficial to different groups of queries, which in turn allows CSTC to achieve a higher NDCG.

The optimization of the loss above was made possible through the use of a mixed-norm relaxation of the $L_0$ norm, allowing the inference-time cost penalty term to be continuous and differentiable. Later, \citet{kusner2014feature} observe that the CSTC problem is NP-hard and that \citet{Xu:2013:CTC:3042817.3042834} developed an approximate solution through the mixed-norm relaxation technique. It turns out that the mixed-norm relaxation is slow to train and requires hyper-parameter tuning.
 To remedy these problems, \citet{kusner2014feature} propose an alternative relaxation using approximate submodularity, called \emph{Approximately Submodular Tree of Classifiers} (ASTC), which casts the objective as an approximate submodular set function optimization problem. This new relaxation proved much simpler to implement, yield equivalent results without the need for hyperparameter tuning.

The authors report the results of an empirical evaluation of ASTC on the Yahoo! Learning to Rank dataset as well as three other non-cost-sensitive datasets: Forest (tree type), CIFAR (image classification), and MiniBooNE (particle identification). Their experiments show that ASTC performs just as well---and sometimes slightly better---than the state-of-the-art CSTC, while its training is up to two orders of magnitude faster.

\section{Mixed optimization strategies of inference efficiency}
In contrast to the methods we reviewed in the previous section which modify the learning process only,
mixed strategies can take as input a model that has already been trained to
optimize effectiveness and improve its inference efficiency through some form
of post-processing without significantly degrading effectiveness.
Alternatively, these algorithms could interleave training and
pruning-based optimization to achieve the same objective.
We study some of these methods in this section.

One work that highlighted the efficacy of post-hoc optimization is a comparative investigation by~\citet{asadi2013training}
of a cost-aware method to train GBRTs and a post-learning pruning of decision trees.
The core idea behind their methods was inspired by the simple observation that the cost of traversing
a decision tree is proportional to its depth and is a function of its structure,
and therefore  compact, shallow, and balanced trees must yield faster predictions.
They then proposed two solutions as manifestations of that basic idea.

Their first method belongs, in fact, to the previous category of algorithms: Optimization of inference efficiency by
modifying the training algorithm. Named ``cost-sensitive tree induction,'' it modifies the splitting strategy used in
learning individual decision trees. The idea is that, while growing a decision tree, instead of splitting the node that
leads to the maximal gain, $G^\ast$, in the effectiveness loss function, we consider the set of all possible splits
whose split gain is at least $(1-\tau)G^\ast$ for some $\tau\in[0,1]$.
Of the splits in this set, the algorithm chooses the split that results in the shallowest depth.
The hyper-parameter $\tau$ trades off efficiency for effectiveness:
when $\tau = 0$ the algorithm ignores tree depth and is reduced to effectiveness-maximizing tree learning,
but as $\tau$ increases splits that are possibly less effective but that do not add to the tree depth are selected.

The other realization of their idea, which is called ``pruning while boosting,'' belongs to the second class.
Once a decision tree in a GBRT forest is learnt using the standard tree learning algorithm, a post-processing
step prunes some of the nodes so as to reduce the tree depth and create a shallower and more {\em balanced} tree.
The pruning algorithm works as follows: Let $|t|$ be the number of nodes in the tree (including leaf nodes)
and $d_t$ be the depth of the tree. The greedy algorithm selects the two deepest leaf nodes and collapses them
into their parent node so long as $|t|\geq \alpha \left( 2^{d_t+1} -1 \right)$, for some hyper-parameter $\alpha\in[0,1]$
which controls the balance between efficiency for effectiveness. When $\alpha=0$ no pruning occurs, while when $\alpha=1$
the output tree is perfectly balanced. Note that the pruning process is oblivious with respect to the loss function.
The rationale behind this pruning is that the loss increment due to the pruning can be counter-balanced by
subsequent trees that will be learnt in the GBRT forest.

\citet{asadi2013training} present experiments on the \msnsmall{} dataset and show that the pruning approach is much more effective.
Cost-sensitive tree induction has a modest effect on efficiency (of approximately $1\%$): trees are slightly less deep,
but more trees are generated in the forest and the total number of nodes in the GBRT ensemble is unaffected.
The pruning approach, on the other hand, leads to an increase in the number of trees in the ensemble, but both
the depth and the total number of nodes in the ensemble is halved. With proper tuning of $\alpha$, it is possible to
reduce the evaluation cost of the forest by approximately $40\%$ while maintaining the same \ndcg@5.

As noted earlier, the work of \citet{asadi2013training} shows the efficacy of post-processing approaches.
The pruning method utilizes the full power of state-of-the-art learning algorithms to learn an effective model,
while also reducing their inference cost by inducing a desired model structure.
Despite this success, the core idea behind this work has a major caveat: Taking the left or right branch in a decision
tree depends on the input data and, as such, paths are taken with different, non-uniform probabilities.
Rather than creating balanced trees that minimize the depth of every path, it may make more sense to minimize
the depth of the most likely paths. Additionally, as with \citep{SIGIR2015,ye2018rapidscorer}, inference cost
may be a non-trivial function of the tree topology, where the tree depth may not be neatly correlated with efficiency.

\subsubsection{Pruning at the ensemble level}

While the pruning technique of \citet{asadi2013training} works at the tree level,
\citet{Lucchese:2016:POT:2911451.2914763} propose \cleaver{} which instead operates at the ensemble level:
instead of pruning nodes, entire decision trees are removed from the ensemble.
The motivation is that the iterative GBRT learning algorithm may generate similar and redundant trees,
especially when the learning rate is small. It must thus be possible to identify a subset of decision trees
that contribute more significantly to the effectiveness.

To that end, given a forest of $n$ trees and with the goal of producing a leaner forest of $p$ trees,
\citet{Lucchese:2016:POT:2911451.2914763} consider six pruning strategies:
\begin{itemize}
	\item \algo{Random}: A subset $p$ is selected at random;
	\item \algo{Last}: The last $n-p$ trees of the ensemble are discarded;
	\item \algo{Skip}: One tree every $\lceil\frac{n}{p}\rceil$ trees is kept;
	\item \algo{Low-Weights}: As learning algorithms typically generate a weighted ensemble,
	    the $p$ trees with the largest weights are kept;
	\item \algo{Score-Loss}: The $p$ trees that contribute the most to the prediction score of the ensemble are kept; and,
	\item \algo{Quality-Loss}: Given a ranking quality metric such as {\ndcg}, the algorithm computes for each tree
	the degradation in quality if that tree were to be discarded from the ensemble, and subsequently selects the
	$p$ trees that result in the smallest decrease in quality.
\end{itemize}

The hyper-parameter $p$ controls the balance between reaching efficiency and effectiveness: When $p = n$, no pruning
is performed and the original effectiveness is maintained. When $p < n$, fewer trees will remain in the ensemble but
at the risk of degrading effectiveness.
As a way to compensate for the possible degradation in quality due to the removal of the $n-p$ trees,
the algorithm performs a tree re-weighting step to assign new weights to the surviving $p$ tree.
To compute new weights, the algorithm simply uses line search as it allows to locally optimize any given quality measure.

Experiments with \lmart{} on both \msnlarge{} and \istellasmall{} show a dramatic reduction of the inference cost.
Authors choose $p$ so as to produce the smallest model that provides at least the same effectiveness as the full \lmart{} model.
In their experiments, the largest model evaluated on \msnlarge{} has 737 trees. The size of the forest was reduced to 369 trees
using the \algo{Quality-Loss} pruning protocol, with a speed-up factor in the inference time of 1.9.
Similarly, using \algo{Skip}, a model of 736 trees on \istellasmall{} was cut down to 368 trees with a speed-up factor of 1.8.
Overall, \citet{Lucchese:2016:POT:2911451.2914763} found that \algo{Quality-Loss} is the most stable strategy across all
experiments, due to the fact that it is aware of the target quality metric during the pruning process.
The observation that a strategy as simple and cheap as \algo{Skip} also provides decent boost to inference efficiency
suggests that ensembles are typically very redundant, and therefore pruning followed by a re-weighting of trees
is a successful way to improve the efficiency of such complex models.

We highlight that the post-processing strategies seen so far---pruning while boosting of \citep{asadi2013training}
and \cleaver{} of \citep{Lucchese:2016:POT:2911451.2914763}---provide about the same savings in inference costs and result in
an approximately $2.0$ speed-up factor. While the two ideas are orthogonal, their combined effect on inference efficiency has not yet been explored.

In a follow-up work, \citet{xcleaver} proposed \xcleaver{}, an iterative meta-algorithm that is able to learn more efficient and effective ranking ensembles. \xcleaver{} interleaves the iterations of a given gradient boosting learning algorithm with pruning and re-weighting phases. First, redundant trees are removed from the given ensemble, and then the weights of the remaining trees are fine-tuned by optimizing the desired ranking quality metric. The authors propose and analyze several pruning strategies, with a subset borrowed from \citep{Lucchese:2016:POT:2911451.2914763}. They assess the benefits of interleaving the pruning and re-weighting phases during learning instead of applying it as a single post-learning optimization step. Experiments on the \msnlarge{} and \istellasmall{} datasets show that \xcleaver{} can be successfully applied to several \ac{ltr} algorithms and optimizes the effectiveness of the learnt ensembles, thereby obtaining more compact forests, making them more efficient at scoring time.

Interestingly, reducing the size of a complex model also brings the advantage of reducing the risk of {\em over-specialization}. This side of the problem is well investigated in \citep{pmlr-v38-korlakaivinayak15}. Authors show empirically that later trees learnt by a GBRT algorithm influence the prediction of a very limited set of training instances and provide a negligible contribution to the rest. This {\em over-fitting} behavior, it is argued, affects the generalization power of the model to unseen test instances.

To overcome this limitation, the authors borrow the idea of {\em dropout} from deep learning \citep{hinton2012improving}. When dropout is applied to a layer in a neural network, it suppresses a random subset of neurons in that layer during training, with the intuition that the remaining active neurons cannot rely on a limited set of connections to compute their output.
In the context of GBRTs, this translates into muting some of the previously learnt trees while learning a new decision tree. Once the new tree is learnt and added to the forest, the algorithm computes its weight as follows: if $k$ trees are muted during training, then the predictions of the newly learnt tree are rescaled by a factor of $1/(k+1)$. The intuition is that, the new tree will likely learn to make large predictions to compensate for the absence of the muted $k$ trees, and that scaling its predictions down attenuates their contribution to the final prediction. Finally, the muted trees are rescaled by a factor of $k/(k+1)$ so as to suppress their contribution to account for the presence of the new tree. In their experiments, \citet{pmlr-v38-korlakaivinayak15} show that the proposed algorithm, called {\dart} (Dropouts meet Multiple Additive Regression Trees), can improve the performance of {\lmart} for the ranking task on the \msnsmall{} dataset.

This idea was further explored by \citet{Lucchese:2017:XBD:3077136.3080725}
who propose to replace the {\em muting} strategy with {\em pruning}.
Their algorithm, named \xdart, uses the same dropout approach with a crucial difference:
At each iteration, either the muted trees are restored into the forest,
or are removed from the forest. In particular, when the newly added tree leads to
better quality metrics than the $k$ muted trees, the muted trees are dropped,
resulting in a forest with $k-1$ fewer trees.
Together with an adaptive strategy to fine-tune $k$, \xdart{} produces impressive performance:
On the \msnlarge{} dataset, an \xdart{} model with $500$ trees yields the same \ndcg@10 as
a \lmart{} forest with $1200$ trees. Similarly, \ndcg@10 of $500$ \xdart{} trees is on par with
a \lmart{} forest of 1500 trees on the \istellasmall{} dataset.
This line of work that was inspired by dropout reduces the size of a ranking forest,
and improves its efficiency up to a factor of $3$ without any loss in accuracy.

\subsubsection{Knowledge distillation}
Both \xdart{} and \cleaver{} rely on the idea that given a large and effective model,
we can find a smaller model that achieves improved efficiency while retaining the same effectiveness.
In particular, these algorithms prune an existing model and find a subset of the model
with the largest contribution to the final quality metric.
But the idea that a small model can approximate a larger one is not new,
and it is at the core of {\em knowledge distillation}~\citep{ba2014deep} in the deep learning literature.

In knowledge distillation, a complex {\em teacher} model is first trained
on a given dataset, then a simpler {\em student} model is
trained by minimizing the deviation of its prediction both from the
ground-truth labels and from the teacher's predictions.
The intuition is that, the teacher model, thanks to its greater complexity,
captures complex relationships in the data and potentially removes noise,
thereby providing a ``cleaner'' training signal to the student model.
In the deep learning scenario, training a new student model from scratch is
a more natural choice than pruning layers or other components of a deep neural network.
Nevertheless, we might consider \xdart{} and \cleaver{} as knowledge distillation
approaches specifically tailored to decision tree forests.

The idea of knowledge distillation has been applied to \ac{ltr} in other ways too.
In what we regard as {\em homogeneous} distillation, such as the work of~\citet{tang2018ranking},
the teacher and student models are both from the same hypothesis class. But~\citet{cohen2018universal}
show that it may be more appropriate for the two models to come from different classes,
which we call {\em heterogeneous} distillation. Let us dissect these two works as examples
of knowledge distillation in \ac{ltr}.

Ranking Distillation by \citet{tang2018ranking} is an example of homogeneous knowledge
distillation in the context of \ac{ltr}. While in that work, the authors fall back to the
usual (pointwise or pairwise) logistic loss to train the teacher model, the novelty of their approach
rests in the way they train the student model, which has fewer parameters than the teacher:
Rather the working on the full training set, the student model is allowed to evaluate only the $k$ documents
that received the highest ranking by the teacher model.
The prediction error---again, based on the logistic loss---of those $k$ documents
is weighted by their rank, and it is adaptively tuned at each training iteration
in a manner similar to boosting. Experiments on a recommendation task show that the student model
performs even better than the teacher both in the case of convolutional neural networks and matrix factorization.
In addition, the student model is about twice as fast as the teacher model.
Interestingly, training a student model on the original training dataset provides
significantly worse performance figures.

In an interesting turn, \citet{cohen2018universal} argue that,
in the context of \ac{ltr}, decision tree ensembles like \lmart{} are a better choice for the teacher model,
but forests may not be all that appropriate for a student model because of the inherent efficiency challenges
during inference. Instead, they propose to use a neural network as student model
for the following two reasons: neural networks should be able to learn the predictions of a decision forest
thanks to the Universal Approximation Theorem, and modern hardware is highly optimized for the inference of neural networks.
While deep networks are still too expensive, the authors show that a {\em medium}-sized
network as a student can be effective and efficient.

In order to make this work, \citet{cohen2018universal} create a new training dataset that consists of
the original training instances used in learning a teacher \lmart{}, and a set of randomly generated
instances whose feature values are chosen so as to lie in between the different splitting points of the \lmart{} forest.
The labels of both types of training instances is the output of the teacher \lmart{} model.
Intuitively, the enhanced training dataset helps the student model approximate the behavior of the
teacher on the original training instances as well as points close to its {\em discontinuities}.

Experiments on the {\msnlarge} dataset show that a feed-forward network with two fully connected hidden layers of 500 and 100 neurons, achieves results that are on par with the teacher model with any observed difference being statistically insignificant.
Authors explore both CPU and GPU for inference and find that GPUs achieve up to 100$\times$ speed-up.
While a fair comparison between a multi-threaded implementation of a neural network against
a multi-threaded implementation of a forest traversal algorithm was not presented,
this work overcomes the limitations of back-propagation for ranking by {\em mirroring} a \lmart{} forest,
and, in turn, benefits from advances in hardware that is typically used for computations in neural networks.

Finally,~\citet{9716821} explore the additivity of knowledge distillation, pruning, and fast matrix multiplication in bringing about inference efficiency. The authors first use the knowledge distillation framework to train shallow neural networks from an ensemble of regression trees. They additionally apply neural network pruning to the learnt network so as to induce more sparsity in its most computationally-intensive layers.
The sparse, shallow network is then executed with a optimized sparse matrix multiplication algorithm.
Their experiments on two public \ac{ltr} datasets show that sparse neural networks produced
with this approach are competitive at every point of the effectiveness-efficiency trade-off
when compared with tree-based ensembles, leading to 4$\times$ speed-up during inference
without adversely affecting ranking quality.

\section{Open challenges and future directions}

\begin{table}[ht]
\caption{\label{tab:learning-summary} Highlights of cost-aware learning methods.}
\small
\begin{tabular}{p{.22\textwidth}p{.20\textwidth}p{.30\textwidth}p{.12\textwidth}}
\toprule
\textsc{Method} & \textsc{Category} & \textsc{Strategy} & \textsc{Inference Speed-up}\\
\midrule

Joint optimization \citep{WangSIGIR10} & Cost-sensitive learning & Learn linear functions with a novel metric (EET) mixing efficiency and effectiveness &  $2\times$\\
 & & & \\

CSTC \citep{Xu:2013:CTC:3042817.3042834} & Cost-sensitive learning & Tree of classifiers reducing the number of features extracted per instance &  $2\times$ \\
 & & & \\

Submodular trees of classifiers \citep{kusner2014feature} & Cost-sensitive learning & Tree of classifiers reducing the number of features extracted per instance. Reduced time for learning the model w.r.t \citep{Xu:2013:CTC:3042817.3042834} & up to $119\times$ (training time) \\
 & & & \\

Pruning while boosting \citep{asadi2013training} & Node pruning &
 Collapse leaves to reduce tree depth. & $1.6\times$\\
 & & & \\

{\cleaver} \citep{Lucchese:2016:POT:2911451.2914763} & Tree pruning &
 Remove trees and tune weights. & $2.6\times$ \\
 & & & \\

{\xdart} \citep{Lucchese:2017:XBD:3077136.3080725} & Tree pruning &
 Mute trees and remove them when appropriate & $3\times$ \\
 & & & \\

{Ranking Distillation} \citep{tang2018ranking} & Homogeneous knowledge distillation
 & 
 Student trained on top-ranking instances by teacher. & $2\times$ \\
& & & \\

{Fast neural networks from tree-based ensembles} \citep{9716821,tang2018ranking,cohen2018universal} & Heterogeneous knowledge distillation &  Feed-forward networks learned as students approximating {\lmart} &
 4$\times$ (CPU), $100\times$ (GPU) \\

\bottomrule
\end{tabular}
\end{table}

Table~\ref{tab:learning-summary} summarizes the methods discussed in this chapter.
We observe two main research directions pursued over the recent years.
The first includes novel methods for learning regression forests wherein
not only is an effectiveness metric maximized, but their inference cost is also taken into consideration.
We identify two different ways of realizing this idea: (1) methods that quantify inference efficiency and
optimize it while learning an \ac{ltr} model, (2) methods that introduce a more efficient organization of
an \ac{ltr} model so as to reduce the overall feature computation cost.
We believe these works can be foundations for new, more complex approaches that learn cost-aware models for web search.
Many of these ideas, for example, may be extended to directly learning complex ranking cascades.

The second class of algorithms attempt to condense an existing, effective model into a {\em smaller} model
that is more efficient during inference but just as effective.
This reduction in size not only affects a model's inference cost, but as highlighted by \citet{pmlr-v38-korlakaivinayak15},
it can help to prevent over-fitting.
Indeed, finding a model of the smallest complexity that achieves high effectiveness is not just an important
question in ranking, it is a great challenge in machine learning in general. We expect that research in this area can not only contribute to the specific task of efficient ranking, but that it can also lead to contributions
of a wider scope to the data mining community.

We saw that it is possible to borrow and translate ideas from the deep learning literature to design novel learning algorithms
(c.f., \dart) or use simple feed-forward networks to achieve effective ranking while benefiting from efficient hardware.
It is very likely that novel advances in deep learning will affect how we think about \ac{ltr} and help us design better
algorithms. Though, we note that information retrieval systems often have additional requirements and constraints
(such as on scale and time complexity) that create non-trivial challenges.

An interesting and relevant line of research which we did not cover in this chapter is that of online \ac{ltr}.
When we need to promptly explore user feedback and fine-tune or re-train a ranking model, the time constraints on
the training procedure are even more strict. \citet{oosterhuis2017balancing} consider this problem and propose a cascading model,
where a {\em fast-to-train} model is first optimized to provide reasonable effectiveness from a small number
of user interactions, and only when this model converges, it is used to initialize a second {\em expressive} model.
This direction of research highlights, once again, the importance of efficiency and creates new challenges for an \ac{ltr} system.

\chapter{Efficient Inference of Tree-based Models}
\label{chap:scoring}

In the previous chapter, we described ideas that either learn a model that is efficient during inference
or reduce the efficiency cost of an already-trained model. An orthogonal idea is to improve the efficiency of
model evaluation itself by reducing its computational complexity.
This is particularly important in the context of decision forests because, in order
to make a prediction, we must traverse each decision tree in the forest from root to leaf nodes, which
involves the evaluation of a decision at every intermediate node and branching to the appropriate sub-tree.
While methods from the previous chapter can help reduce the complexity and the size of a decision forest,
we are nonetheless faced with the challenge of evaluating the leaner but still large model.

We study that second problem in this chapter and review data structures and algorithms that help lower the costs
of model evaluation. This includes improved traversal of decision trees, producing approximate predictions, and
cascading models. The following sections explore these ideas in more depth.

\section{Efficient traversal of decision forests}
\label{sec:scoring-efficient}

Let us depart from the realm of theory, and go back to the basics and consider how we may implement a decision tree from scratch.
The most straightforward and perhaps na\"ive way of materializing a tree is by using
some form of if-then-else blocks of code or a conditional or ternary operator depending on the language used.
That is, at each node, we compare a feature value against a threshold
and decide if we need to take the right or left branch. We continue executing similar decisions until we reach a leaf node.

While this implementation may be the most readable, it is computationally intensive.
Yes, the resulting code may be compiled with whatever optimization strategy a compiler can muster,
but regardless the size of the resulting code is proportional to the total number of nodes in the ensemble,
and it is impossible to successfully leverage the instruction cache due to frequent branching in the code.
Conditional blocks have proven to be efficient when the feature set is small \citep{LinTKDE},
but it still suffers from \emph{control hazard}, defined as instruction dependencies introduced by conditional branches.

\citet{LinTKDE} improve this implementation in a data structure they call \algo{Struct+}.
This data structure stores the feature id of intermediate nodes and the threshold used in the split,
and has pointers to the left and right children. The traversal of the tree starts from the root
and moves down to leaves according to the result of a boolean expression on the traversed nodes.
But different from the na\"ive implementation, \algo{Struct+} uses an optimized memory layout that
linearizes the tree nodes via a breadth-first traversal of the tree.

While the improved memory layout brings about advantages, \algo{Struct+} still has a few notable drawbacks.
Importantly, the next node to be processed is known only after the boolean decision is evaluated and as such
does not address the frequent control hazard. Its efficiency thus is a function of the branch mis-prediction rate.
Another caveat is that, due to the unpredictability of the path visited by a given test instance,
tree traversal has low temporal and spatial locality, leading to low cache hit ratio and poor CPU cache utilization.

How may we work around these limitations? \citet{LinTKDE} offer an algorithm called \algo{Pred} that
rearranges the computation such that control hazard is replaced with \emph{data hazard}, defined as
data dependencies introduced when one instruction requires the result of another.
The algorithm works by first replacing pointers with node indices within a contiguous array of memory,
then using the output of the binary expression directly to compute the index of the next node.
The traversal of a tree of depth $d$ is then statically ``un-rolled'' into $d$ operations,
starting from the root node to the leaves.
Leaf nodes are encoded so that their indexes generate self loops.
At the end of the traversal, the algorithm identifies the leaf node and uses
a look-up table to retrieve the predicted value of the tree.

\algo{Pred} removes control hazards because the next instruction to be executed is always known.
However, it now introduces data dependencies because the output of one instruction is necessary
to execute the subsequent one. The algorithm does not address poor memory access patterns of \algo{Struct+}
either because the path traversed depends on the test instance.
Finally, \algo{Pred} creates yet a new source of overhead:
for a tree of depth $d$, even if a test instance ends in a shallower leaf, the algorithm executes all $d$ instructions anyway.
\citet{LinTKDE} remedy some of these limitations by creating a vectorized version of the algorithm, named \algo{VPred},
which interleaves the evaluation of a small batch of documents. \algo{VPred} was shown to be $25$\% to $70$\%
faster than \algo{Pred} on synthetic data, and to outperform other methods we discussed so far.

\subsubsection{Feature-major traversal}
The algorithms we discussed so far have taken a node-at-a-time view to evaluate a decision tree:
When a test instance enters the decision tree at the root, these algorithms evaluate decisions in each
recursively until they reach a leaf node. They then move onto evaluating the next document.
\citet{SIGIR2015} offer a different traversal pattern in {\quickscorer} by devising a feature-wise evaluation
of decision trees.

{\quickscorer} traverses a complete forest by evaluating all the nodes that make a decision using the first feature,
then all the nodes branching off of the second feature, and so on. Note that, the order in which features
are selected is immaterial and may be arbitrary.
The algorithm then creates a bit vector for each tree, called \emph{leafindex},
which has as many bits as there are leaves in that tree, and updates it to mark
the subset of leaves than will never be reached for the instance under evaluation.
That some leaves will never be visited happens because of the structure of the tree:
if an intermediate node that evaluates to \emph{false}, then its right subtree is not visited.
Each intermediate node too has a bit vector associated with it, which is called a \emph{nodemask}.
This bit vector encodes the leaves that are never reached should the condition in that node evaluate to false.

Given these bit vectors and the outcome of the feature-wise evaluation,
{\quickscorer} takes the nodes whose condition evaluated to false, and
performs a logical AND of the \emph{leafindex} vector with that node's \emph{nodemask}.
After all false nodes have been processed, the \emph{leafindex} bit-vector identifies
the exit leaf for the test instance, which the algorithm uses to retrieve a value from
a look-up table.

Due to this reorganization of the computation, {\quickscorer}'s data structure
can be implemented as a set of contiguous arrays, enabling fast linear scans and bit-wise operations.
Overall, because of these properties, {\quickscorer} exhibits a cache-efficient behavior.
But \citet{TangJY14} show that cache utilization can further improve by what is called \emph{blocking};
partitioning the tree ensemble into subsets of limited size, so that each subset can be processed
entirely in cache. One can tailor the block sizes based on the different levels of CPU
cache~\citep{Jin:2016:CCB:2911451.2911520}.
\citet{SIGIR2015} apply these ideas to {\quickscorer} with different flavors of
blocking~\citep{Dato:2016:FRA:3001595.2987380}, and make further improvements through
vectorization over multiple documents~\citep{Lucchese:2016:ECS:2911451.2914758}, multi-core and GPU parallelism~\citep{DBLP:journals/tpds/LettichLNOPTV19}. More recently, ~\citet{10.1007/978-3-030-99736-6_18} and \citet{MOLINA202138} propose a novel design of of the {\quickscorer} algorithm and the application of binning or quantization techniques to tree ensembles to fully leverage novel, energy-efficient field-programmable gate arrays (FPGAs).

\citet{ye2018rapidscorer} take the data structure in {\quickscorer} and make it more compact
in their algorithm, \algo{RapidScorer}. The first observation was that \emph{nodemasks}
are two sequences of $1$'s separated by a sequence of $0$'s (i.e., $1^a0^b1^c$ for some $a, b, c \geq 0$),
and that only the sequence $0^b$ is relevant for the logical AND operations.
The second observation was that, node tests may be repeated several times throughout a forest,
leading to duplication in the original {\quickscorer} data structure.
Consider for instance the case of a binary feature, where the only valid test is $x_i\leq 0$,
and this can be repeated hundreds or thousands of times in a large forest.
Finally, similar to vectorized {\quickscorer}~\citep{Lucchese:2016:ECS:2911451.2914758},
one may use SIMD instructions to evaluate multiple documents in parallel.

Following these observations, \citet{ye2018rapidscorer} propose a more compact representation of
\emph{nodemasks}; a merging mechanism to store and execute repeated node tests only once;
and a suitable data structure to allow efficient use of SIMD instructions that operate on $256$ (or larger)
bit-wide registers. Together this added compactness and parallelism reduces the algorithm's memory footprint
and number of operations.
These improvements boost inference speed by a factor of $3.5$ over {\quickscorer} on various datasets
with up to $870$ features, and for a variety of models with up to $400$ leaves and $20$,$000$ trees.

\section{Approximate prediction by partial evaluation}

In the previous section---and indeed the previous chapter---we insisted on making exact predictions
and evaluating all nodes in a forest. But what if we relaxed this strict requirement and allowed the
inference algorithm to produce an \emph{approximate} prediction instead? If we are able to produce
inexact scores faster but in such a way that the final quality remains unaffected, then this
approximation would be acceptable.

That question motivated the work of~\citet{cambazoglu10early}. Their work started with the
observation that two properties of web search allow one to potentially short-circuit the scoring process in additive ensembles.
First, that document relevance follows a skewed distribution: for most queries, there are very few highly relevant documents,
but many non-relevant documents. Second, that most users view only the first few top-ranking documents, and therefore,
it may be possible to terminate the scoring of documents that are unlikely to be ranked within the top $k$.
Given these observations, the question before the authors was whether it is possible to terminate the inference midway
through the forest (i.e., without consuming every tree) and yet maintain high quality among the top-$k$ documents.

To answer that question we need to first consider the ways in which a tree prediction algorithm scores a set of documents
with respect to a query. One obvious approach is the ``document-ordered traversal'' (DOT) strategy where
documents are scored separately by computing predictions from all trees in the forest.
Alternatively, we may evaluate each tree (or \emph{scorer}) on all documents at once,
resulting in the ``scorer-ordered traversal'' strategy (SOT). In SOT, we must accumulate and keep track of partial scores
for all documents until the entire ensemble has been evaluated. Both strategies have advantages and disadvantages
in terms of memory footprint and cache friendliness. We note that, the vectorization methods of {\quickscorer}
and \algo{RapidScorer} belong to the DOT class, but where small batches of documents are evaluated with a SOT strategy.

\citet{cambazoglu10early} place \emph{exit} points at fixed positions in a given forest (e.g., every $100$ trees)
and introduce four \emph{early-exit} algorithms that decide at each exit point whether the evaluation of a document
can be terminated early. These algorithms are as follows:

\begin{itemize}
	\item Early Exits Using Score Thresholds (EST): This simple approach filters documents on the basis of a pre-computed
	threshold and drops documents whose score does not exceed it along the way.
	\item Early Exits Using Capacity Thresholds (ECT): A more adaptive solution that maintains a heap with the highest scores
	and drops documents that do not fit in the heap.
	\item Early Exits Using Rank Thresholds (ERT): Similar to the EST method, except that thresholds are applied to the rank
	of documents.
	\item Early Exits Using Proximity Thresholds (EPT): Preserves the top-$k$ documents but additionally keeps all
	 documents whose score is within a range $p$ from the $k$-th document's, where $p$ is learnt and computed offline.
\end{itemize}

We highlight that, EST is applicable to both SOT and DOT whereas ECT applies only to the DOT method.
ERT and EPT, on the other hand, operate only within the SOT scheme. Experiments showed that the EPT
strategy led to the largest gains in inference efficiency (with a speed-up factor of 4$\times$)
with only a negligible loss in precision.

\citet{cambazoglu10early} used statistical information from document scores and ranks
to decide when to exit early. In contrast, \citet{DBLP:conf/sigir/BusolinLN00T21}
introduced a learnt technique, called LEAR, that uses a classifier to predict whether a document
should trigger early termination if it is unlikely to be ranked among the final top-$k$ results.
The early exit decision occurs at a \emph{sentinel} point (i.e.,
after having evaluated a limited number of trees) with the partial scores determining
if documents should exit. Their experimental evaluation on two
public datasets shows that LEAR has a significant impact on the efficiency
of the query processing with a speedup of up to 5$\times$ with a negligible loss in NDCG@10.

Separately, \citet{10.1145/3397271.3401256} investigate the problem of
query-level early-exit strategies, where the decision to exit depends on
the partial scores of all candidate documents for a query.
The main finding of the work is that queries exhibit different behaviors as
scores are accumulated during the traversal of the ensemble and
that query-level early stopping can remarkably improve ranking quality with an
overall gain of up to $7.5\%$ in terms of NDCG@10 and a query processing speedup
of up to $2.2\times$.

\section{Efficient cascades}

In the early-exit strategies discussed above, we relied on a partial evaluation of an ensemble to decide whether
or not to exclude a document from further evaluation. That detail is analogous to the idea of multi-stage
rankers, which we reviewed in Chapter~\ref{chap:costs}. Indeed, early-exiting is similar to having
multiple rankers that rank a set of documents sequentially and pass along to subsequent rankers the top-ranking subset.

\citet{Wang:2011:CRM:2009916.2009934} took that idea and fused together such a cascading model using
an additive ranking model. Each ranker in the model is also coupled with a \emph{pruning} function that
removes the least promising documents before passing them on to the next ranker.
This ranker-pruner pair constitutes one stage in the multi-stage ranker,
and documents that are kept by the pruning function keep accumulating partial scores from stages along the way
until they end up in the final top-$k$ set or are dropped in later stages.

\citet{Wang:2011:CRM:2009916.2009934} implement an instance of such a multi-stage ranker following the principles behind
\algo{AdaRank}~\citep{xu2007adarank}, where each ranker operates on a single feature only.
Interestingly, the pruners and rankers are trained jointly with a hyper-parameter that facilitates
fine-tuning the balance between quality and efficiency. Their experiments show that this jointly optimized
cascade model reduces the inference cost with limited impact on quality. We should note that the gains in efficiency
are not as substantial as those achieved by score approximation approaches.

\citet{culpepper2016dynamic} take a slightly different approach. They design a multi-stage ranker composed
of binary classifiers, where the number of classifiers is equal to the number of relevance grades in the training set.
They train the classifier at stage $S_i$ to detect documents with relevance $\leq i$.
When a classifier at stage $S_i$ predicts the probability $Pr_i(d)$ for document $d$,
then $d$ \emph{exits} the inference process with label $i$ if $Pr_i(d)< t$,
and otherwise moves to the next stage.
This design follows the intuition that a different number of candidates may be required for different queries
and that these classifiers can help detect the subset of candidates adaptively.
Experiments show that the best configuration of this cascade design can speed up inference by a factor of
2$\times$ without an adverse effect on the ranking quality.

Another advantage of the cascade design is that the computation of expensive features can be delayed to
later stages, where fewer documents are evaluated.
On the other hand, effective features should be used as early as possible in the cascade so that a larger number of documents
can be filtered early on. That is precisely what \citet{Roi_SIGIR17} explore
in their work. Through extensive experiments, they show that a three-stage cascade with {\lmart} in each stage
is most effective with a cost reduction of about 50\%.

\section{Open challenges and future directions}
\label{sec:scoring-future}

\begin{table}[ht]
\caption{\label{tab:scoring-summary} Highlights of inference methods.}
\small
\begin{tabular}{p{.28\textwidth}p{.14\textwidth}p{.30\textwidth}p{.14\textwidth}}
\toprule
\textsc{Method} & \textsc{Category} & \textsc{Strategy} & \textsc{Inference Speed-up}\\
\midrule

Runtime optimizations for tree-based machine learning models \citep{LinTKDE} & Efficient Traversal & Predication and interleaved multi-document evaluation &  $2\times$\\
 & & & \\

QuickScorer \citep{SIGIR2015} & Efficient Traversal & Feature-wise traversal and cache-aware data layout &  additional $6.5\times$\\
& & & \\

RapidScorer \citep{ye2018rapidscorer} & Efficient Traversal & Compact data layout &  $3.5\times$ (over QuickScorer) \\
& & & \\

Early Exit \citep{cambazoglu10early} & Approximate Scoring & Terminate ensemble traversal early & $4\times$\\
& & & \\

Dynamic cutoff prediction \citep{culpepper2016dynamic} & Cascade & Query-based prediction of the number of candidate documents & $2\times$\\
& & & \\

A Cascade Ranking Model \citep{Wang:2011:CRM:2009916.2009934} & Cascade & Joint learning of pruning and ranker stage & tunable trade-off\\
& & & \\

Cost-Aware Cascade Ranking \citep{Roi_SIGIR17} & Cascade & Expensive features are moved to later stages & $2\times$\\
& & & \\

\bottomrule
\end{tabular}
\end{table}

We summarize in Table~\ref{tab:scoring-summary} the methods reviewed in this chapter which fall naturally
into two major research directions. The first covers efficient algorithms for the traversal of decision tree ensembles.
This research culminated in the {\quickscorer} and \algo{RapidScorer} algorithms which today are
the \emph{de facto} standard tree traversal implementation, not just for the ranking task but in regression
and classification too. But while the existing implementations achieve incredible efficiency in standard computing
environments, investigating the inference of complex models in embedded devices remains an open challenge, especially
with the rapid rise in the use of machine learning models in resource-constrained devices. Additionally, such non-standard
environments define new and unique dimensions of efficiency such as strict bounds on energy consumption
among other factors.

The second line of research is an investigation of cascade models which includes early-exit strategies and
\emph{ad hoc} training of a multi-stage cascade. There remains a lot in this area that warrants further investigation.
The optimal number of stages in a cascade architecture or the value of their hyper-parameters, for example,
have proven difficult to determine, which, in turn, limit the applicability of cascade models.
Moreover, the observed impact on efficiency is smaller than the gains from efficient traversal techniques.
Despite these challenges, we believe that research into cascade models and understanding the trade-offs inherent
in their design are promising directions in \ac{ltr}.

Cascade models, for example, can delay heavy computation to
later stages or leverage the benefits of complementary models (e.g., neural and tree models). They may also provide
\emph{stability} to an \ac{ltr} system, where only a few stages may need to be re-trained or updated to improve
effectiveness. We thus believe that the construction of an efficient cascade that takes into account
feature computation costs, personalization, online training, or offline updates
remains an exciting and potentially impactful research direction.

\chapter{Neural Learning to Rank}
\label{chap:neural}

In Chapter~\ref{chap:supervised}, we wrote about the various statistical signals
which exist in queries and documents that a ranking model can use to estimate relevance.
There, and indeed throughout the past chapters, we took for granted
that query-document pairs are given to us in the form of
vectors of $k$ pre-computed features that in some way quantize those signals,
and instead focused on learning a ranking function from them.
Let us now take a step back and reconsider feature vectors.

A feature vector is effectively a function that maps queries and documents
to a $k$-dimensional space. Typically, a subset of these features can be viewed
as a function of the query alone ($\phi_q: \mathcal{Q} \rightarrow \mathbb{R}^{k_q}$),
another of the document alone ($\phi_d: \mathcal{D} \rightarrow \mathbb{R}^{k_d}$),
and the rest form a joint function of query and document pairs
($\phi_{q,d}: \mathcal{Q} \times \mathcal{D} \rightarrow \mathbb{R}^{k_{q,d}}$).
These functions map their input to a vector of real values,
together making up $k=k_q + k_d + k_{q,d}$ features as the
representation of a query-document pair:
$(\phi_q(\cdot), \phi_d(\cdot), \phi_{q,d}(\cdot, \cdot))$.

If no feature vectors exist and all we have is the raw data,
the thinking goes, we must define and build our own input-to-feature mappings.
This process of defining and computing mappings from
our input space to feature values is known as \emph{feature engineering}.
It is often laborious and costly, involving meticulous analysis of the data and
making judgment calls on the usefulness of individual features to a machine learning model \citep{10.1145/1277741.1277811,10.1145/2970398.2970433}.
Furthermore, our model's ability to learn an effective ranking function from
engineered features is tied to and bounded by their richness and discriminative
power, which may be limited because we design features following our own
intuition and often incomplete understanding of the problem.

Can we avoid the costs and pitfalls of feature engineering and
find features that are more helpful to our model? In other words,
instead of constructing them by hand, can we learn the functions
$\phi_q(\cdot)$, $\phi_d(\cdot)$, and $\phi_{q,d}(\cdot, \cdot)$?

The question above hints at one of the primary reasons behind
the emergence of deep learning in information retrieval. After
researchers in other communities demonstrated the success of
deep neural networks in learning rich representations from raw
images and natural texts, it was only natural to consider their
application to text ranking. The promise deep learning held
for ranking was that it would obviate the need for extensive
feature engineering and, instead, it would automatically learn features
that give the model the necessary power to estimate relevance.

There are three major directions in the information retrieval literature
that explore the role of deep models in learning a representation
of queries and documents. We review these briefly in the remainder
of this chapter.

\begin{figure}[t]
    \centering
    \centerline{
        \subfloat[Representation-based]{
        \includegraphics[width=0.5\linewidth]{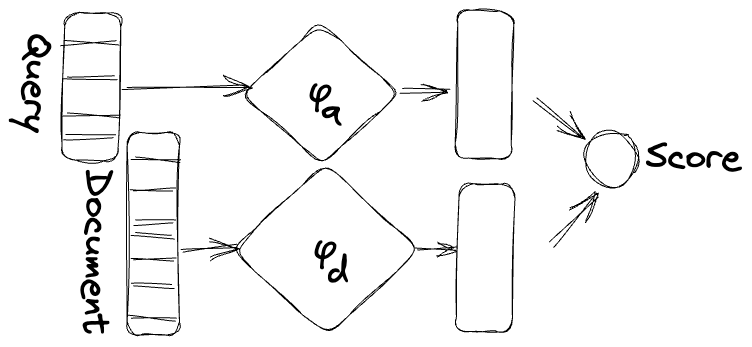}
        \label{fig:07:nir-models:representation}}
        \subfloat[Interaction-based]{
        \includegraphics[width=0.5\linewidth]{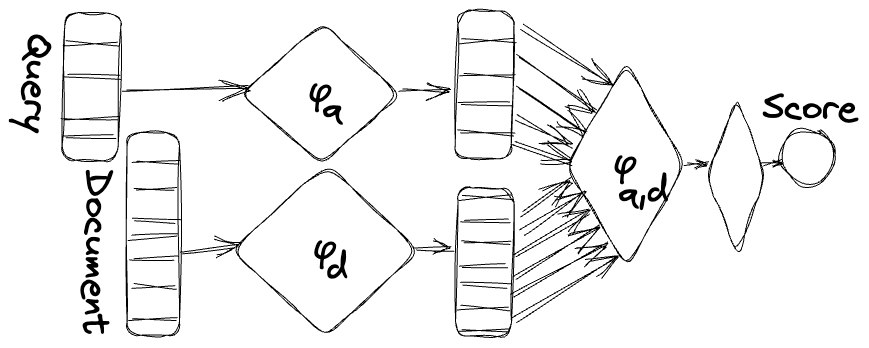}
        \label{fig:07:nir-models:interaction}}
    }
    \caption{Schematic illustration of pre-Transformer neural rankers. In (a), the two functions $\phi_q$ and $\phi_d$ learn
    representations of query and document such that relevant documents stay ``closer'' to the query than non-relevant ones,
    where closeness is measured by a vector similarity function such as cosine similarity.
    In (b), the representations of query and document terms \emph{interact} with each other in $\phi_{q,d}$
    and their relevance is estimated by a subsequent function.}
    \label{fig:07:nir-models}
\end{figure}

\section{Representation-based models}

The first wave of deep learning models for ranking,
aptly known as \emph{representation-based}
models, focus squarely on learning $\phi_q(\cdot)$ and $\phi_d(\cdot)$
(with $k_q=k_d$) from queries and documents such that the representations
of relevant documents are ``closer'' than non-relevant ones to the representation of queries.
Closeness is typically determined with a simple metric such as inner product or
cosine similarity. We illustrate a generic version of this approach in
Figure~\subref*{fig:07:nir-models:representation}.

An early iteration of this idea called
Deep Structure Semantic Model (DSSM)~\citep{huang2013learning}, for example,
uses feed-forward networks to learn $\phi_q$ and $\phi_d$ from character $n$-grams
of queries and documents. Subsequent works in this space extend the same idea
in different ways. \citet{shen2014learning}, for example, uses
convolutional neural networks instead to capture contextual features.
Dual Embedding Space Model (DESM) \citep{mitra2016dual} takes as input the
pre-trained word2vec~\citep{word2vec} representations instead of character $n$-grams.

\section{Interaction-based models}

What is left out of the representation-based models is the joint
query-document function $\phi_{q,d}(\cdot, \cdot)$; no component
of these models captures the interactions between query terms and document terms.
Modeling $\phi_{q,d}$ motivated another class of neural rankers
that are often known as \emph{interaction-based} models.
Specifically, as depicted in Figure~\subref*{fig:07:nir-models:interaction},
these models create an ``interaction'' matrix of the representations
of query terms and document terms, often in the form of a similarity matrix.
The interaction matrix then becomes an input to another function that estimates relevance.
Methods in this class include DRMM~\citep{Guo2016DRMM}, KNRM~\citep{xiong2017knrm},
and ConvKNRM~\citep{dai2018convkrnm}.

The distinction between representation- and interaction-based methods
lies not just in what features each is capable of learning which affects
their effectiveness, but also in their computational efficiency.
Because representation-based models learn to map documents to
representations with a function $\phi_d$ that is distinct from
$\phi_q$, we can store learnt document representations to enable efficient inference.
Interaction-based models, in contrast, are more expensive because
we must compute $\phi_{q,d}$ during inference as its output
depends jointly on the query as well as the document.
Despite these differences, the two ideas are not mutually exclusive.
In fact, models such as DUET~\citep{mitra2017learning},
incorporate elements of representation- and
interaction-based models to learn all three mappings
$\phi_q$, $\phi_d$, and $\phi_{q,d}$ for effective and efficient ranking.

Foregoing feature engineering for representation learning also makes it
possible to explore functions beyond the three mappings above. If we can
model the interactions between query terms and document terms, for example,
why stop there and not capture the interactions among the set of documents
being ranked too? In other words, we may extract additional features as a joint
function of a set of $m$ documents $\phi_{q, \bm{d}}: \mathcal{Q} \times \mathcal{D}^m \rightarrow \mathbb{R}^{k_{q,\bm{d}}}$. Different flavors of this idea were investigated
by~\citet{ai2019groupwsie} and~\citet{pang2019setrank}, where the main challenge
is in ensuring that $\phi_{q, \bm{d}}$ is permutation-invariant
(i.e., the output of the function does not depend on the order in which documents
are presented to the function).

Our brief review above only sketches an outline of ideas in
the early years of neural ranking and leaves out a great deal of
details. We refer the interested reader to existing surveys
on representation- and interaction-based neural rankers for a more comprehensive
review and analysis of these methods~\citep{Onal2018NeuralIR,mitra2017neural,Guo2020NeuralRanking}. But even from this outline emerges a clear picture: models
grew more and more complex as we sought to enrich the representations of queries
and documents. That trend continues to date, with a notable jump
in model complexity when rankers based on the Attention mechanism in
Transformers~\citep{vaswani2017attention} dwarfed many early
models.\footnote{We often refer to neural rankers that are based on the Attention
mechanism as ``Transformer-based'' rankers. However, we recognize that
``Transformer'' implies an encoder \emph{and} a generative decoder neural module,
with the latter playing no role in the vast majority of retrieval and ranking systems.}

\section{Transformer-based models}
\begin{figure}[t]
    \centering
    \centerline{
        \subfloat[monoBERT]{
        \includegraphics[width=0.5\linewidth]{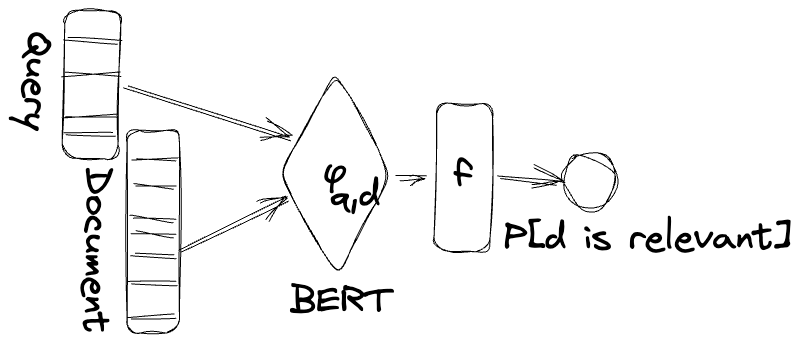}
        \label{fig:07:bert:mono}}
        \subfloat[duoBERT]{
        \includegraphics[width=0.5\linewidth]{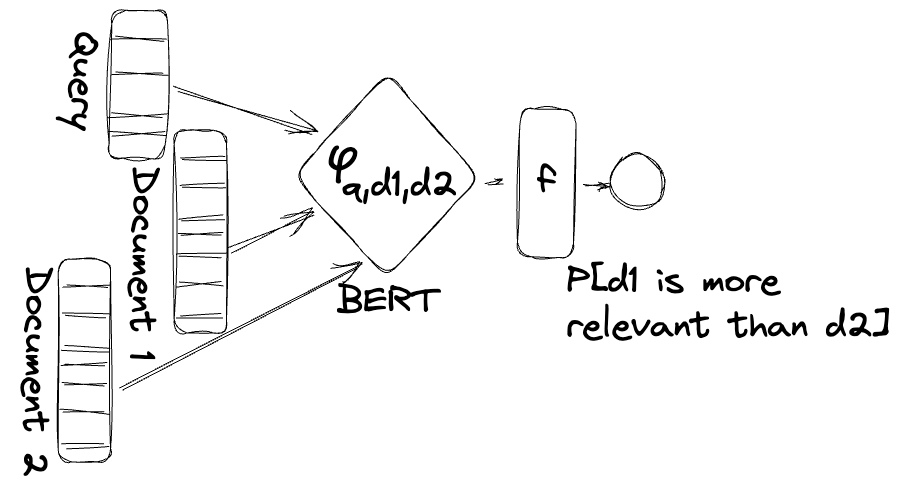}
        \label{fig:07:bert:duo}}
    }
    \caption{Illustration of the (a) monoBERT and (b) duoBERT rankers. MonoBERT predicts a
    relevance score of a single query-document pair---a familiar scoring machinery in \ac{ltr}.
    DuoBERT takes a pair of documents and predicts whether the first document is more relevant to
    the query than the second document. Note that, the scoring function itself takes two documents as input,
    in contrast to \emph{pairwise} \ac{ltr} where the \emph{loss} function takes the scores of two documents.}
    \label{fig:07:bert}
\end{figure}

That began when~\citet{nogueira2020passage} reported a dramatic jump in ranking quality
by applying Bidirectional Encoder Representations from Transformers
(BERT)~\citep{devlin2019bert} to the MS MARCO~\citep{nguyen2016msmarco}
passage re-ranking task, where short passages are to be ranked with
respect to a text query. Their model was later named ``monoBERT.''
In the language of our discussion here, the BERT component in
monoBERT serves as the joint function $\phi_{q,d}$,
producing a representation for a query-document pair. From that, a simple
feed-forward network learns to estimate relevance by optimizing a pointwise loss function.
This is illustrated in Figure~\subref*{fig:07:bert:mono}.
The remarkable effectiveness of this network architecture generated much
excitement in the community and led to a flurry of research activity.

Many subsequent works~\citep{yilmaz2019birch,Li2020PARADEPR,dai2019bertmaxp,macavaney2019cedr}
seek to address monoBERT's limitation on input size (capped at $512$ tokens),
which was enough to rank short passages but not sufficiently large to apply
to long documents. Others~\citep{nogueira2019multi} extended the model to
learn representations for pairs of documents; that is, $\phi_{q,\bm{d}}$ where $\bm{d}$ is a set of two documents.
The resulting model is known as ``duoBERT'' and is shown in Figure~\subref*{fig:07:bert:duo}.
Yet others~\citep{nogueira2020monot5,pradeep2021expandomonoduo} go beyond BERT
and use a sequence-to-sequence model such as Text-to-Text
Transfer Transformer (T5)~\citep{raffel2020t5} for ranking.
We refer the interested reader to a recent survey by~\citet{lin2021pretrained}
on Transformer-based rankers for a detailed discussion of each method and
their many existing variants.

As is often the case, this march from basic feed-forward networks
to gargantuan stacks of Transformer-based neural modules with millions of parameters
has made stunning improvements in quality possible only at the expense
of training and inference efficiency. Better accuracy through ever-increasing
complexity once again presents a new but familiar challenge: How do we
balance the two competing objectives of efficiency and effectiveness?
This question has gained even more significance due to the sheer scale
of neural rankers and the the multitude of additional efficiency dimensions they
introduced, such as sample- and energy-efficiency. Additionally, scaling
these models to long documents (as opposed to short ``passages'') introduces
another efficiency challenge that is largely unique to neural rankers~\citep{lin2021pretrained}.

In the next chapter, we will review some of the ideas that explore
this trade-off in the context of neural rankers, many of which will,
unsurprisingly, look familiar to the reader, just as the question above did.
So as we present our summary of each class of methods, we highlight
their connection to the first half of this manuscript.

\chapter{Efficiency in Neural Learning to Rank}
\label{chap:nir}

We have argued in this monograph that neural rankers are just another instantiation of
the general \ac{ltr} framework, where the hypothesis class is the set of deep
neural networks. It is therefore not surprising that the ideas that were introduced in
previous chapters to manage inefficiency in decision forest models carry over to neural \ac{ltr}
at a high level.

This portability of ideas is easy to see in the case of the \emph{multi-stage} and \emph{cascade} architecture of retrieval and ranking
because their general setup is agnostic to the specific choice of models in each stage.
In other words, by trimming the candidate list in the first stage (or first few stages),
we can reduce the volume of candidates that must be re-ranked by an expensive, neural ranker, thereby improving
the inference and training efficiency of the end-to-end ranking system.

This general method was first investigated by~\citet{nogueira2019multi} in the context of neural rankers.
\citet{nogueira2019multi} observe that
duoBERT, which learns to score pairs of documents jointly as explained in Chapter~\ref{chap:neural},
is more effective than monoBERT but at a much higher inference cost. That is because, given $k$ candidates,
duoBERT performs inference on $k(k - 1)$ pairs of documents $d_i$ and $d_j$ to estimate the probability $p_{i,j}$
of $d_i$ being more relevant than the other.
It then aggregates the probabilities $p_{i,\ast}$ using one of the many proposed aggregation functions (e.g., sum)
to arrive at a single relevance score for individual documents. It is clear that duoBERT, due to its more
computationally intensive inference, would fare better for smaller values of $k$ relative to monoBERT.

\begin{figure}[t]
    \centering
    \includegraphics[width=\linewidth]{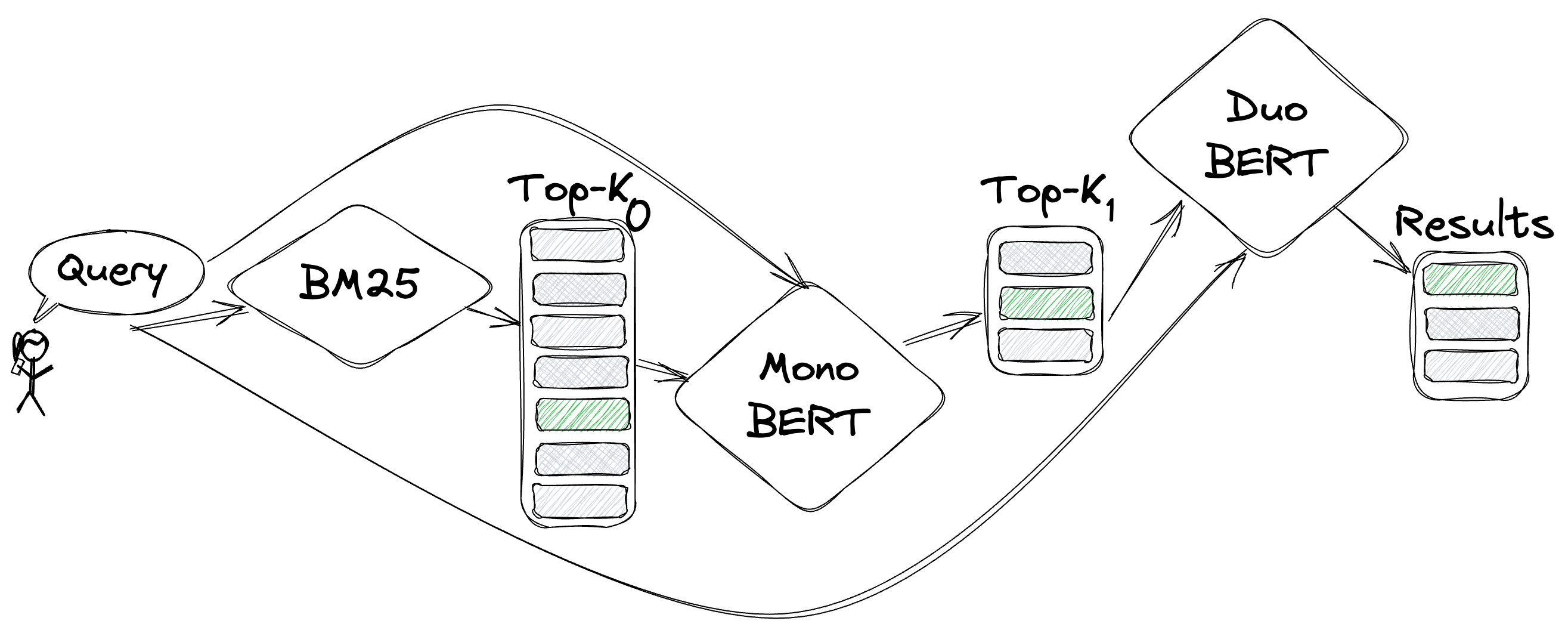}
    \caption{Multi-stage search pipeline of~\citet{nogueira2019multi} consisting of
    BM25, monoBERT, and duoBERT.}
    \label{fig:08-mono-duo-multistage}
\end{figure}

That prompted the authors to consider a multi-stage ranking pipeline illustrated in
Figure~\ref{fig:08-mono-duo-multistage} where candidates, generated by BM25,
are first ranked by monoBERT and only then the top candidates are rearranged by a duoBERT ranker.
\citet{nogueira2019multi} then explore the trade-offs such a setup offers between ranking quality
and inference latency, by studying the interplay between quality and the number of candidates retrieved with BM25,
along with the number of candidates passed from monoBERT to duoBERT.

Among the many interesting observations,
they found that providing a larger pool of candidates to monoBERT helps ranking quality---up to a point,
beyond which we see diminishing returns. That indicates that documents with a relatively low BM25 score
can indeed be relevant to the query and be placed at higher ranks with monoBERT.
Interestingly, it is often enough to apply duoBERT to a handful of top documents ranked by monoBERT
to obtain the highest achievable ranking quality.

In fact, this last point turns into a rather surprising phenomenon in a follow-up study.
\citet{pradeep2021expandomonoduo} extend the multi-stage pipeline of Figure~\ref{fig:08-mono-duo-multistage}
and add one more stage right before BM25: The authors use doc2query-T5~\citep{doc2query-t5},
a \emph{generative} model, to expand documents with \emph{predicted} queries, prior to the BM25 stage.
They also replace BERT with an adaptation of T5~\citep{raffel2020t5},
a sequence-to-sequence model, to the ranking task. They find that, for some but not all monoT5's
aggregation functions, passing more candidates between monoT5 and duoT5 re-rankers leads to a drop in quality.
\citet{pradeep2021expandomonoduo} do not articulate if the quality degradation is statistically
significant, nor do they explain why they observe this rather counter-intuitive behavior.
It is therefore unclear if this points to a weakness of the duoT5 model itself, or
the aggregation functions used to produce relevance scores from probabilities.

Some small details aside, the general observations reported in these works are consistent
with the prior literature on multi-stage ranking systems which attests to the robustness of
this general and rather intuitive design. In fact, the idea is so natural that others have
also investigated similar setups in different contexts, e.g.,~\citep{matsubara2020multistage,zhang-etal-2021-ltrmuppets}.

In the remainder of this chapter, we summarize other solutions for
efficiency in neural \ac{ltr} whose connection to the earlier literature is less obvious, and discuss other new problems.
We revisit, for example, \emph{early-exit strategies} and show how this simple technique can be baked into a neural \ac{ltr}
model. We show how \emph{distillation} can be used to find a small, more efficient model given a large, more effective ranker.
Finally, we review the literature on dense or semantic retrieval and describe the challenges this new problem introduces.

\section{Early exit strategies}
What motivated~\citet{cambazoglu10early} to only partially evaluate a gradient-boosted decision forest for some documents, thereby
exiting the inference algorithm early, was the hypothesis that trees that come later in the forest are there to refine the ranking
among the top candidates; the scores of the vast majority of candidates, especially those that are obviously non-relevant,
should not change dramatically after the evaluation of the first few trees. So using some form of thresholding, we can
exit early and prevent certain candidates from going through an entire forest.

In effect, a gradient-boosted forest can itself be understood as a cascade ranker where each stage is a decision
tree. We can therefore trim the candidate list every few stages in the cascade and progressively reduce the cost
of inference for any query.

Interestingly, one can view a stack of Transformers much the same way: Each Transformer
layer in a multi-layered model such as BERT is akin to a stage in a cascade
ranking system! As a candidate list bubbles up the stack of Transformers, analogous to the decision forest scenario,
the score of non-relevant documents should change less and less substantially and their position in the ranked list
should move up or down less dramatically. As such, it may be possible to exit inference early for a subset of candidates
and avoid evaluating the full model on the entire candidate list, all without noticeable impact on effectiveness.

\begin{figure}
    \centering
    \includegraphics[width=\linewidth]{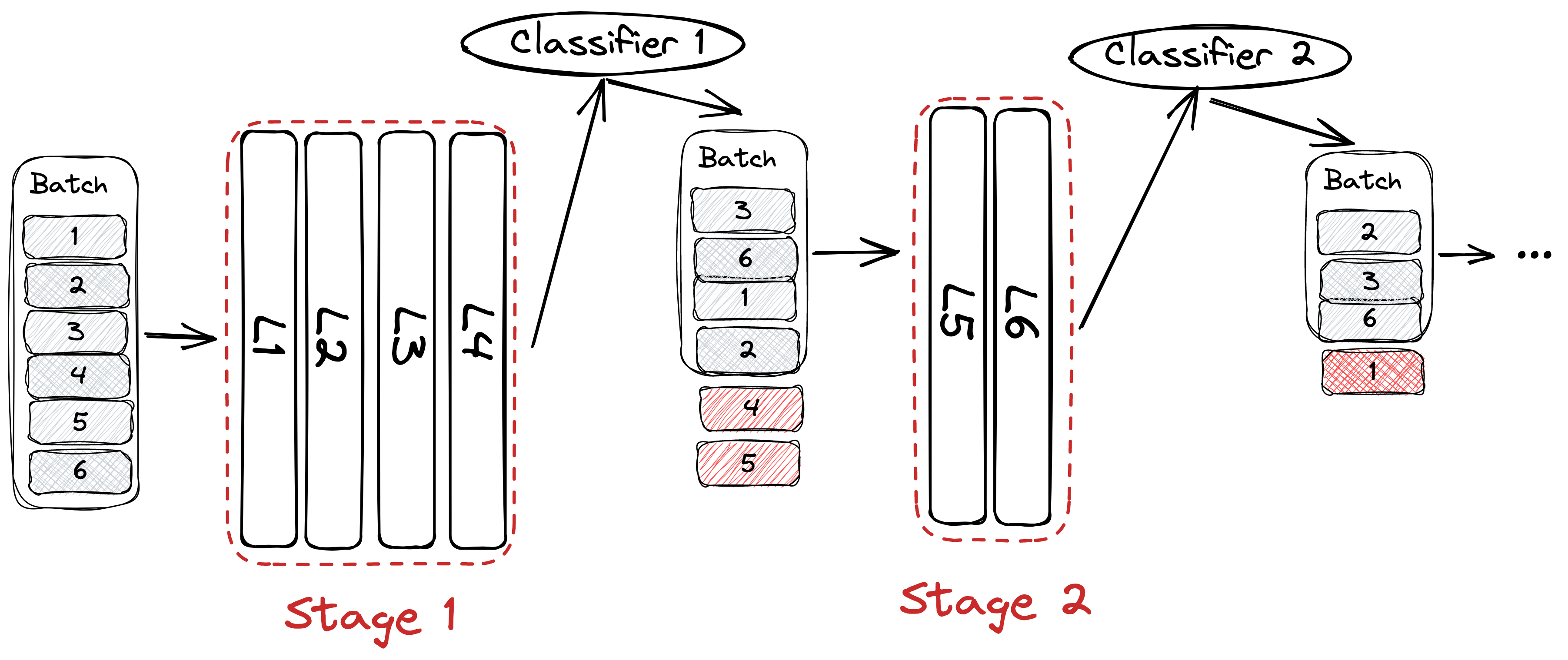}
    \caption{Illustration of the Cascade Transformer~\citep{Soldaini2020TheCT}.
    Every group of Transformer layers forms a stage of the cascade and is followed by
    a classifier, whose output is used to rank candidates in a batch.
    Between consecutive stages, the model discards $\alpha\%$ of candidates.}
    \label{fig:08-cascade-transformer}
\end{figure}

That is, in fact, one major idea that has emerged in the neural information retrieval literature to improve
inference efficiency. In the context of an Answer Sentence Selection task for Question Answering, for example,
\citet{Soldaini2020TheCT} apply that idea to the multi-layered Transformer model---which
they call Monolithic Transformer---to obtain the Cascade Transformer.
In particular, they intersperse lightweight classifiers between every few
layers of the model, so that layers $4$, $6$, $8$, $10$, and $12$ become exit points in the cascade.
This is illustrated in Figure~\ref{fig:08-cascade-transformer}.

Conceptually, one can think of each classifier as predicting whether a document should continue to be evaluated
or whether it can safely exit the inference altogether. The concrete logic is quite straightforward:
Given the predictions of one classifier on a batch of candidates,
the cascade ranks the candidates according to their scores and discard $\alpha\%$ of them,
before passing the rest onto the next stage in the cascade.
This logic repeats until the very last classifier, which produces the final predictions for the remaining candidates.
For example, if there are $128$ documents in the initial batch and $\alpha=30\%$, the first
exit point drops $38$ documents, with the remaining $90$ documents moving onto the next exit point.
The authors justify this trimming strategy by noting that, discarding a fixed number of items from a batch
allows the cascade to know the batch size in each stage \emph{a priori}---a desirable
behavior in today's neural network inference engines. Were they to discard candidates by a
fixed score or rank threshold as~\citet{cambazoglu10early} did, in contrast,
batch sizes at each stage of the cascade would become dynamic,
possibly leading to sub-optimal throughput.

Another notable difference with earlier work in the context of decision forests is that
the model and classifiers can be learnt jointly in an end-to-end manner: At each training iteration,
the training procedure selects one stage in the pipeline uniformly at random, computes the value of the loss
function, and back-propagates the error throughout earlier stages. This training schedule,
the authors claim, makes the later stages of the cascade more robust to noise that stems
from high variance in earlier stages.

\citet{Soldaini2020TheCT} evaluate the Cascade Transformer in terms of quality and inference cost reduction
with respect to a Monolithic Transformer on a number of Question Answering datasets. The results present
no surprises: Choosing a larger value of $\alpha$ and discarding candidates more aggressively between stages
of the cascade degrades end effectiveness but leads to a larger inference cost reduction ($-37\%$ when $\alpha=30\%$).

While the Cascade Transformer was tailored to improving \emph{throughput}---hence the emphasis on
fixed batch sizes throughout the cascade---other works target the inference \emph{latency}
(i.e., batch size of $1$) using very similar ideas. For example,~\citet{xin-etal-2020-deebert}
explored a design that inserts a classifier after every layer. At each exit point (called ``off-ramp''
in their work), its classifier is evaluated and, depending on the confidence in the classifier's prediction,
inference terminates early or the sample goes on to be evaluated by later layers.
In a follow-on study, \citet{xin-etal-2021-berxit} present an evolution of this idea
where, instead of deciding to exit early based on the classifiers' confidence,
the decision to exit or not is itself learnt at every exit point.

These and other contemporary works on early-exit strategies in Transformer-based
models~\citep{liu-etal-2020-fastbert,schwartz-etal-2020-right} are a natural application of the idea of
partial evaluation of~\citet{cambazoglu10early} to neural rankers,
often resulting in similar trade-offs between effectiveness and efficiency.

\section{Knowledge distillation and neural compression}

We introduced knowledge distillation in Chapter~\ref{chap:learning}, where the idea was to train a small and efficient
``student'' model that attempts to imitate the behavior and predictions of a larger and effective ``teacher'' model.
The intuition was that, because the teacher model has a greater complexity, it
captures nontrivial relationships in the data and cleans up noisy input. In this way, the student
model learns from a ``clean'' version of the data and can focus on learning what is more germane to the task.
In that chapter, we also reviewed examples of what we argued may be seen as knowledge distillation in the decision tree-based \ac{ltr} literature.
This included \xdart{}, \cleaver{}, and \xcleaver{}, which arrive at a student model by \emph{compressing} an existing tree ensemble.

Compression is easy to translate to the world of neural networks:
Individual connections or entire layers can be removed, reset in a network
if their existence is not ``salient'' according to some definition of salience,
or quantized.
The resulting network may become leaner in size or its weight matrix sparser,
leading to different and often more efficient computational patterns.
We refer the reader to existing surveys~\citep{xu2022survey} that cover this growing literature in the
context of large language models for a detailed discussion of these methods.

Knowledge distillation~\citep{ba2014deep} in the deep learning literature works not by pruning a large model, but by
learning a new, more compact model instead.
Perhaps the most relevant and pivotal studies from the knowledge distillation literature
are TinyBERT~\citep{jiao-etal-2020-tinybert} and DistilBERT~\citep{sanh2020distilbert}. While the specifics of these two
works are different, the general idea is rather similar: Given a large pre-trained BERT language model, the two studies
explore strategies to train a more compact student model that uses the same Transformer architecture but has fewer parameters.
In TinyBERT, for example, the student model learns to fit the individual weight matrices from the attention modules of
a large BERT model, with the intuition that linguistic knowledge can be transferred to the student in this way.
Additionally, the output of Transformer layers, the embedding layers, and the predictions of the model too become
objectives for the student model to attain. As one would expect, distillation leads to models that are several factors
faster and lighter than the large teacher models, with little to no loss in effectiveness.

In the context of \ac{ltr}, we have already called out three works~\citep{tang2018ranking,cohen2018universal,9716821}
in Chapter~\ref{chap:learning} that demonstrate the usefulness of knowledge distillation for balancing efficiency and effectiveness of ranking models.

Building on that foundation,~\citet{gao2020distillation} explore how \emph{ranking-specific} knowledge can be transferred between
a teacher and a student BERT-based model.
\citet{gao2020distillation} study several distillation strategies. In one, dubbed ``Ranker Distill,''
they train monoBERT over the MS MARCO~\citep{nguyen2016msmarco} dataset,
then randomly initialize a smaller student model and ask it to
reproduce the teacher model's ranking behavior.
In ``LM Distill + Fine-tuning'', they transfer knowledge
from a pre-trained, general-purpose BERT model (but not monoBERT) to the student model,
and only then fine-tune the student model for ranking.
Finally, in a hybrid of the two methods, they have the student
model learn from a general-purpose BERT model first, followed by distillation of
ranking behavior from mono-BERT.

Of the three distillation strategies,~\citet{gao2020distillation} find that the hybrid approach
can produce a student ranking model that is just as effective as monoBERT but that is up to $9$ times
faster during inference. This is an important finding that goes to suggest that large models may 
be over-parameterized and that knowledge distillation can substantially reduce inference cost
in exchange for additional training of a more compact, student model.

One can even take the idea of distillation a bit further and distill the collective knowledge of an \emph{ensemble} of
teacher models into a compact, student model. This is the idea~\citet{zhuang2021ictir} studied for ranking,
which they claim leads to a model that is more efficient during inference and that displays more stable predictions.

As the authors articulate, one can either have the student predict an \emph{aggregated}, ensemble-level teacher label,
or solve a multi-objective optimization problem by predicting all model-level labels simultaneously.
In either case, we need to define what the model- or ensemble-level labels are.
\citet{zhuang2021ictir} experiment with two model-level labels: (1) the raw score of rankers in the ensemble;
and, (2) the reciprocal rank (i.e., $1/\big( \alpha + \pi_i \big)$, where $\alpha$ is a constant and $\pi_i$ is
the rank of document $i$). They define the ensemble-level label as the mean of model-level labels.

Their experiments on the MS MARCO passage ranking dataset appears to corroborate the authors' hypothesis
that distilling knowledge from multiple teachers can indeed produce a high-quality student model.
It turns out that optimizing a single loss defined on ensemble-level labels is just as good as
a multi-objective formulation; and that, using raw scores versus reciprocal ranks makes little difference in the end.

While the idea in this article is interesting in and of itself, it is unclear how generalizable the
empirical findings are---as the authors themselves point out too.
Perhaps what gives one pause is that the student model is itself a BERT ranker
and that the individual rankers in the teacher ensemble are also BERT rankers that are all trained in
the same manner but starting out from different initial weights.
The authors justify this by suggesting that evaluating a single BERT ranker is cheaper than
executing an ensemble of BERT rankers.
That may be true, but that small detail may explain the observed insensitivity to the
choice of labels and the distillation strategy.
More importantly, one wonders if, in this instance, knowledge distillation is not
simply related to variance reduction as the student model learns the average of predictions of teacher models.

\section{Dense retrieval}
One of the major innovations in neural \ac{ltr} research was the evolution of ``cross-encoders'' such as monoBERT
and duoBERT to two-tower or ``bi-encoder'' models. Instead of learning a parameterized function (like BERT)
that takes a query-document pair (or multiple documents) as input and simultaneously learns
representations (i.e., features) and predicts their relevance,
bi-encoders disentangle the relevance prediction function from representation learning, and simplify the former
as much as possible. The idea is to push much of
the complex, time-consuming inference operations offline and thereby speed up query processing.

\begin{figure}
    \centering
    \includegraphics[width=0.7\linewidth]{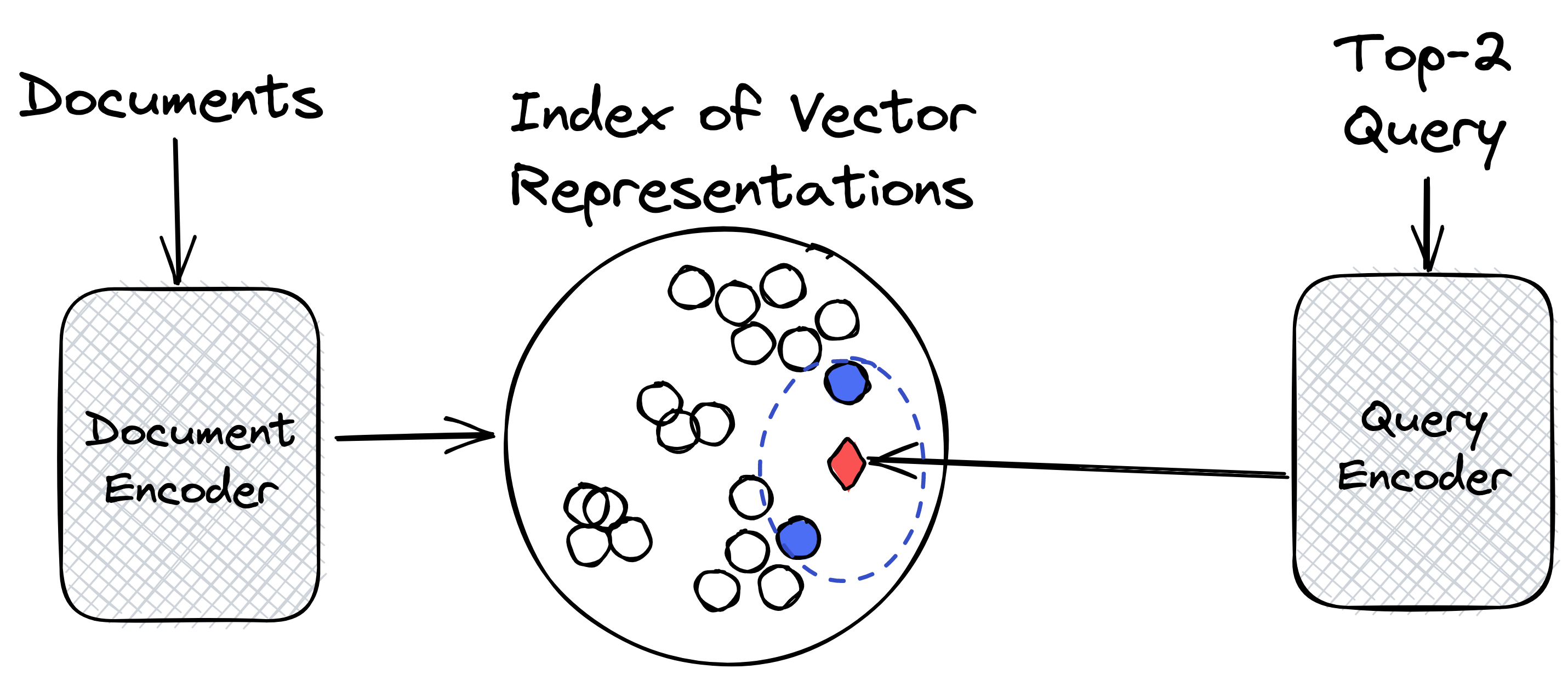}
    \caption{The most basic form of dense retrieval produces vector representations (circles) for a collection of
    documents offline. During inference, the query is similarly transformed into a vector (red rhombus)
    in the same representational space. An approximate nearest neighbor algorithm then finds the closest
    document vectors to the query vector (blue circles). By training the encoders in the right way, we can structure
    the representational space such that closeness in the space implies semantic similarity.}
    \label{fig:08-bi-encoder-dense-retrieval}
\end{figure}

In its most basic form, this design resembles the representation-based
neural ranker architecture of Figure~\subref*{fig:07:nir-models:representation},
but where $\phi_d$ and $\phi_q$ are pre-trained large language models that may be further fine-tuned.
Because the representation of documents is independent of queries,
we may store the document vectors in an offline index and, during retrieval,
compute the representation of the query and find its closest document vectors.
This paradigm is often referred to as ``dense retrieval'' or ``semantic search,'' which is typically solved
using a $k$ nearest neighbor (NN) search or $k$ approximate nearest neighbor (ANN) search algorithm.
We illustrate this procedure in Figure~\ref{fig:08-bi-encoder-dense-retrieval}.

The feasibility of this basic idea was demonstrated by several
works~\citep{repbert2020dr,karpukhin-etal-2020-dense,ma2021replication-of-dpr,qu-etal-2021-rocketqa}.
\citet{karpukhin-etal-2020-dense} present what they call Dense Passage Retrieval
(DPR) and show that dense vector representations can indeed be used to perform
the retrieval task following the recipe above.
Interestingly, effective representations can be learnt from a small number
of questions and passages (in a question-answering task)
by a simple bi-encoder framework. They evaluate DPR on a wide range of
open-domain QA datasets, with the results showing that
DPR outperforms a strong BM25 system by 9\%--19\% absolute points
in terms of top-$20$ passage retrieval accuracy.

In the general dense retrieval framework of Figure~\ref{fig:08-bi-encoder-dense-retrieval},
there are two key factors that contribute to the overall efficiency and effectiveness:
the ANN algorithm itself that performs the search over millions or billions of vectors, and, as we will discuss,
the nature and quality of the learnt vector representations.
In the rest of this section, we pay particular attention to the latter and review
a subset of methods that encode queries and documents for dense
retrieval.\footnote{We reiterate that our monograph is primarily focused on \emph{ranking}.
While recent works on dense retrieval blur the line between retrieval and ranking,
we still believe that a deeper discussion of dense retrieval and efficiency and effectiveness
trade-offs in that literature is beyond the scope of this monograph.}

Existing encoding models fall into one of two categories based on the granularity of the representations they produce:
per-document encoders (\emph{single-vector}) and per-term encoders (\emph{multi-vector}). The former class, which includes DPR, learns a single vector representation for a document, whereas the latter produces a contextualized
representation for every term in a document.

\subsubsection{Single-vector dense retrieval}

Intuitively, for a single-vector dense retrieval method to work well,
the structure of the representational space must tightly preserve the
semantic similarity between pieces of text. In other words,
vectors that are close to each other according to some vector distance function,
should be semantically similar to each other and vice versa.

A key factor that contributes to the quality of representations and thus the structure
of the space is the quality of the negative examples used to train an encoder.
This is a challenge because, unlike in standard \ac{ltr} where the assumption is that
a short list of documents with relevant and possibly non-relevant documents exist,
in dense retrieval the set of possibly non-relevant documents is extremely large and diverse;
everything in a collection minus the very few positive documents is non-relevant to any arbitrary query.
Choosing which documents to present to the model as negative examples can
have a profound effect on the effectiveness of the final model and its representational space.

In DPR~\citep{karpukhin-etal-2020-dense}, for example, negative documents are selected using three different strategies:
randomly; by retrieving top-$k$ documents with BM25 and selecting those that
do not contain the answer but have significant lexical overlap with the query; or,
positive documents for other queries that appear in the same training batch.

ANCE~\citep{xiong2021approximate}, which like DPR is a single-vector encoder,
continually finds negative examples based on the current structure of the vector space.
The way it realizes this idea is by asynchronously updating an ANN index
\emph{while} the dense retrieval model is being trained,
and retrieving the top-$k$ set for training queries from this index.
Every document that appears in the top-$k$ set that is not a positive example
is deemed a ``hard'' negative and used to further train the model.
This training procedures continues until convergence.

The authors show that ANCE outperforms other competitive dense and sparse retrievers,
with substantial margins in long document retrieval task.
Perhaps more importantly, by measuring the success rate of an ANN algorithm
in finding relevant documents, they claim that sampling negative examples as done in ANCE
improves the quality of the final vector representations.
In fact, an ANCE-based dense retrieval model approaches the
effectiveness of a cross-encoder, BERT re-ranker in a multi-stage setup.
These observations,~\citet{xiong2021approximate} state, cast doubt on
``a previously-held belief that modeling term-level interactions is necessary in search.''
Comparing their inference with a BERT re-ranker, they conclude that
ANCE brings about a speedup of 100$\times$ due to higher quality representations.
While some of these claims are not entirely supported and the speedup may be exaggerated,
the fact is that sampling negative examples in this way appears to be more
effective than previous strategies.

While the relative inference efficiency and effectiveness of ANCE is rather impressive,
its training is resource-intensive and time-consuming.
This is because an ANN index must be updated with the latest representations
and negative examples must be retrieved from this index periodically.
To lessen the training cost, \citet{lindgren2021efficient} suggest to maintain a
cache of possibly-stale negative examples.
Caching, as shown in experiments, allows the training procedure to work more efficiently and scale
to a larger pool of negative examples with a lower memory and computational footprint.
It turns out, as the authors show in their theoretical analysis, updating a very small
fraction of the cache at each iteration ensures fast convergence.

Up to this point, the marriage of ANN search algorithms and learnt vectors
from general-purpose language models and encoders led to much success in retrieval quality.
\citet{gao2021condenser}, however, question the suitability of the latter:
Are general-purpose language models optimal for dense retrieval?

\citet{gao2021condenser} argue that existing language models are
suboptimal and inefficient because of the way they aggregate and condense information
into a single vector, which is designed for tasks (such as next sentence prediction)
that are removed from the objectives of dense retrieval.
This is because, how information is aggregated and what signals are
encoded are determined by how the ``attention'' mechanism works.
In particular, the representation of the CLS token---a
token that is prepended to a sequence and whose representation is typically
taken as the final vector representation of a query or
document---has weak interactions with other document terms in early layers,
and uses too broad of an attention span over document terms in later layers.

Having made this observation, \citet{gao2021condenser} tailor the Transformer
architecture in a model they call Condenser by customizing how it \emph{attends}
to document terms. Notably, the ``head'' of the model takes the CLS representation
from the later layers as in a Transformer but, additionally,
takes \emph{term representations} from early layers.
The authors experimentally evaluate their model on two public datasets
and show that the use of Condenser improves over standard language models
by large margins on several text retrieval and similarity tasks.

In a follow-up study,~\citet{gao2021unsupervised} argue further that for the CLS representation to be effective, it must be
transformed by a head, a typically non-linear function. This is different from what takes place in ANN search, where we
simply find the closest neighbors with respect to dot product (or other simple distances). There is therefore a disconnect
between how representations are generated and how they are used to perform dense retrieval.
They then propose augmenting the pre-training loss with an unsupervised corpus-level contrastive loss to ``warm start'' the embedding
space; in effect, the pre-training stage asks the model to learn that similar passages (spans within a document) should
have closer representations and dissimilar passages (spans from different documents) should instead be positioned farther.
They show the effectiveness of this regime and, as a side-effect, show that it renders unnecessary
heavy data engineering efforts such as augmentation, synthesis, and filtering.

The single-vector dense retrieval literature is vast and growing still
with many other works that straddle the literature on the ANN as well as
the modeling pieces. We do highlight the works of~\citet{zhan2021joint}
and~\citet{reconc2022wsdm} who investigate a joint optimization of vector representation learning
\emph{and} the construction of the ANN index for more efficient and
effective overall retrieval and a more compact index.

\subsubsection{Multi-vector dense retrieval}

The methods we reviewed so far learn a single vector representation for a document.
\citet{10.1145/3397271.3401075} question whether a simple ANN search over single-vector
representations is sufficient to ensure quality and whether relevance estimation
would improve by replacing the simple distance function (e.g., dot product)
with a more complex function.

ColBERT~\citep{10.1145/3397271.3401075,santhanam-etal-2022-colbertv2} approaches the encoding and similarity estimation problems in two steps: A first step uses a language model (e.g., BERT) to encode query terms and document terms (separately).
The output of this step is a sequence of vectors representing terms in a query or document.
A subsequent step uses a ``late-interaction'' function to estimate the overall
similarity of query and document terms. This function may be the norm
of a matrix whose entry at row $i$ and column $j$ is the inner product of
the query's $i^\mathit{th}$ vector and a document's $j^\mathit{th}$ vector.

As a result of this two-step process, ColBERT can leverage the expressiveness of
deep language models and, at the same time, enable us to pre-compute document
representations offline.
The authors comprehensively evaluate ColBERT
using two passage search datasets (MS MARCO Ranking and TREC Complex Answer Retrieval)
and show that it is more effective than
non-BERT baselines and competitive with existing BERT-based models,
but that ColBERT is two orders of magnitude faster
and requires up to four orders of magnitude fewer FLOPs per query.

It should come as no surprise that ColBERT's higher effectiveness comes at a cost.
One new challenge is the inflated size of the index: rather than storing a single vector for every document
in our index, we must now find room for term-level vector representations.
\citet{santhanam-etal-2022-colbertv2} offer a solution to remedy this particular cost
by reducing the overall memory footprint of the representations.

Another added cost that pits multi-vector representations against single-vector encoders
is that we can no longer perform retrieval in a single step by using an exiting ANN search algorithm:
With single-vector representations, once documents are encoded,
all we must do to retrieve the top-$k$ documents is to ask for the $k$ nearest neighbors from an ANN index.
\citet{tonellotto2021colbert} proposed to rank terms by their importance
and compute the similarity score for query-document pairs only using a subset of query terms instead.
\citet{lin2020distilling} ask, instead, if knowledge distillation
makes it possible to learn a model that is just as effective as ColBERT but offers a single-step search like earlier works.
With the intuition that tight coupling between the teacher and student models may enable more flexible
distillation strategies that yield better representations, the authors show that their distilled model, called TCT-ColBERT,
does indeed improve query latency and greatly reduces memory usage with a limited reduction in effectiveness relative to
ColBERT.

\section{Open challenges and future directions}
\label{sec:neural-future}

Table~\ref{tab:neural-summary} summarizes the methods we reviewed in this chapter.
They naturally fall into three major research directions.
The first covers early exit strategies to speed up the inference of Transformer-based neural rankers.
We discussed the connection to the literature on tree-based \ac{ltr},
where early-exit strategies were applied successfully to ensembles of regression trees.
With Transformer-based rankers, the approach is similar to tree ensembles:
A stack of Transformer layers is equipped with classifiers, placed at different points of the network.
These classifiers are in charge of deciding when to stop the inference of a given document.
Several works contributed to this direction by proposing how to position the different classifiers
and how to decide when to stop the inference.

The second line of research investigates the use of knowledge distillation for ranking,
where we observe two main classes of ideas. The first focuses on applying knowledge distillation
to ensemble of regression trees to distill their ``knowledge'' into a small, more efficient neural networks.
The second concerns Transformer-based rankers and attempts to derive networks that
are faster during inference without loss in accuracy.

The third category is the literature on dense retrieval methods
that concern the efficiency-effectiveness trade-offs with Transformer-based networks.
Most proposals use a pre-trained language model to learn representations of
documents that can be pre-computed and quickly searched through at query processing time with fast similarity operations.
We touched on two main approaches in the literature, single-vector vs. multiple-vector representations, and
reviewed how they induce specific time-space trade-offs involving approximate nearest neighbors search.

We note that the dense retrieval literature is still evolving rapidly with new innovative methods being developed actively
to learn higher-quality representations and to search for approximate nearest neighbors more efficiently and effectively.
As we stated earlier, we believe dense retrieval and the topic of efficiency and effectiveness trade-offs in this specialty
deserves its own, more comprehensive survey than what we delivered in our ranking-focused monograph.
We therefore refer the reader to a recent survey by~\citet{zhao2022dense-retrieval} on this topic for
a complete treatment.

\begin{table}[ht]
\caption{\label{tab:neural-summary} Highlights of neural learning to rank.}
\scriptsize
\begin{tabular}{p{.35\textwidth}p{.10\textwidth}p{.45\textwidth}}
\toprule
\textsc{Method} & \textsc{Category} & \textsc{Strategy} \\
\midrule

{Early-exit on cascade Transformers}~\citep{Soldaini2020TheCT} & Early exit & Discard fixed-size subset of documents at each exit point. \\

{Early-exit on Transformers}~\citep{schwartz-etal-2020-right} & Early exit & Early-exit inference of ``easy'' documents in Transformer-based networks. More layers are executed for ``difficult'' documents. \\

{Per-layer early-exit for Transformers}~\citep{xin-etal-2021-berxit,xin-etal-2020-deebert} & Early exit & One ``off-ramp'' classifier for each layer that terminates inference. Decision taken based on the confidence in the classification~\citep{xin-etal-2020-deebert}. In a later work, the decision of the classifier (i.e., to ``exit'' or not) is learnt at every exit point of the network~\citep{xin-etal-2021-berxit}. \\

{Adaptive inference for distilling fast Transformer-based networks}~\citep{liu-etal-2020-fastbert} & Early exit, Distillation & Student-teacher model applied to BERT. The student model uses classifiers to enable early exits based on confidence. \\

{Distilling smaller models from BERT}~\citep{gao2020distillation,} & Distillation &  Smaller models learnt from BERT and monoBERT. Hybrid distillation strategies (i.e., a student model learns from a general-purpose BERT model first, followed by distillation of ranking behavior from monoBERT) perform as well as monoBERT but are up to 9$\times$ faster at inference. \\

{Single-vector dense retrieval}~\citep{karpukhin-etal-2020-dense,xiong2021approximate,lindgren2021efficient} & Dense retrieval & Documents and passages are encoded as single vectors. Several contributions concern the negative selection strategy that is crucial to generating effective dense representations. While DPR~\citep{karpukhin-etal-2020-dense} selects negative examples in the beginning of the training process, ANCE~\citep{xiong2021approximate} continually finds negative examples based on the current structure of the vector space by using an ANN index as the model is being trained. To limit the computational burden of the learning process, \citet{lindgren2021efficient} propose to cache negative results during training so to work more efficiently. \\

{Multiple-vector dense retrieval}~\citep{10.1145/3397271.3401075,santhanam-etal-2022-colbertv2} & Dense retrieval & Each term of the documents and passages are encoded as a vector. Finer granularity leads to improved performance. Inference is often a two-step process: (1) use a language model to encode query and document terms (separately), (2) use a ``late-interaction'' function to estimate the overall similarity of query and document terms. \\

\bottomrule
\end{tabular}
\end{table}

\chapter{Discussion and Open Challenges}
\label{chap:challenges}

The preceding chapters offered a review of \ac{ltr} and the many ideas put forward in the literature to
understand the efficiency and effectiveness aspects of \ac{ltr} methods. We reviewed tree-based methods
separately from neural network-based methods, but showed how some of the ideas carry from one area to the other.
In this chapter, we conclude our monograph by looking ahead and identifying the problems within this space that
we anticipate will require significant attention from and research by the community in the coming years.

\section{Stochastic cascades}
Conventionally, ranking functions are deterministic: given a query-document pair, the output of an \ac{ltr} function
is a score that captures the relevance between the input query and document. By sorting candidates by this relevance score,
we obtain a final, unique ranked list. It turns out that one may view the set of relevance scores for a list of
candidates together as defining a \emph{distribution} from which a ranked list may be \emph{sampled}. This stochastic
view of ranking scores, first proposed by~\citet{Bruch:wsdm:2020}, has proven to be a principled perspective and has
already led to a flurry of research and many innovations~\citep{oosterhuis2021plrank,diaz2020exposure,Zamani2022cascade}
due to its flexibility and theoretical properties.

One notable application of this idea that is relevant to the discussion on efficiency and effectiveness is the work
of~\citet{Zamani2022cascade} where the authors take the cascade architecture introduced in
Chapter~\ref{chap:scoring} and theoretically analyze the connection between the first-stage retrieval and a second-stage ranker.
In particular, by viewing retrieval and subsequent ranking as a stochastic process, they show that, contrary to conventional
wisdom, it is not enough for the first-stage retrieval to return a candidate list that maximizes recall. Instead, the retrieved
set must maximize \emph{precision}. One implication of this analysis is that retrieving the same number of candidates for all queries in a cascade
architecture is not appropriate, and, in fact, individual queries may require a shorter or a longer list of candidates.

The conclusions of~\citet{Zamani2022cascade} are reminiscent of the work by~\citet{Wang:2011:CRM:2009916.2009934} and help reaffirm
the idea of a simultaneous ranking and \emph{pruning} of the candidate list in each stage of the cascade. But more importantly,
their work lays the foundation for a more principled construction of cascade ranking models where its end-to-end
efficiency and effectiveness may be modeled and optimized. Is it, for example, feasible to construct a cascade system
with improved efficiency (by way of pruning candidate lists between stages) \emph{and} enhanced quality (by maximizing
precision in early stages)? Can we learn the parameters of such a cascade efficiently?
As we stated in our concluding remarks in Chapter~\ref{chap:scoring}, we believe an exploration of this question to be
important and consequential for efficiency and effectiveness in retrieval and ranking systems.

Going one step further, we ask what implications, if any, this stochastic view of ranking systems has for cascade-like
rankers such as decision forests and layered Transformer models. While we often place early exit methods in a category
separate from post-hoc pruning algorithms (of nodes, trees, or neural connections), can we unify these methods instead
by casting ranking as a stochastic process? If so, what opportunities does such a unified framework bring about insofar
as the trade-offs between efficiency and effectiveness? These are open questions that we believe can help shape the future
of this topic.

\section{Retrieval of hybrid vectors}
Throughout this monograph, we emphasized the role of cascade architectures in enabling efficient and effective ranking systems.
But one thread that has emerged in recent years is whether it is feasible for a cascade ranking system to collapse into
a single stage. Can we achieve effectiveness and efficiency (in all its senses) by applying a single function to an entire
collection of documents and directly obtain a ranked list? Indeed, this is one of the motivating factors behind the research
on ``dense retrieval'' methods.

As explained earlier, in most dense retrieval methods, we project documents into a vector space where each coordinate is dense (i.e.,
every coordinate is almost surely non-zero) to obtain a vector representation (or ``embedding''). During inference, queries
too are projected into the same vector space. Finding a ranked list of documents that are the most relevant to a query is then
equivalent to finding the document vectors that are closest to the query vector. This problem can often be solved efficiently
using an Approximate Nearest Neighbor Search algorithm such as FAISS~\citep{Johnson2021faiss} or
Hierarchical Navigable Small World Graphs~\citep{malkov2016hnsw}.

While dense retrieval methods produce high-quality ranked lists, they are typically much more inefficient than their
inverted index-based counterparts such as BM25. This observation has led researchers to explore \emph{sparse representations}.
The crux of the idea is to learn \emph{sparse} representations in a space that has as many dimensions as there are terms
in the vocabulary, where each coordinate encodes the ``importance'' of the corresponding term in the context of a query or document \citep{10.1145/3397271.3401262}.
By regularizing the model to encourage sparsity in its output, we can create vector representations that have very few
non-zero coordinates relative to the total number of dimensions. Given this sparsity, the thinking goes, we may leverage
traditional inverted index-based algorithms for efficient retrieval. Examples of this research include the works
of~\citep{lassance2022sigir,formal2022splade,tilde,zhuang2022reneuir} among others.

While retrieval over learnt sparse representations is often more efficient than dense retrieval, and there is ongoing
research on making ``sparse retrieval'' algorithms more efficient~\citep{mallia2022sigir}, many challenges still remain.
For example, if certain coordinates of document vectors are non-zero for a large portion of the collection, the retrieval
algorithm will need to visit more documents to obtain the top-$k$ candidates, thereby creating scalability and efficiency
issues. Given that existing retrieval algorithms such as~\citep{broder2003wand} and its variants, generally assume that
queries are much shorter than documents, we face similar scalability and efficiency challenges if a query has a large
number of non-zero coordinates in its sparse representation. More research is therefore needed in developing data structures
and algorithms that can operate over sparse representations.

Furthermore, there is increasing evidence that a hybrid retrieval framework---where we \emph{fuse} dense and sparse
retrieval to obtain a final candidate list---brings about substantial gains in retrieval and ranking
quality~\citep{thakur2021beir,luan-etal-2021-sparse,wang2021bert,chen2022ecir,bruch2022fusion}.
While existing studies only consider BM25 for the sparse (also known as ``lexical'' part),
it is in theory possible to extend hybrid retrieval to learnt dense and sparse representations,
resulting in hybrid vectors for queries and documents. In fact, as explained in the previous paragraph,
learnt sparse representations can themselves be dense in some subspace, thereby taking on a hybrid form in practice.
It is as yet unclear how this joint retrieval problem should be addressed and what trade-offs exist
in this regime. We believe these research questions to be important to the discussion on efficiency and effectiveness.

\section{A multi-faceted view of efficiency}
Efficiency has historically been taken to mean space- or time-efficiency, primarily
in the context of inference. But we should not forget the other factors that contribute to the overall efficiency
of a system. For instance, \citet{scells2022sigir-green-ir} show through an extensive comparison of a
range of models from bag-of-words to decision trees to large language model-based rankers, that complex neural models
are unsurprisingly energy-hungry, especially during training. This increased energy consumption coupled with the need
for larger and larger datasets present new challenges to retrieval and \ac{ltr}, especially considering the
environmental impact of this research.

These new challenges underline the importance of broadening the definition of efficiency to encompass
not just time- and space-efficiency as before, but also other related facets such as \emph{sample}-efficiency (i.e., the amount of
data required to train an effective model), \emph{resource}-efficiency (e.g., the amount of computational resources
needed to train a model), and \emph{energy}-efficiency (i.e., the emissions produced during the course of model training).

This expansion requires the development of formal definitions and standardized metrics for measuring and reporting
the efficiency of a retrieval and ranking system. To that end, research is needed to design efficiency-oriented
evaluation protocols and guidelines that can help researchers assess the merits of an approach and better understand the trade-offs between various methods.
For example, if a work improves efficiency in certain dimensions, but not others, all at the cost of effectiveness,
how should we evaluate and interpret the empirical results. This additionally highlights the importance of developing
appropriate benchmark datasets. We believe these research questions to be instrumental to the future of efficiency
within neural retrieval and ranking.

\section{Designing multidimensional leaderboards}
Existing leaderboards and open challenges in information retrieval that draw much attention and competition
from the research community have historically been centered on measures of quality or effectiveness.
For example, the MS MARCO~\citep{nguyen2016msmarco} leaderboard orders submitted systems
for its various tasks in decreasing order of ranking quality such as MRR$@10$.

These leaderboards have demonstrably contributed to the progress we have witnessed over the years:
MRR$@10$ for the MS MARCO passage retrieval task, as a representative example, has remarkably gained over $24$
points since its debut! But as~\citet{santhanam2022moving} argue, the emphasis on quality hides the
fact that some ranked lists are much more expensive to obtain than others. The authors show this
by conducting a \emph{post-hoc} comparison of published works as well as an in-depth cost analysis of representative
methods (BM25, Dense Passage Retrieval, SPLADE, and ColBERTv2) to arrive at conclusions that are
broadly consistent with the observations around model inference of~\citet{scells2022sigir-green-ir}.

\citet{santhanam2022moving} use this fact to encourage the adoption of multidimensional leaderboards
and motivate research on metrics that capture the overall utility of a retrieval or ranking method
in a single quantity. They point to the \emph{Dynascores} proposed by~\citet{Ma2021DynaboardAE} as
one such measure that allows for a single ranking of a collection of systems. For example,
they evaluate the four retrieval methods above in terms of their query latency, accuracy, and
dollar cost (as measured on different cloud-based
hardware platforms per million queries). By assigning different weights to each dimension (a ``policy'')
and combining the measurements using Dynascores according to the policy,
they order retrieval systems by their utility in the context of the given policy.

We too encourage the development of multidimensional leaderboards to incentivize research
into efficient and effective systems. In fact, while~\citet{santhanam2022moving} argue for
leaderboards that capture inference efficiency, we believe training efficiency too must be
reflected in the overall utility of a retrieval and ranking system.
In spite of arguments that training a model incurs a cost that is amortized and thus
comparably insignificant, we note that retrieval and ranking models have a relatively short lifetime:
As the data distribution shifts, models must often be re-trained or fine-tuned on fresh samples.
By incorporating these costs into model evaluation and comparison, a leaderboard could encourage
reusability and recyclability of models. How these costs may be measured and factored into a ranking
on a leaderboard, however, is an open question.

\begin{acknowledgements}
We are grateful to the three anonymous reviewers who perused an earlier
version of this monograph meticulously and gave us constructive feedback.
This manuscript benefited greatly from their thorough and thoughtful suggestions.

We drew inspiration from discussions we had with participants of the Workshop
on Reaching Efficiency in Neural Information Retrieval (ReNeuIR) at ACM SIGIR 2022.
We thank them for the topics they brought to our attention and their insight into
all aspects of efficiency.

Finally, we extend our sincere gratitude to Maarten de Rijke 
for his patience, encouragement, and invaluable feedback as we prepared this manuscript.

This research has been partly funded by PNRR - M4C2 - Investimento 1.3, Partenariato Esteso PE00000013 - ``FAIR - Future Artificial Intelligence Research'' - Spoke 1 ''Human-centered AI'', funded by the European Commission under the NextGeneration EU programme.
\end{acknowledgements}

\backmatter

\printbibliography

@misc{gomes_2017, 
	title={Our Latest Quality Improvements for Search}, 
	url={https://www.blog.google/products/search/our-latest-quality-improvements-search/}, 
	journal={Google}, 
	publisher={Google}, 
	author={Gomes, Ben}, 
	year={2017}, 
	month={4},
	note = {Accessed: 2022-01-24}
}

@misc{the_guardian_2017, 
	title={Google tells Army of 'Quality Raters' to Flag Holocaust denial},
	url={https://www.theguardian.com/technology/2017/mar/15/google-quality-raters-flag-holocaust-denial-fake-news}, 
	author={{The Guardian}}, 
	publisher={Guardian News and Media}, 
	year={2017}, 
	month={3},
	note = {Accessed: 2022-01-24}
}

@inproceedings{santhanam-etal-2022-colbertv2,
	title = "{C}ol{BERT}v2: Effective and Efficient Retrieval via Lightweight Late Interaction",
	author = "Santhanam, Keshav  and
	Khattab, Omar  and
	Saad-Falcon, Jon  and
	Potts, Christopher  and
	Zaharia, Matei",
	booktitle = "Proceedings of the 2022 Conference of the North American Chapter of the Association for Computational Linguistics: Human Language Technologies",
	month = jul,
	year = "2022",
	pages = "3715--3734",
}

@article{LUCCHESE2020100614,
	title = {RankEval: Evaluation and Investigation of Ranking Models},
	journal = {SoftwareX},
	volume = {12},
	pages = {100614},
	year = {2020},
	issn = {2352-7110},
	author = {Claudio Lucchese and Cristina Ioana Muntean and Franco Maria Nardini and Raffaele Perego and Salvatore Trani},
}

@InProceedings{10.1007/978-3-030-99739-7_38,
	author="MacAvaney, Sean
	and Macdonald, Craig
	and Ounis, Iadh",
	title="Streamlining Evaluation with ir-measures",
	booktitle="Advances in Information Retrieval",
	year="2022",
	publisher="Springer International Publishing",
}

@inproceedings{10.1145/3077136.3084140,
	author = {Lucchese, Claudio and Muntean, Cristina Ioana and Nardini, Franco Maria and Perego, Raffaele and Trani, Salvatore},
	title = {RankEval: An Evaluation and Analysis Framework for Learning-to-Rank Solutions},
	year = {2017},
	booktitle = {Proceedings of the 40th International ACM SIGIR Conference on Research and Development in Information Retrieval},
	pages = {1281–1284},
	location = {Shinjuku, Tokyo, Japan},
}

@inproceedings{10.1145/3397271.3401093,
	author = {MacAvaney, Sean and Nardini, Franco Maria and Perego, Raffaele and Tonellotto, Nicola and Goharian, Nazli and Frieder, Ophir},
	title = {Efficient Document Re-Ranking for Transformers by Precomputing Term Representations},
	year = {2020},
	booktitle = {Proceedings of the 43rd International {ACM} {SIGIR} Conference on Research and Development in Information Retrieval},
	pages = {49–58},
	location = {Virtual Event, China},
}

@article{xcleaver,
	author = {Lucchese, Claudio and Nardini, Franco Maria and Orlando, Salvatore and Perego, Raffaele and Silvestri, Fabrizio and Trani, Salvatore},
	title = {X-CLEaVER: Learning Ranking Ensembles by Growing and Pruning Trees},
	year = {2018},
	issue_date = {November 2018},
	publisher = {Association for Computing Machinery},
	address = {New York, NY, USA},
	volume = {9},
	number = {6},
	journal = {ACM Transactions on Intelligent Systems and Technology},
	month = oct,
	articleno = {62},
	numpages = {26},
}

@article{MOLINA202138,
	title = {Efficient Traversal of Decision Tree Ensembles with FPGAs},
	journal = {Journal of Parallel and Distributed Computing},
	volume = {155},
	pages = {38-49},
	year = {2021},
	author = {Romina Molina and Fernando Loor and Veronica Gil-Costa and Franco Maria Nardini and Raffaele Perego and Salvatore Trani},
}

@InProceedings{10.1007/978-3-030-99736-6_18,
	author="Gil-Costa, Veronica
	and Loor, Fernando
	and Molina, Romina
	and Nardini, Franco Maria
	and Perego, Raffaele
	and Trani, Salvatore",
	title="Ensemble Model Compression for Fast and Energy-Efficient Ranking on FPGAs",
	booktitle="Advances in Information Retrieval",
	year="2022",
	publisher="Springer",
	pages="260--273",
}

@inproceedings{DBLP:conf/sigir/BusolinLN00T21,
	author    = {Francesco Busolin and
			  Claudio Lucchese and
			  Franco Maria Nardini and
			  Salvatore Orlando and
			  Raffaele Perego and
			  Salvatore Trani},
	title     = {Learning Early Exit Strategies for Additive Ranking Ensembles},
	booktitle = {Proceedings of the 44th International {ACM} {SIGIR} Conference on Research
			  and Development in Information Retrieval, Virtual Event, Canada},
	pages     = {2217--2221},
	year      = {2021},
}

@inproceedings{10.1145/3397271.3401256,
	author = {Lucchese, Claudio and Nardini, Franco Maria and Orlando, Salvatore and Perego, Raffaele and Trani, Salvatore},
	title = {Query-Level Early Exit for Additive Learning-to-Rank Ensembles},
	year = {2020},
	publisher = {ACM},
	address = {New York, NY, USA},
	booktitle = {Proceedings of the 43rd International {ACM} {SIGIR} Conference on Research and Development in Information Retrieval},
	pages = {2033–2036},
	numpages = {4},
	location = {Virtual Event, China},
	series = {SIGIR '20}
}

@inproceedings{10.1145/3397271.3401262,
	author = {MacAvaney, Sean and Nardini, Franco Maria and Perego, Raffaele and Tonellotto, Nicola and Goharian, Nazli and Frieder, Ophir},
	title = {Expansion via Prediction of Importance with Contextualization},
	year = {2020},
	booktitle = {Proceedings of the 43rd International {ACM} {SIGIR} Conference on Research and Development in Information Retrieval},
	pages = {1573–1576},
}

@ARTICLE{9716821,
	author={Nardini, Franco Maria and Rulli, Cosimo and Trani, Salvatore and Venturini, Rossano},
	journal={IEEE Transactions on Knowledge and Data Engineering}, 
	title={Distilled Neural Networks for Efficient Learning to Rank}, 
	year={2022},
	volume={},
	number={},
	pages={1-1},
}

@article{gao2021unsupervised,
	title={Unsupervised Corpus Aware Language Model Pre-training for Dense Passage Retrieval},
	author={Gao, Luyu and Callan, Jamie},
	eprint = {2108.05540},
	archivePrefix = {arXiv},
	primaryClass = {cs.IR},
	year={2021}
}

@article{gao2021condenser,
	title={Condenser: a Pre-training Architecture for Dense Retrieval},
	author={Gao, Luyu and Callan, Jamie},
	eprint = {2104.08253},
	archivePrefix = {arXiv},
	primaryClass = {cs.CL},
	year={2021}
}

@article{lindgren2021efficient,
	title={Efficient Training of Retrieval Models using Negative Cache},
	author={Lindgren, Erik and Reddi, Sashank and Guo, Ruiqi and Kumar, Sanjiv},
	journal={Advances in Neural Information Processing Systems},
	volume={34},
	pages={4134--4146},
	year={2021}
}

@article{lin2020distilling,
	title={Distilling Dense Representations for Ranking using Tightly-coupled Teachers},
	author={Lin, Sheng-Chieh and Yang, Jheng-Hong and Lin, Jimmy},
	eprint={2010.11386},
	archivePrefix={arXiv},
	primaryClass={cs.IR},
	year={2020}
}

@inproceedings{10.1145/3397271.3401075,
	author = {Khattab, Omar and Zaharia, Matei},
	title = {ColBERT: Efficient and Effective Passage Search via Contextualized Late Interaction over BERT},
	year = {2020},
	booktitle = {Proceedings of the 43rd International ACM SIGIR Conference on Research and Development in Information Retrieval},
	pages = {39–48},
	numpages = {10},
	location = {Virtual Event, China},
	organization={ACM}
}

@incollection{Buckley:2005,
	author      = "Chris Buckley and Ellen Voorhees",
	title       = "Retrieval System Evaluation",
	booktitle   = "TREC: Experiment and Evaluation in Information Retrieval",
	publisher   = "MIT Press",
	year        = 2005,
	chapter     = 3,
}

@inproceedings{karpukhin-etal-2020-dense,
	title = "Dense Passage Retrieval for Open-Domain Question Answering",
	author = "Karpukhin, Vladimir  and
	Oguz, Barlas  and
	Min, Sewon  and
	Lewis, Patrick  and
	Wu, Ledell  and
	Edunov, Sergey  and
	Chen, Danqi  and
	Yih, Wen-tau",
	booktitle = "Proceedings of the 2020 Conference on Empirical Methods in Natural Language Processing (EMNLP)",
	month = nov,
	year = "2020",
	pages = "6769--6781",
}

@inproceedings{Wang:2010:RUT:1871437.1871452,
	author = {Wang, Lidan and Metzler, Donald and Lin, Jimmy},
	title = {Ranking Under Temporal Constraints},
	booktitle = {Proceedings of the 19th ACM International Conference on Information and Knowledge Management},
	year = {2010},
	location = {Toronto, ON, Canada},
	pages = {79--88},
	numpages = {10},
}

@inproceedings{kohavi2013online,
	title={Online Controlled Experiments at Large Scale},
	author={Kohavi, Ron and Deng, Alex and Frasca, Brian and Walker, Toby and Xu, Ya and Pohlmann, Nils},
	booktitle={Proceedings of the 19th ACM SIGKDD international conference on Knowledge discovery and data mining},
	pages={1168--1176},
	year={2013},
}

@inproceedings{dredze2007learning,
	title={Learning Fast Classifiers for Image Spam.},
	author={Dredze, Mark and Gevaryahu, Reuven and Elias-Bachrach, Ari},
	booktitle={CEAS},
	pages={2007--487},
	year={2007}
}

@article{efron2004least,
	title={Least Angle Regression},
	author={Efron, Bradley and Hastie, Trevor and Johnstone, Iain and Tibshirani, Robert and others},
	journal={The Annals of Statistics},
	volume={32},
	number={2},
	pages={407--499},
	year={2004},
	publisher={Institute of Mathematical Statistics}
}

@inproceedings{mohan2011web,
	title={Web-search Ranking with Initialized Gradient Boosted Regression Trees},
	author={Mohan, Ananth and Chen, Zheng and Weinberger, Kilian},
	booktitle={Proceedings of the learning to rank challenge},
	pages={77--89},
	year={2011}
}

@inproceedings{chen2012classifier,
	title={Classifier cascade for minimizing feature evaluation cost},
	author={Chen, Minmin and Xu, Zhixiang and Weinberger, Kilian and Chapelle, Olivier and Kedem, Dor},
	booktitle={Artificial Intelligence and Statistics},
	pages={218--226},
	year={2012}
}

@article{chapelle2011boosted,
	title={Boosted Multi-task Learning},
	author={Chapelle, Olivier and Shivaswamy, Pannagadatta and Vadrevu, Srinivas and Weinberger, Kilian and Zhang, Ya and Tseng, Belle},
	journal={Machine learning},
	volume={85},
	number={1-2},
	pages={149--173},
	year={2011},
	publisher={Springer}
}

@inproceedings{zheng2008general,
	title={A General Boosting Method and its Application to Learning Ranking Functions for Web Search},
	author={Zheng, Zhaohui and Zha, Hongyuan and Zhang, Tong and Chapelle, Olivier and Chen, Keke and Sun, Gordon},
	booktitle={Advances in Neural Information Processing Systems},
	pages={1697--1704},
	year={2008}
}

@inproceedings{ba2014deep,
	title={Do Deep Nets Really Need to be Deep?},
	author={Ba, Jimmy and Caruana, Rich},
	booktitle={Advances in neural information processing systems},
	pages={2654--2662},
	year={2014}
}

@article{hinton2012improving,
	title={Improving Neural Networks by Preventing Co-adaptation of Feature Detectors},
	author={Hinton, Geoffrey E and Srivastava, Nitish and Krizhevsky, Alex and Sutskever, Ilya and Salakhutdinov, Ruslan R},
	eprint = {1207.0580},
	archivePrefix = {arXiv},
	primaryClass = {cs.NE},
	year={2012}
}

@inproceedings{ye2018rapidscorer,
	title={RapidScorer: Fast Tree Ensemble Evaluation by Maximizing Compactness in Data Level Parallelization},
	author={Ye, Ting and Zhou, Hucheng and Zou, Will Y and Gao, Bin and Zhang, Ruofei},
	booktitle={Proceedings of the 24th ACM SIGKDD International Conference on Knowledge Discovery and Data Mining},
	pages={941--950},
	year={2018},
}

@inproceedings{oosterhuis2017balancing,
	title={Balancing Speed and Quality in Online Learning to Rank for Information Retrieval},
	author={Oosterhuis, Harrie and de Rijke, Maarten},
	booktitle={Proceedings of the 2017 ACM on Conference on Information and Knowledge Management},
	pages={277--286},
	year={2017},
	organization={ACM}
}

@inproceedings{liu2017cascade,
	title={Cascade Ranking for Operational E-commerce Search},
	author={Liu, Shichen and Xiao, Fei and Ou, Wenwu and Si, Luo},
	booktitle={Proceedings of the 23rd ACM SIGKDD International Conference on Knowledge Discovery and Data Mining},
	pages={1557--1565},
	year={2017},
	organization={ACM}
}

@inproceedings{kusner2014feature,
	title={Feature-Cost Sensitive Learning with Submodular Trees of Classifiers},
	author={Kusner, Matt J and Chen, Wenlin and Zhou, Quan and Xu, Zhixiang Eddie and Weinberger, Kilian Q and Chen, Yixin},
	booktitle={AAAI},
	pages={1939--1945},
	year={2014}
}

@inproceedings{cohen2018universal,
	title={Universal Approximation Functions for Fast Learning to Rank: Replacing Expensive Regression Forests with Simple Feed-forward Networks},
	author={Cohen, Daniel and Foley, John and Zamani, Hamed and Allan, James and Croft, W Bruce},
	booktitle={The 41st International ACM SIGIR Conference on Research \& Development in Information Retrieval},
	pages={1017--1020},
	year={2018},
	organization={ACM}
}

@inproceedings{mackenzie2018query,
	title={Query Driven Algorithm Selection in Early Stage Retrieval},
	author={Mackenzie, Joel and Culpepper, J Shane and Blanco, Roi and Crane, Matt and Clarke, Charles LA and Lin, Jimmy},
	booktitle={Proceedings of the Eleventh ACM International Conference on Web Search and Data Mining},
	pages={396--404},
	year={2018},
	organization={ACM}
}

@inproceedings{kveton2015cascading,
	title={Cascading Bandits: Learning to Rank in the Cascade Model},
	author={Kveton, Branislav and Szepesvari, Csaba and Wen, Zheng and Ashkan, Azin},
	booktitle={International Conference on Machine Learning},
	pages={767--776},
	year={2015}
}

@inproceedings{culpepper2016dynamic,
	title={Dynamic Cutoff Prediction in Multi-stage Retrieval Systems},
	author={Culpepper, J Shane and Clarke, Charles LA and Lin, Jimmy},
	booktitle={Proceedings of the 21st Australasian Document Computing Symposium},
	pages={17--24},
	year={2016},
	organization={ACM}
}

@inproceedings{tang2018ranking,
	title={Ranking Distillation: Learning Compact Ranking Models With High Performance for Recommender System},
	author={Tang, Jiaxi and Wang, Ke},
	booktitle={Proceedings of the 24th ACM SIGKDD International Conference on Knowledge Discovery and Data Mining},
	pages={2289--2298},
	year={2018},
}

@inproceedings{ponte1998language,
	title={A Language Modeling Approach to Information Retrieval},
	author={Ponte, Jay M and Croft, W Bruce},
	booktitle={Proceedings of the 21st annual international ACM SIGIR conference on Research and development in information retrieval},
	pages={275--281},
	year={1998},
}

@inproceedings{robertson2004simple,
	title={Simple BM25 Extension to Multiple Weighted Fields},
	author={Robertson, Stephen and Zaragoza, Hugo and Taylor, Michael},
	booktitle={Proceedings of the 13th ACM International Conference on Information and Knowledge Management},
	pages={42--49},
	year={2004},
}

@article{jones2000probabilistic,
	title={A Probabilistic Model of Information Retrieval: Development and Comparative Experiments: Part 2},
	author={Jones, K Sparck and Walker, Steve and Robertson, Stephen E.},
	journal={Information processing \& management},
	volume={36},
	number={6},
	pages={809--840},
	year={2000},
	publisher={Elsevier}
}

@article{salton1988term,
	title={Term-weighting approaches in automatic text retrieval},
	author={Salton, Gerard and Buckley, Christopher},
	journal={Information Processing \& Management},
	volume={24},
	number={5},
	pages={513--523},
	year={1988},
	publisher={Elsevier}
}

@article{sparck1972statistical,
	title={A Statistical Interpretation of Term Specificity and its Application in Retrieval},
	author={Sparck Jones, Karen},
	journal={Journal of documentation},
	volume={28},
	number={1},
	pages={11--21},
	year={1972},
	publisher={MCB UP Ltd}
}

@inproceedings{chapelle2011yahoo,
	title={Yahoo! Learning to Rank Challenge Overview},
	author={Chapelle, Olivier and Chang, Yi},
	booktitle={Proceedings of the Learning to Rank Challenge},
	pages={1--24},
	year={2011}
}

@inproceedings{lucchese2015speeding,
	title={Speeding up Document Ranking with Rank-based Features},
	author={Lucchese, Claudio and Nardini, Franco Maria and Orlando, Salvatore and Perego, Raffaele and Tonellotto, Nicola},
	booktitle={Proceedings of the 38th International ACM SIGIR Conference on Research and Development in Information Retrieval},
	pages={895--898},
	year={2015},
	organization={ACM}
}

@article{macdonald2013whens,
	title={The Whens and Hows of Learning to Rank for Web Search},
	author={Macdonald, Craig and Santos, Rodrygo LT and Ounis, Iadh},
	journal={Information Retrieval},
	volume={16},
	number={5},
	pages={584--628},
	year={2013},
	publisher={Springer}
}

@InProceedings{pmlr-v38-korlakaivinayak15,
	title = 	 {{DART: Dropouts meet Multiple Additive Regression Trees}},
	author = 	 {Rashmi Korlakai Vinayak and Ran Gilad-Bachrach},
	booktitle = 	 {Proceedings of the 18th International Conference on Artificial Intelligence and Statistics},
	pages = 	 {489--497},
	year = 	 {2015},
	editor = 	 {Guy Lebanon and S. V. N. Vishwanathan},
	volume = 	 {38},
	series = 	 {Proceedings of Machine Learning Research},
	address = 	 {San Diego, California, USA},
	month = 	 {5},
	publisher = 	 {PMLR},
}

@inproceedings{Lucchese:2017:XBD:3077136.3080725,
	author = {Lucchese, Claudio and Nardini, Franco Maria and Orlando, Salvatore and Perego, Raffaele and Trani, Salvatore},
	title = {X-DART: Blending Dropout and Pruning for Efficient Learning to Rank},
	booktitle = {Proceedings of the 40th International ACM SIGIR Conference on Research and Development in Information Retrieval},
	year = {2017},
	location = {Shinjuku, Tokyo, Japan},
	pages = {1077--1080},
	numpages = {4},
}

@inproceedings{ling2017model,
	title={Model Ensemble for Click Prediction in Bing Search Ads},
	author={Ling, Xiaoliang and Deng, Weiwei and Gu, Chen and Zhou, Hucheng and Li, Cui and Sun, Feng},
	booktitle={Proceedings of the 26th International Conference on World Wide Web Companion},
	pages={689--698},
	year={2017},
}

@inproceedings{he2014practical,
	title={Practical Lessons from Predicting Clicks on Ads at Facebook},
	author={He, Xinran and Pan, Junfeng and Jin, Ou and Xu, Tianbing and Liu, Bo and Xu, Tao and Shi, Yanxin and Atallah, Antoine and Herbrich, Ralf and Bowers, Stuart and others},
	booktitle={Proceedings of the 8th International Workshop on Data Mining for Online Advertising},
	pages={1--9},
	year={2014},
}

@inproceedings{sorokina2016amazon,
	title={Amazon Search: The Joy of Ranking Products},
	author={Sorokina, Daria and Cant{\'u}-Paz, Erick},
	booktitle={Proceedings of the 39th International ACM SIGIR Conference on Research and Development in Information Retrieval},
	pages={459--460},
	year={2016},
}

@article{hofmann2013balancing,
	title={Balancing Exploration and Exploitation in Listwise and Pairwise Online Learning to Rank for Information Retrieval},
	author={Hofmann, Katja and Whiteson, Shimon and de Rijke, Maarten},
	journal={Information Retrieval},
	volume={16},
	number={1},
	pages={63--90},
	year={2013},
	publisher={Springer}
}

@inproceedings{hofmann2013reusing,
	title={Reusing Historical Interaction Data for Faster Online Learning to Rank for IR},
	author={Hofmann, Katja and Schuth, Anne and Whiteson, Shimon and de Rijke, Maarten},
	booktitle={Proceedings of the 6th ACM International Conference on Web Search and Data Mining},
	pages={183--192},
	year={2013},
}

@article{yue2012k,
	title={The k-armed Dueling Bandits Problem},
	author={Yue, Yisong and Broder, Josef and Kleinberg, Robert and Joachims, Thorsten},
	journal={Journal of Computer and System Sciences},
	volume={78},
	number={5},
	pages={1538--1556},
	year={2012},
	publisher={Elsevier}
}

@inproceedings{yue2009interactively,
	title={Interactively Optimizing Information Retrieval Systems as a Dueling Bandits Problem},
	author={Yue, Yisong and Joachims, Thorsten},
	booktitle={Proceedings of the 26th Annual International Conference on Machine Learning},
	pages={1201--1208},
	year={2009},
}

@inproceedings{radlinski2008learning,
	title={Learning Diverse Rankings with Multi-armed Bandits},
	author={Radlinski, Filip and Kleinberg, Robert and Joachims, Thorsten},
	booktitle={Proceedings of the 25th International Conference on Machine Learning},
	pages={784--791},
	year={2008},
}

@inproceedings{szummer2011semi,
	title={Semi-supervised Learning to Rank with Preference Regularization},
	author={Szummer, Martin and Yilmaz, Emine},
	booktitle={Proceedings of the 20th ACM International Conference on Information and Knowledge Management},
	pages={269--278},
	year={2011},
}

@inproceedings{borisov2016neural,
	title={A Neural Click Model for Web Search},
	author={Borisov, Alexey and Markov, Ilya and de Rijke, Maarten and Serdyukov, Pavel},
	booktitle={Proceedings of the 25th International Conference on World Wide Web},
	pages={531--541},
	year={2016},
}

@inproceedings{DBLP:conf/sigir/DehghaniZSKC17,
	author    = {Mostafa Dehghani and
			  Hamed Zamani and
			  Aliaksei Severyn and
			  Jaap Kamps and
			  W. Bruce Croft},
	title     = {Neural Ranking Models with Weak Supervision},
	booktitle = {Proceedings of the 40th International {ACM} {SIGIR} Conference on
			  Research and Development in Information Retrieval},
	location = {Tokyo, Japan},
	pages     = {65--74},
        year = {2017},
}

@inproceedings{mitra2017learning,
	title={Learning to Match using Local and Distributed Representations of Text for Web Search},
	author={Mitra, Bhaskar and Diaz, Fernando and Craswell, Nick},
	booktitle={Proceedings of the 26th International Conference on World Wide Web},
	pages={1291--1299},
	year={2017},
}

@article{mitra2016dual,
	title={A Dual Embedding Space Model for Document Ranking},
	author={Mitra, Bhaskar and Nalisnick, Eric and Craswell, Nick and Caruana, Rich},
	archivePrefix = {arXiv},
	eprint = {1602.01137},
	primaryClass = {cs.IR},
	year={2016}
}

@inproceedings{huang2013learning,
	title={Learning Deep Structured Semantic Models for Web Search using Clickthrough Data},
	author={Huang, Po-Sen and He, Xiaodong and Gao, Jianfeng and Deng, Li and Acero, Alex and Heck, Larry},
	booktitle={Proceedings of the 22nd ACM International Conference on Information \& Knowledge Management},
	pages={2333--2338},
	year={2013},
	organization={ACM}
}

@inproceedings{xgboost,
	author = {Chen, Tianqi and Guestrin, Carlos},
	title = {XGBoost: A Scalable Tree Boosting System},
	booktitle = {Proceedings of the 22nd ACM SIGKDD International Conference on Knowledge Discovery and Data Mining},
	year = {2016},
	location = {San Francisco, California, USA},
	pages = {785--794},
	numpages = {10},
}

@inproceedings{ganjisaffar2011bagging,
	title={Bagging Gradient-boosted Trees for High Precision, Low Variance Ranking Models},
	author={Ganjisaffar, Yasser and Caruana, Rich and Lopes, Cristina Videira},
	booktitle={Proceedings of the 34th international ACM SIGIR conference on Research and development in Information Retrieval},
	pages={85--94},
	year={2011},
}

@article{tax2015cross,
	title={A Cross-benchmark Comparison of 87 Learning to Rank Methods},
	author={Tax, Niek and Bockting, Sander and Hiemstra, Djoerd},
	journal={Information Processing \& Management},
	volume={51},
	number={6},
	pages={757--772},
	year={2015},
	publisher={Elsevier}
}

@article{qin2010general,
	title={A General Approximation Framework for Direct Optimization of Information Retrieval Measures},
	author={Qin, Tao and Liu, Tie-Yan and Li, Hang},
	journal={Information Retrieval},
	volume={13},
	number={4},
	pages={375--397},
	year={2010},
}

@inproceedings{bruch2019approxndcg,
	author = {Bruch, Sebastian and Zoghi, Masrour and Bendersky, Michael and Najork, Marc},
	title = {Revisiting Approximate Metric Optimization in the Age of Deep Neural Networks},
	year = {2019},
	booktitle = {Proceedings of the 42nd International ACM SIGIR Conference on Research and Development in Information Retrieval},
	pages = {1241–1244},
	numpages = {4},
	location = {Paris, France},
}

@inproceedings{bruch2019xendcg,
	author = {Bruch, Sebastian and Wang, Xuanhui and Bendersky, Michael and Najork, Marc},
	title = {An Analysis of the Softmax Cross Entropy Loss for Learning-to-Rank with Binary Relevance},
	year = {2019},
	booktitle = {Proceedings of the 2019 ACM SIGIR International Conference on Theory of Information Retrieval},
	pages = {75–78},
	numpages = {4},
	location = {Santa Clara, CA, USA},
}

@inproceedings{bruch2021xendcg,
	author = {Bruch, Sebastian},
	title = {An Alternative Cross Entropy Loss for Learning-to-Rank},
	year = {2021},
	booktitle = {Proceedings of the Web Conference 2021},
	pages = {118–126},
	numpages = {9},
	location = {Ljubljana, Slovenia},
}

@inproceedings{taylor2008softrank,
	author = {Taylor, Michael and Guiver, John and Robertson, Stephen and Minka, Tom},
	title = {SoftRank: Optimizing Non-Smooth Rank Metrics},
	year = {2008},
	booktitle = {Proceedings of the 2008 International Conference on Web Search and Data Mining},
	pages = {77–86},
	numpages = {10},
	location = {Palo Alto, California, USA},
}

@inproceedings{xia2008listmle,
	author = {Xia, Fen and Liu, Tie-Yan and Wang, Jue and Zhang, Wensheng and Li, Hang},
	title = {Listwise Approach to Learning to Rank: Theory and Algorithm},
	year = {2008},
	pages = {1192–1199},
	numpages = {8},
	location = {Helsinki, Finland},
	booktitle = {Proceedings of the 25th International Conference on Machine Learning},
}

@inproceedings{cao2007learning,
	title={Learning to Rank: from Pairwise Approach to Listwise Approach},
	author={Cao, Zhe and Qin, Tao and Liu, Tie-Yan and Tsai, Ming-Feng and Li, Hang},
	booktitle={Proceedings of the 24th International Conference on Machine learning},
	pages={129--136},
	year={2007},
	organization={ACM}
}

@article{freund2003efficient,
	title={An Efficient Boosting Algorithm for Combining Preferences},
	author={Freund, Yoav and Iyer, Raj and Schapire, Robert E and Singer, Yoram},
	journal={Journal of Machine Learning Research},
	volume={4},
	number={Nov},
	pages={933--969},
	year={2003}
}

@article{chapelle2012large,
	title={Large-scale Validation and Analysis of Interleaved Search Evaluation},
	author={Chapelle, Olivier and Joachims, Thorsten and Radlinski, Filip and Yue, Yisong},
	journal={ACM Transactions on Information Systems},
	volume={30},
	number={1},
	pages={6},
	year={2012},
	publisher={ACM}
}

@article{moffat2008rank,
	title={Rank-biased Precision for Measurement of Retrieval Effectiveness},
	author={Moffat, Alistair and Zobel, Justin},
	journal={ACM Transactions on Information Systems},
	volume={27},
	number={1},
	pages={2},
	year={2008},
}

@inproceedings{chapelle2009expected,
	title={Expected Reciprocal Rank for Graded Relevance},
	author={Chapelle, Olivier and Metlzer, Donald and Zhang, Ya and Grinspan, Pierre},
	booktitle={Proceedings of the 18th ACM conference on Information and knowledge management},
	pages={621--630},
	year={2009},
}

@inproceedings{jarvelin2000ir,
	title={IR Evaluation Methods for Retrieving Highly Relevant Documents},
	author={J{\"a}rvelin, Kalervo and Kek{\"a}l{\"a}inen, Jaana},
	booktitle={Proceedings of the 23rd International ACM SIGIR Conference on Research and Development in Information Retrieval},
	pages={41--48},
	year={2000},
}

@inproceedings{yilmaz2009deep,
	title={Deep versus Shallow Judgments in Learning to Rank},
	author={Yilmaz, Emine and Robertson, Stephen},
	booktitle={Proceedings of the 32nd international ACM SIGIR Conference on Research and Development in Information Retrieval},
	pages={662--663},
	year={2009},
}

@inproceedings{long2010active,
	title={Active Learning for Ranking through Expected Loss Optimization},
	author={Long, Bo and Chapelle, Olivier and Zhang, Ya and Chang, Yi and Zheng, Zhaohui and Tseng, Belle},
	booktitle={Proceedings of the 33rd International ACM SIGIR Conference on Research and Development in Information Retrieval},
	pages={267--274},
	year={2010},
}

@inproceedings{radlinski2005query,
	title={Query Chains: Learning to Rank from Implicit Feedback},
	author={Radlinski, Filip and Joachims, Thorsten},
	booktitle={Proceedings of the 11th ACM SIGKDD International Conference on Knowledge Discovery in Data Mining},
	pages={239--248},
	year={2005},
	organization={ACM}
}

@inproceedings{joachims2002optimizing,
	title={Optimizing Search Engines using Clickthrough Data},
	author={Joachims, Thorsten},
	booktitle={Proceedings of the 8th ACM SIGKDD International Conference on Knowledge Discovery and Data Mining},
	pages={133--142},
	year={2002},
}

@inproceedings{joachims2005accurately,
	title={Accurately Interpreting Clickthrough Data as Implicit Feedback},
	author={Joachims, Thorsten and Granka, Laura and Pan, Bing and Hembrooke, Helene and Gay, Geri},
	booktitle={Proceedings of the 28th International ACM SIGIR Conference on Research and Development in Information Retrieval},
	pages={154--161},
	year={2005},
}

@article{carterette2008here,
	title={Here or there},
	author={Carterette, Ben and Bennett, Paul and Chickering, David and Dumais, Susan},
	journal={Advances in Information Retrieval},
	pages={16--27},
	year={2008},
	publisher={Springer}
}

@article{rasolofo2003term,
	title={Term Proximity Scoring for Keyword-based Retrieval Systems},
	author={Rasolofo, Yves and Savoy, Jacques},
	journal={Advances in Information Retrieval},
	pages={79--79},
	year={2003},
	publisher={Springer}
}

@inproceedings{shen2014learning,
	title={Learning Semantic Representations using Convolutional Neural Networks for Web Search},
	author={Shen, Yelong and He, Xiaodong and Gao, Jianfeng and Deng, Li and Mesnil, Gr{\'e}goire},
	booktitle={Proceedings of the 23rd International Conference on World Wide Web},
	pages={373--374},
	year={2014},
	organization={ACM}
}

@article{henzinger2000link,
	title={Link Analysis in Web Information Retrieval},
	author={Henzinger, Monika Rauch and others},
	journal={IEEE Data Engineering Bulletin},
	volume={23},
	number={3},
	pages={3--8},
	year={2000}
}

@article{lucchese2013discovering,
	title={Discovering Tasks from Search Engine Query Logs},
	author={Lucchese, Claudio and Orlando, Salvatore and Perego, Raffaele and Silvestri, Fabrizio and Tolomei, Gabriele},
	journal={ACM Transactions on Information Systems},
	volume={31},
	number={3},
	pages={14},
	year={2013},
	publisher={ACM}
}

@inproceedings{bennett2010classification,
	title={Classification-enhanced Ranking},
	author={Bennett, Paul N and Svore, Krysta and Dumais, Susan T},
	booktitle={Proceedings of the 19th International Conference on World Wide Web},
	pages={111--120},
	year={2010},
}

@article{jiang2016query,
	title={Query Intent Mining with Multiple Dimensions of Web Search Data},
	author={Jiang, Di and Leung, Kenneth Wai-Ting and Ng, Wilfred},
	journal={Proceedings of the 25th International Conference on World Wide Web},
	volume={19},
	number={3},
	pages={475--497},
	year={2016},
}

@article{metzler2007linear,
	title={Linear Feature-based Models for Information Retrieval},
	author={Metzler, Donald and Croft, W Bruce},
	journal={Information Retrieval},
	volume={10},
	number={3},
	pages={257--274},
	year={2007},
	publisher={Springer}
}

@inproceedings{burges2005learning,
	title={Learning to Rank using Gradient Descent},
	author={Burges, Chris and Shaked, Tal and Renshaw, Erin and Lazier, Ari and Deeds, Matt and Hamilton, Nicole and Hullender, Greg},
	booktitle={Proceedings of the 22nd international conference on Machine learning},
	pages={89--96},
	year={2005},
	organization={ACM}
}

@inproceedings{joachims2017unbiased,
	title={Unbiased Learning-to-rank with Biased Feedback},
	author={Joachims, Thorsten and Swaminathan, Adith and Schnabel, Tobias},
	booktitle={Proceedings of the 10th ACM International Conference on Web Search and Data Mining},
	pages={781--789},
	year={2017},
}

@inproceedings{SIGIR2015ltr-shorttext,
	title={Learning to Rank Short Text Pairs with Convolutional Deep Neural Networks},
	author={Severyn, Aliaksei and Moschitti, Alessandro},
	booktitle={Proceedings of the 38th International ACM SIGIR Conference on Research and Development in Information Retrieval},
	pages={373--382},
	year={2015},
	organization={ACM}
}

@article{Friedman01,
	title={Greedy Function Approximation: a Gradient Boosting Machine},
	author={Friedman, Jerome H},
	journal={Annals of Statistics},
	pages={1189--1232},
	year={2001},
	publisher={JSTOR}
}

@article{Liu08,
	title={Learning to Rank for Information Retrieval},
	author={Liu, Tie-Yan},
	journal={Foundations and Trends in Information Retrieval},
	volume={3},
	number={3},
	pages={225--331},
	year={2009},
	publisher={Now Publishers Inc.}
}

@inproceedings{Roi_SIGIR17,
	author = {Chen, Ruey-Cheng and Gallagher, Luke and Blanco, Roi and Culpepper, J. Shane},
	title = {Efficient Cost-Aware Cascade Ranking in Multi-Stage Retrieval},
	booktitle = {Proceedings of the 40th International ACM SIGIR Conference on Research and Development in Information Retrieval},
	year = {2017},
	location = {Shinjuku, Tokyo, Japan},
	pages = {445--454},
	numpages = {10},
}

@article{DBLP:journals/tpds/LettichLNOPTV19,
	author    = {Francesco Lettich and
			  Claudio Lucchese and
			  Franco Maria Nardini and
			  Salvatore Orlando and
			  Raffaele Perego and
			  Nicola Tonellotto and
			  Rossano Venturini},
	title     = {Parallel Traversal of Large Ensembles of Decision Trees},
	journal   = {{IEEE} Transactions on Parallel and Distributed Systems},
	volume    = {30},
	number    = {9},
	pages     = {2075--2089},
	year      = {2019},
}

@inproceedings{Lucchese:2016:ECS:2911451.2914758,
	author = {Lucchese, Claudio and Nardini, Franco Maria and Orlando, Salvatore and Perego, Raffaele and Tonellotto, Nicola and Venturini, Rossano},
	title = {Exploiting CPU SIMD Extensions to Speed-up Document Scoring with Tree Ensembles},
	booktitle = {Proceedings of the 39th International ACM SIGIR Conference on Research and Development in Information Retrieval},
	year = {2016},
	location = {Pisa, Italy},
	pages = {833--836},
	numpages = {4},
}

@inproceedings{relyahoo,
	author = {Dawei Yin and Yuening Hu and  Jiliang Tang and  Tim Daly Jr. and  Mianwei Zhou and  Hua Ouyang and  Jianhui Chen and  Changsung Kang and  Hongbo Deng and  Chikashi Nobata and  Jean-Marc Langlois and  Yi Chang},
	title = {Ranking Relevance in Yahoo Search},
	booktitle = {Proceedings of the 22nd ACM SIGKDD Conference on Knowledge Discovery and Data Mining},
	year = {2016},
}

@inproceedings{Lucchese:2016:POT:2911451.2914763,
	author = {Lucchese, Claudio and Nardini, Franco Maria and Orlando, Salvatore and Perego, Raffaele and Silvestri, Fabrizio and Trani, Salvatore},
	title = {Post-Learning Optimization of Tree Ensembles for Efficient Ranking},
	booktitle = {Proceedings of the 39th International ACM SIGIR Conference on Research and Development in Information Retrieval},
	year = {2016},
	location = {Pisa, Italy},
	pages = {949--952},
	numpages = {4},
}

@article{Dato:2016:FRA:3001595.2987380,
	author = {Dato, Domenico and Lucchese, Claudio and Nardini, Franco Maria and Orlando, Salvatore and Perego, Raffaele and Tonellotto, Nicola and Venturini, Rossano},
	title = {Fast Ranking with Additive Ensembles of Oblivious and Non-Oblivious Regression Trees},
	journal = {ACM Transactions on Information Systems},
	issue_date = {December 2016},
	volume = {35},
	number = {2},
	month = dec,
	year = {2016},
	pages = {15:1--15:31},
	articleno = {15},
	numpages = {31},
}

@article{capannini2016quality,
	author = {Capannini, Gabriele and Lucchese, Claudio and Nardini, Franco Maria and Orlando, Salvatore and Perego, Raffaele and Tonellotto, Nicola},
	title = {Quality Versus Efficiency in Document Scoring with Learning-to-rank Models},
	journal = {Information Processing \& Management},
	issue_date = {November 2016},
	volume = {52},
	number = {6},
	month = nov,
	year = {2016},
	pages = {1161--1177},
	numpages = {17},
}

@inproceedings{SIGIR2015,
	author = {Lucchese, Claudio and Nardini, Franco Maria and Orlando, Salvatore and Perego, Raffaele and Tonellotto, Nicola and Venturini, Rossano},
	title = {QuickScorer: A Fast Algorithm to Rank Documents with Additive Ensembles of Regression Trees},
	booktitle = {Proceedings of the 38th International ACM SIGIR Conference on Research and Development in Information Retrieval},
	year = {2015},
	pages = {73--82},
	numpages = {10},
}

@TechReport{export:132652,
	author       = {Christopher J.C. Burges},
	month        = {6},
	number       = {MSR-TR-2010-82},
	title        = {From RankNet to LambdaRank to LambdaMART: An Overview},
	year         = {2010},
}

@inproceedings{asadi2013training,
	author = {Asadi, Nima and Lin, Jimmy},
	title = {Training Efficient Tree-Based Models for Document Ranking},
	year = {2013},
	booktitle = {Proceedings of the 35th European Conference on Advances in Information Retrieval},
	pages = {146–157},
	numpages = {12},
	location = {Moscow, Russia},
}

@article{LinTKDE,
	author              = {Nima Asadi and Jimmy Lin and Arjen P. de Vries},
	title               = {Runtime Optimizations for Tree-Based Machine Learning Models.},
	journal             = {IEEE Transactions on Knowledge and Data Engineering},
	year                = {2014},
	pages               = {2281-2292},
	volume				= {26},
	number				= {9},
}

@article{hoerl1970ridge,
	title={Ridge Regression: Biased Estimation for Nonorthogonal Problems},
	author={Hoerl, Arthur E and Kennard, Robert W},
	journal={Technometrics},
	volume={12},
	number={1},
	pages={55--67},
	year={1970},
	publisher={Taylor \& Francis}
}

@article{tseng1988coordinate,
	title={Coordinate Ascent for Maximizing Nondifferentiable Concave Functions},
	author={Tseng, Paul and others},
	year={1988},
	publisher={Massachusetts Institute of Technology, Laboratory for Information and~…}
}

@inproceedings{Wang:2011:CRM:2009916.2009934,
	author = {Wang, Lidan and Lin, Jimmy and Metzler, Donald},
	title = {A Cascade Ranking Model for Efficient Ranked Retrieval},
	booktitle = {Proceedings of the 34th International ACM SIGIR Conference on Research and Development in Information Retrieval},
	year = {2011},
	location = {Beijing, China},
	pages = {105--114},
	numpages = {10},
}

@inproceedings{WangSIGIR10,
	author    = {Lidan Wang and Jimmy J. Lin and Donald Metzler},
	title     = {Learning to Efficiently Rank},
	booktitle = {Proceeding of the 33rd International {ACM} {SIGIR} Conference on Research
			  and Development in Information Retrieval},
	year      = {2010},
	pages     = {138-145}
}

@incollection{ecir13,
	title={Two-Stage learning to rank for information retrieval},
	author={Dang, Van and Bendersky, Michael and Croft, W Bruce},
	booktitle={Advances in Information Retrieval},
	pages={423--434},
	year={2013},
	publisher={Springer}
}

@inproceedings{Xu:2013:CTC:3042817.3042834,
	author = {Xu, Zhixiang and Kusner, Matt J. and Weinberger, Kilian Q. and Chen, Minmin},
	title = {Cost-sensitive Tree of Classifiers},
	booktitle = {Proceedings of the 30th International Conference on International Conference on Machine Learning - Volume 28},
	year = {2013},
	location = {Atlanta, GA, USA},
	pages = {I-133--I-141},
}

@inproceedings{xu2007adarank,
	title={AdaRank: a Boosting Algorithm for Information Retrieval},
	author={Xu, Jun and Li, Hang},
	booktitle={Proceedings of the 30th International ACM SIGIR Conference on Research and Development in Information Retrieval},
	pages={391--398},
	year={2007}
}

@inproceedings{cambazoglu10early,
	author = {Cambazoglu, Berkant Barla and Zaragoza, Hugo and Chapelle, Olivier and Chen, Jiang and Liao, Ciya and Zheng, Zhaohui and Degenhardt, Jon},
	title = {Early Exit Optimizations for Additive Machine Learned Ranking Systems},
	booktitle = {Proceedings of the 3rd International Conference on Web Search and
			  Web Data Mining ({WSDM})},
	year = {2010},
	pages = {411--420},
	publisher = {ACM},
}

@inproceedings{TangJY14,
	author              = {Xun Tang and Xin Jin and Tao Yang},
	title               = {Cache-conscious Runtime Optimization for Ranking Ensembles},
	booktitle           = {{P}roceedings of the 37th {A}nnual {I}nternational {ACM SIGIR} {C}onference on {R}esearch and {D}evelopment in {I}nformation {R}etrieval ({SIGIR})},
	year                = {2014},
	pages               = {1123-1126}
}

@inproceedings{Jin:2016:CCB:2911451.2911520,
	author = {Jin, Xin and Yang, Tao and Tang, Xun},
	title = {A Comparison of Cache Blocking Methods for Fast Execution of Ensemble-based Score Computation},
	booktitle = {Proceedings of the 39th International ACM SIGIR Conference on Research and Development in Information Retrieval},
	year = {2016},
	location = {Pisa, Italy},
	pages = {629--638},
	numpages = {10},
}

@inproceedings{blocking,
	author = {Tang, Xun and Jin, Xin and Yang, Tao},
	title = {Cache-conscious Runtime Optimization for Ranking Ensembles},
	booktitle = {Proceedings of the 37th International ACM SIGIR Conference on Research and Development in Information Retrieval},
	year = {2014},
	location = {Gold Coast, Queensland, Australia},
	pages = {1123--1126},
	numpages = {4},
}

@inproceedings{10.1145/1277741.1277811,
	author = {Geng, Xiubo and Liu, Tie-Yan and Qin, Tao and Li, Hang},
	title = {Feature Selection for Ranking},
	booktitle = {Proceedings of the 30th Annual International ACM SIGIR Conference on Research and Development in Information Retrieval},
	year = {2007},
	location = {Amsterdam, The Netherlands},
	pages = {407--414},
	numpages = {8},
}

@inproceedings{10.1145/2970398.2970433,
	author = {Gigli, Andrea and Lucchese, Claudio and Nardini, Franco Maria and Perego, Raffaele},
	title = {Fast Feature Selection for Learning to Rank},
	year = {2016},
	booktitle = {Proceedings of the 2016 ACM International Conference on the Theory of Information Retrieval},
	pages = {167–170},
	numpages = {4},
	location = {Newark, Delaware, USA},
}

@inproceedings{lightgbm,
	author = {Ke, Guolin and Meng, Qi and Finley, Thomas and Wang, Taifeng and Chen, Wei and Ma, Weidong and Ye, Qiwei and Liu, Tie-Yan},
	title = {LightGBM: A Highly Efficient Gradient Boosting Decision Tree},
	year = {2017},
	booktitle = {Proceedings of the 31st International Conference on Neural Information Processing Systems},
	pages = {3149–3157},
	numpages = {9},
	location = {Long Beach, California, USA},
}

@article{macavaney:arxiv2020-abnirml,
	author = {MacAvaney, Sean and Feldman, Sergey and Goharian, Nazli and Downey, Doug and Cohan, Arman},
	title = {ABNIRML: Analyzing the Behavior of Neural IR Models},
	year = {2020},
	url = {https://arxiv.org/abs/2011.00696},
	journal = {arXiv},
	volume = {abs/2011.00696}
}

@inproceedings{catboost,
	author = {Prokhorenkova, Liudmila and Gusev, Gleb and Vorobev, Aleksandr and Dorogush, Anna Veronika and Gulin, Andrey},
	title = {CatBoost: Unbiased Boosting with Categorical Features},
	year = {2018},
	booktitle = {Proceedings of the 32nd International Conference on Neural Information Processing Systems},
	pages = {6639–6649},
	numpages = {11},
	location = {Montr\'{e}al, Canada},
}

@article{Onal2018NeuralIR,
	author = {Onal, Kezban Dilek and Zhang, Ye and Altingovde, Ismail Sengor and Rahman, Md Mustafizur and Karagoz, Pinar and Braylan, Alex and Dang, Brandon and Chang, Heng-Lu and Kim, Henna and Mcnamara, Quinten and Angert, Aaron and Banner, Edward and Khetan, Vivek and Mcdonnell, Tyler and Nguyen, An Thanh and Xu, Dan and Wallace, Byron C. and Rijke, Maarten and Lease, Matthew},
	title = {Neural Information Retrieval: At the End of the Early Years},
	year = {2018},
	issue_date = {June 2018},
	publisher = {Kluwer Academic Publishers},
	address = {USA},
	volume = {21},
	number = {2–3},
	journal = {Information Retrieval},
	month = {6},
	pages = {111–182},
	numpages = {72},
}

@misc{mitra2017neural,
	title={Neural Models for Information Retrieval}, 
	author={Bhaskar Mitra and Nick Craswell},
	year={2017},
	eprint={1705.01509},
	archivePrefix={arXiv},
	primaryClass={cs.IR}
}

@article{Guo2020NeuralRanking,
	title = {A Deep Look into Neural Ranking Models for Information Retrieval},
	journal={Information Processing \& Management},
	volume = {57},
	number = {6},
	year = {2020},
	author = {Jiafeng Guo and Yixing Fan and Liang Pang and Liu Yang and Qingyao Ai and Hamed Zamani and Chen Wu and W. Bruce Croft and Xueqi Cheng},
}

@inproceedings{Guo2016DRMM,
	author = {Guo, Jiafeng and Fan, Yixing and Ai, Qingyao and Croft, W. Bruce},
	title = {A Deep Relevance Matching Model for Ad-Hoc Retrieval},
	year = {2016},
	booktitle = {Proceedings of the 25th ACM International on Conference on Information and Knowledge Management},
	pages = {55–64},
	numpages = {10},
	location = {Indianapolis, Indiana, USA},
}

@inproceedings{xiong2017knrm,
	author = {Xiong, Chenyan and Dai, Zhuyun and Callan, Jamie and Liu, Zhiyuan and Power, Russell},
	title = {End-to-End Neural Ad-Hoc Ranking with Kernel Pooling},
	year = {2017},
	booktitle = {Proceedings of the 40th International ACM SIGIR Conference on Research and Development in Information Retrieval},
	pages = {55–64},
	numpages = {10},
	location = {Shinjuku, Tokyo, Japan},
}

@misc{word2vec,
	title	= {Efficient Estimation of Word Representations in Vector Space},
	author	= {Tomas Mikolov and Kai Chen and Greg S. Corrado and Jeffrey Dean},
	year	= {2013},
	eprint = {1301.3781},
	archivePrefix = {arXiv},
	primaryClass = {cs.CL},
}

@inproceedings{dai2018convkrnm,
	author = {Dai, Zhuyun and Xiong, Chenyan and Callan, Jamie and Liu, Zhiyuan},
	title = {Convolutional Neural Networks for Soft-Matching N-Grams in Ad-Hoc Search},
	year = {2018},
	booktitle = {Proceedings of the 11th ACM International Conference on Web Search and Data Mining},
	pages = {126–134},
	numpages = {9},
	location = {Marina Del Rey, CA, USA},
}

@inproceedings{ai2019groupwsie,
	author = {Ai, Qingyao and Wang, Xuanhui and Bruch, Sebastian and Golbandi, Nadav and Bendersky, Michael and Najork, Marc},
	title = {Learning Groupwise Multivariate Scoring Functions Using Deep Neural Networks},
	year = {2019},
	booktitle = {Proceedings of the 2019 ACM SIGIR International Conference on Theory of Information Retrieval},
	pages = {85–92},
	numpages = {8},
	location = {Santa Clara, CA, USA},
}

@inproceedings{pang2019setrank,
	title={SetRank: Learning a Permutation-Invariant Ranking Model for Information Retrieval},
	author={Liang Pang and Jun Xu and Qingyao Ai and Yanyan Lan and Xueqi Cheng and Jirong Wen},
	booktitle = {Proceedings of the 43rd International ACM SIGIR Conference on Research and Development in Information Retrieval},
	year = {2020},
}

@inproceedings{vaswani2017attention,
	author = {Vaswani, Ashish and Shazeer, Noam and Parmar, Niki and Uszkoreit, Jakob and Jones, Llion and Gomez, Aidan N. and Kaiser, \L{}ukasz and Polosukhin, Illia},
	title = {Attention is All You Need},
	year = {2017},
	booktitle = {Proceedings of the 31st International Conference on Neural Information Processing Systems},
	pages = {6000–6010},
	numpages = {11},
	location = {Long Beach, California, USA},
}

@misc{nogueira2020passage,
	title={Passage Re-ranking with BERT}, 
	author={Rodrigo Nogueira and Kyunghyun Cho},
	year={2020},
	eprint={1901.04085},
	archivePrefix={arXiv},
	primaryClass={cs.IR}
}

@inproceedings{devlin2019bert,
	title = "{BERT}: Pre-training of Deep Bidirectional Transformers for Language Understanding",
	author = "Devlin, Jacob  and
	Chang, Ming-Wei  and
	Lee, Kenton  and
	Toutanova, Kristina",
	booktitle = "Proceedings of the 2019 Conference of the North {A}merican Chapter of the Association for Computational Linguistics: Human Language Technologies, Volume 1 (Long and Short Papers)",
	month = jun,
	year = "2019",
	pages = "4171--4186",
}

@article{nguyen2016msmarco,
	author = {Nguyen, Tri and Rosenberg, Mir and Song, Xia and Gao, Jianfeng and Tiwary, Saurabh and Majumder, Rangan and Deng, Li},
	title = {MS MARCO: A Human Generated MAchine Reading COmprehension Dataset},
	year = {2016},
	month = {11},  
}

@inproceedings{yilmaz2019birch,
	title = "Applying {BERT} to Document Retrieval with Birch",
	author = "Akkalyoncu Yilmaz, Zeynep  and
	Wang, Shengjin  and
	Yang, Wei  and
	Zhang, Haotian  and
	Lin, Jimmy",
	booktitle = "Proceedings of the 2019 Conference on Empirical Methods in Natural Language Processing and the 9th International Joint Conference on Natural Language Processing (EMNLP-IJCNLP): System Demonstrations",
	month = nov,
	year = "2019",
}

@article{Li2020PARADEPR,
	title={PARADE: Passage Representation Aggregation for Document Reranking},
	author={Canjia Li and Andrew Yates and Sean MacAvaney and Ben He and Yingfei Sun},
	archivePrefix={arXiv},
	primaryClass = {cs.IR},
	year={2020},
	eprint={2008.09093},
}

@inproceedings{dai2019bertmaxp,
	author = {Dai, Zhuyun and Callan, Jamie},
	title = {Deeper Text Understanding for IR with Contextual Neural Language Modeling},
	year = {2019},
	booktitle = {Proceedings of the 42nd International ACM SIGIR Conference on Research and Development in Information Retrieval},
	pages = {985–988},
	numpages = {4},
	location = {Paris, France},
}

@inproceedings{macavaney2019cedr,
	author = {MacAvaney, Sean and Yates, Andrew and Cohan, Arman and Goharian, Nazli},
	title = {CEDR: Contextualized Embeddings for Document Ranking},
	year = {2019},
	booktitle = {Proceedings of the 42nd International ACM SIGIR Conference on Research and Development in Information Retrieval},
	pages = {1101–1104},
	numpages = {4},
	location = {Paris, France},
}

@article{nogueira2019multi,
	title={Multi-stage document ranking with BERT},
	author={Nogueira, Rodrigo and Yang, Wei and Cho, Kyunghyun and Lin, Jimmy},
	eprint = {1910.14424},
	archivePrefix = {arXiv},
	primaryClass = {cs.IR},
	year={2019}
}

@article{raffel2020t5,
	author  = {Colin Raffel and Noam Shazeer and Adam Roberts and Katherine Lee and Sharan Narang and Michael Matena and Yanqi Zhou and Wei Li and Peter J. Liu},
	title   = {Exploring the Limits of Transfer Learning with a Unified Text-to-Text Transformer},
	journal = {Journal of Machine Learning Research},
	year    = {2020},
	volume  = {21},
	number  = {140},
	pages   = {1-67},
}

@inproceedings{nogueira2020monot5,
	title = "Document Ranking with a Pretrained Sequence-to-Sequence Model",
	author = "Nogueira, Rodrigo  and
	Jiang, Zhiying  and
	Pradeep, Ronak  and
	Lin, Jimmy",
	booktitle = "Findings of the Association for Computational Linguistics: EMNLP 2020",
	month = nov,
	year = "2020",
	pages = "708--718",
}

@misc{pradeep2021expandomonoduo,
	title={The Expando-Mono-Duo Design Pattern for Text Ranking with Pretrained Sequence-to-Sequence Models}, 
	author={Ronak Pradeep and Rodrigo Nogueira and Jimmy Lin},
	year={2021},
	eprint={2101.05667},
	archivePrefix={arXiv},
	primaryClass={cs.IR}
}

@misc{lin2021pretrained,
	title={Pretrained Transformers for Text Ranking: BERT and Beyond}, 
	author={Jimmy Lin and Rodrigo Nogueira and Andrew Yates},
	year={2021},
	eprint={2010.06467},
	archivePrefix={arXiv},
	primaryClass={cs.IR}
}

@inproceedings{lucchese2018selective,
	author = {Lucchese, Claudio and Nardini, Franco Maria and Perego, Raffaele and Orlando, Salvatore and Trani, Salvatore},
	title = {Selective Gradient Boosting for Effective Learning to Rank},
	year = {2018},
	booktitle = {The 41st International ACM SIGIR Conference on Research and Development in Information Retrieval},
	pages = {155–164},
	numpages = {10},
	location = {Ann Arbor, MI, USA},
}

@Book{CART,
	Title                    = {Classification and Regression Trees},
	Author                   = {Leo Breiman and Jerome Friedman and Charles J. Stone and R.A. Olshen},
	Publisher                = {Chapman and Hall/CRC},
	Year                     = {1984}
}

@inproceedings{asadi2013efficiency,
	author = {Asadi, Nima and Lin, Jimmy},
	title = {Effectiveness/Efficiency Tradeoffs for Candidate Generation in Multi-Stage Retrieval Architectures},
	year = {2013},
	booktitle = {Proceedings of the 36th International ACM SIGIR Conference on Research and Development in Information Retrieval},
	pages = {997–1000},
	numpages = {4},
	location = {Dublin, Ireland},
}

@book{asadi2013phd,
	title={Multi-Stage Search Architectures for Streaming Documents}, 
	author={Nima Asadi},
	year={2013},
	publisher = {University of Maryland}
}

@inproceedings{asadi2012bloom,
	author = {Asadi, Nima and Lin, Jimmy},
	title = {Fast Candidate Generation for Two-Phase Document Ranking: Postings List Intersection with Bloom Filters},
	year = {2012},
	booktitle = {Proceedings of the 21st ACM International Conference on Information and Knowledge Management},
	pages = {2419–2422},
	numpages = {4},
	location = {Maui, Hawaii, USA},
}

@article{asadi2013bloom,
	author = {Asadi, Nima and Lin, Jimmy},
	title = {Fast Candidate Generation for Real-Time Tweet Search with Bloom Filter Chains},
	year = {2013},
	issue_date = {July 2013},
	volume = {31},
	number = {3},
	journal = {ACM Transactions on Information Systems},
	month = {8},
	articleno = {13},
	numpages = {36},
}

@article{nogueira2019document,
	title={Document Expansion by Query Prediction},
	author={Nogueira, Rodrigo and Yang, Wei and Lin, Jimmy and Cho, Kyunghyun},
	eprint={1904.08375},
	archivePrefix = {arXiv},
	primaryClass = {cs.IR},
	year={2019}
}

@inproceedings{xiong2021approximate,
	author = {Xiong, Lee and Xiong, Chenyan and Li, Ye and Tang, Kwok-Fung and Liu, Jialin and Bennett, Paul and Ahmed, Junaid and Overwijk, Arnold},
	title = {Approximate Nearest Neighbor Negative Contrastive Learning for Dense Text Retrieval},
	booktitle = {International Conference on Learning Representations},
	year = {2021},
	month = {4},
}

@article{brown2020language,
	title={Language Models are Few-Shot Learners},
	author={Tom B. Brown and Benjamin Mann and Nick Ryder and Melanie Subbiah and Jared Kaplan and Prafulla Dhariwal and Arvind Neelakantan and Pranav Shyam and Girish Sastry and Amanda Askell and Sandhini Agarwal and Ariel Herbert-Voss and Gretchen Krueger and Tom Henighan and Rewon Child and Aditya Ramesh and Daniel M. Ziegler and Jeffrey Wu and Clemens Winter and Christopher Hesse and Mark Chen and Eric Sigler and Mateusz Litwin and Scott Gray and Benjamin Chess and Jack Clark and Christopher Berner and Sam McCandlish and Alec Radford and Ilya Sutskever and Dario Amodei},
	year={2020},
	eprint={2005.14165},
	archivePrefix={arXiv},
	primaryClass={cs.CL}
}

@inbook{matsubara2020multistage,
	author = {Matsubara, Yoshitomo and Vu, Thuy and Moschitti, Alessandro},
	title = {Reranking for Efficient Transformer-Based Answer Selection},
	year = {2020},
	booktitle = {Proceedings of the 43rd International ACM SIGIR Conference on Research and Development in Information Retrieval},
	pages = {1577–1580},
	numpages = {4}
}

@inproceedings{Soldaini2020TheCT,
	title={The Cascade Transformer: an Application for Efficient Answer Sentence Selection},
	author={Luca Soldaini and Alessandro Moschitti},
	booktitle={ACL},
	year={2020}
}

@inproceedings{xin-etal-2020-deebert,
	title = "{D}ee{BERT}: Dynamic Early Exiting for Accelerating {BERT} Inference",
	author = "Xin, Ji  and
	Tang, Raphael  and
	Lee, Jaejun  and
	Yu, Yaoliang  and
	Lin, Jimmy",
	booktitle = "Proceedings of the 58th Annual Meeting of the Association for Computational Linguistics",
	month = jul,
	year = "2020",
}

@inproceedings{xin-etal-2021-berxit,
	title = "{BER}xi{T}: Early Exiting for {BERT} with Better Fine-Tuning and Extension to Regression",
	author = "Xin, Ji  and
	Tang, Raphael  and
	Yu, Yaoliang  and
	Lin, Jimmy",
	booktitle = "Proceedings of the 16th Conference of the European Chapter of the Association for Computational Linguistics: Main Volume",
	month = apr,
	year = "2021",
	pages = "91--104",
}

@inproceedings{gordon-etal-2020-compressing,
	title = "Compressing {BERT}: Studying the Effects of Weight Pruning on Transfer Learning",
	author = "Gordon, Mitchell  and
	Duh, Kevin  and
	Andrews, Nicholas",
	booktitle = "Proceedings of the 5th Workshop on Representation Learning for NLP",
	month = jul,
	year = "2020",
	pages = "143--155",
}

@misc{mccarley2021structured,
	title={Structured Pruning of a BERT-based Question Answering Model}, 
	author={J. S. McCarley and Rishav Chakravarti and Avirup Sil},
	year={2021},
	eprint={1910.06360},
	archivePrefix={arXiv},
	primaryClass={cs.CL}
}

@inproceedings{lin-etal-2020-pruning,
	title = "Pruning Redundant Mappings in Transformer Models via Spectral-Normalized Identity Prior",
	author = "Lin, Zi  and
	Liu, Jeremiah  and
	Yang, Zi  and
	Hua, Nan  and
	Roth, Dan",
	booktitle = "Findings of the Association for Computational Linguistics: EMNLP 2020",
	month = nov,
	year = "2020",
}

@inproceedings{liu-etal-2021-ebert,
	title = "{EBERT}: Efficient {BERT} Inference with Dynamic Structured Pruning",
	author = "Liu, Zejian  and
	Li, Fanrong  and
	Li, Gang  and
	Cheng, Jian",
	booktitle = "Findings of the Association for Computational Linguistics: ACL-IJCNLP 2021",
	month = aug,
	year = "2021",
	pages = "4814--4823",
}

@inproceedings{jiao-etal-2020-tinybert,
	title = "{T}iny{BERT}: Distilling {BERT} for Natural Language Understanding",
	author = "Jiao, Xiaoqi  and
	Yin, Yichun  and
	Shang, Lifeng  and
	Jiang, Xin  and
	Chen, Xiao  and
	Li, Linlin  and
	Wang, Fang  and
	Liu, Qun",
	booktitle = "Findings of the Association for Computational Linguistics: EMNLP 2020",
	month = nov,
	year = "2020",
}

@misc{sanh2020distilbert,
	title={DistilBERT, a Distilled Version of BERT: Smaller, Faster, Cheaper and Lighter}, 
	author={Victor Sanh and Lysandre Debut and Julien Chaumond and Thomas Wolf},
	year={2020},
	eprint={1910.01108},
	archivePrefix={arXiv},
	primaryClass={cs.CL}
}

@inproceedings{gao2020distillation,
	author = {Gao, Luyu and Dai, Zhuyun and Callan, Jamie},
	title = {Understanding BERT Rankers Under Distillation},
	year = {2020},
	booktitle = {Proceedings of the 2020 ACM SIGIR on International Conference on Theory of Information Retrieval},
	pages = {149–152},
	numpages = {4},
	location = {Virtual Event, Norway},
}

@inbook{mitra2021conformer,
	author = {Mitra, Bhaskar and Hofst\"{a}tter, Sebastian and Zamani, Hamed and Craswell, Nick},
	title = {Improving Transformer-Kernel Ranking Model Using Conformer and Query Term Independence},
	year = {2021},
	booktitle = {Proceedings of the 44th International ACM SIGIR Conference on Research and Development in Information Retrieval},
	pages = {1697–1702},
	numpages = {6}
}

@inproceedings{Hofstaetter2020_sigir,
	author = {Hofst{\"a}tter, Sebastian and Zamani, Hamed and Mitra, Bhaskar and Craswell, Nick and Hanbury, Allan},
	title = {{Local Self-Attention over Long Text for Efficient Document Retrieval}},
	booktitle = {Proceedings of the 43rd International ACM SIGIR Conference on Research and Development in Information Retrieval},
	year = {2020},
	location = {Virtual Event, China},
	pages = {2021–2024},
	numpages = {4},
}

@inproceedings{scells2022sigir-green-ir,
	author = {Scells, Harrisen and Zhuang, Shengyao and Zuccon, Guido},
	title = {Reduce, Reuse, Recycle: Green Information Retrieval Research},
	year = {2022},
	booktitle = {Proceedings of the 45th International ACM SIGIR Conference on Research and Development in Information Retrieval},
	pages = {2825–2837},
	numpages = {13},
	location = {Madrid, Spain},
}

@inproceedings{strubell-etal-2019-energy,
	title = "Energy and Policy Considerations for Deep Learning in {NLP}",
	author = "Strubell, Emma  and
	Ganesh, Ananya  and
	McCallum, Andrew",
	booktitle = "Proceedings of the 57th Annual Meeting of the Association for Computational Linguistics",
	month = jul,
	year = "2019",
	address = "Florence, Italy",
	pages = "3645--3650",
}

@misc{xu2021green-dl,
	author = {Xu, Jingjing and Zhou, Wangchunshu and Fu, Zhiyi and Zhou, Hao and Li, Lei},
	title = {A Survey on Green Deep Learning},
	eprint = {2111.05193},
	primaryClass = {cs.CL},
	archivePrefix = {arXiv},
	year = {2021},
}

@inproceedings{bruch2022reneuir,
	author = {Bruch, Sebastian and Lucchese, Claudio and Nardini, Franco Maria},
	title = {ReNeuIR: Reaching Efficiency in Neural Information Retrieval},
	year = {2022},
	booktitle = {Proceedings of the 45th International ACM SIGIR Conference on Research and Development in Information Retrieval},
	pages = {3462–3465},
	numpages = {4},
	location = {Madrid, Spain},
}

@article{bruch2022reneuir-report,
	author = {Bruch, Sebastian and Lucchese, Claudio and Nardini, Franco Maria},
	title = {Report on the 1st Workshop on Reaching Efficiency in Neural Information Retrieval (ReNeuIR 2022) at SIGIR 2022},
	year = {2023},
	issue_date = {December 2022},
	volume = {56},
	number = {2},
	journal = {SIGIR Forum},
	articleno = {12},
	numpages = {14}
}

@inproceedings{mallia2022sigir,
	author = {Mallia, Antonio and Mackenzie, Joel and Suel, Torsten and Tonellotto, Nicola},
	title = {Faster Learned Sparse Retrieval with Guided Traversal},
	year = {2022},
	booktitle = {Proceedings of the 45th International ACM SIGIR Conference on Research and Development in Information Retrieval},
	pages = {1901–1905},
	numpages = {5},
	location = {Madrid, Spain},
}

@inproceedings{lassance2022sigir,
	author = {Lassance, Carlos and Clinchant, St\'{e}phane},
	title = {An Efficiency Study for SPLADE Models},
	year = {2022},
	booktitle = {Proceedings of the 45th International ACM SIGIR Conference on Research and Development in Information Retrieval},
	pages = {2220–2226},
	numpages = {7},
	location = {Madrid, Spain},
}

@inproceedings{broder2003wand,
	author = {Broder, Andrei Z. and Carmel, David and Herscovici, Michael and Soffer, Aya and Zien, Jason},
	title = {Efficient Query Evaluation Using a Two-Level Retrieval Process},
	year = {2003},
	booktitle = {Proceedings of the 12th International Conference on Information and Knowledge Management},
	pages = {426–434},
	numpages = {9},
	location = {New Orleans, LA, USA},
}

@inproceedings{ding2011bmwand,
	author = {Ding, Shuai and Suel, Torsten},
	title = {Faster Top-k Document Retrieval Using Block-Max Indexes},
	year = {2011},
	booktitle = {Proceedings of the 34th International ACM SIGIR Conference on Research and Development in Information Retrieval},
	pages = {993–1002},
	numpages = {10},
	location = {Beijing, China},
}

@article{mackenzie2021anytime,
	author = {Mackenzie, Joel and Petri, Matthias and Moffat, Alistair},
	title = {Anytime Ranking on Document-Ordered Indexes},
	year = {2021},
	issue_date = {January 2022},
	volume = {40},
	number = {1},
	journal = {ACM Transactions on Information Systems},
	month = {9},
	articleno = {13},
	numpages = {32},
}

@inproceedings{petri2019accelerated,
	author = {Petri, Matthias and Moffat, Alistair and Mackenzie, Joel and Culpepper, J. Shane and Beck, Daniel},
	title = {Accelerated Query Processing Via Similarity Score Prediction},
	year = {2019},
	booktitle = {Proceedings of the 42nd International ACM SIGIR Conference on Research and Development in Information Retrieval},
	pages = {485–494},
	numpages = {10},
	location = {Paris, France},
}

@misc{malkov2016hnsw,
	author = {Malkov, Yu. A. and Yashunin, D. A.},
	title = {Efficient and Robust Approximate Nearest Neighbor Search using Hierarchical Navigable Small World graphs},
	archivePrefix = {arXiv},
	eprint = {1603.09320},
	primaryClass = {cs.DS},
	year = {2016},
}

@article{wang2021graphanns,
	author = {Wang, Mengzhao and Xu, Xiaoliang and Yue, Qiang and Wang, Yuxiang},
	title = {A Comprehensive Survey and Experimental Comparison of Graph-Based Approximate Nearest Neighbor Search},
	year = {2021},
	issue_date = {July 2021},
	publisher = {VLDB Endowment},
	volume = {14},
	number = {11},
	journal = {Proc. VLDB Endow.},
	month = {7},
	pages = {1964–1978},
	numpages = {15}
}

@inproceedings{Bruch:wsdm:2020,
	author = {Bruch, Sebastian and Han, Shuguang and Bendersky, Michael and Najork, Marc},
	title = {A Stochastic Treatment of Learning to Rank Scoring Functions},
	year = {2020},
	booktitle = {Proceedings of the 13th International Conference on Web Search and Data Mining},
	pages = {61–69},
	numpages = {9},
}

@inproceedings{oosterhuis2021plrank,
	Author = {Oosterhuis, Harrie},
	Booktitle = {Proceedings of the 44th International ACM SIGIR Conference on Research and Development in Information Retrieval},
	Organization = {ACM},
	Title = {Computationally Efficient Optimization of Plackett-Luce Ranking Models for Relevance and Fairness},
	Year = {2021}
}

@inproceedings{diaz2020exposure,
	author = {Diaz, Fernando and Mitra, Bhaskar and Ekstrand, Michael D. and Biega, Asia J. and Carterette, Ben},
	title = {Evaluating Stochastic Rankings with Expected Exposure},
	year = {2020},
	booktitle = {Proceedings of the 29th ACM International Conference on Information \& Knowledge Management},
	pages = {275–284},
	numpages = {10},
	location = {Virtual Event, Ireland},
}

@inproceedings{Zamani2022cascade,
	title	= {Stochastic Retrieval-Conditioned Reranking},
	author	= {Hamed Zamani and Mike Bendersky and Donald Metzler and Honglei Zhuang and Marc Najork},
	year	= {2022},
	booktitle	= {Proceedings of the 2022 ACM SIGIR International Conference on the Theory of Information Retrieval},
	location = {Madrid, Spain},
}

@article{Johnson2021faiss,
	title={Billion-Scale Similarity Search with GPUs},
	author={Jeff Johnson and Matthijs Douze and Herv{\'e} J{\'e}gou},
	journal={IEEE Transactions on Big Data},
	year={2021},
	volume={7},
	pages={535-547}
}

@inproceedings{zhuang2022reneuir,
	author = {Zhuang, Shengyao and Zuccon, Guido},
	title = {Fast Passage Re-ranking with Contextualized Exact Term Matching and Efficient Passage Expansion},
	booktitle = {Workshop on Reaching Efficiency in Neural Information Retrieval, the 45th International ACM SIGIR Conference on Research and Development in Information Retrieval},
	year = {2022},
}

@inproceedings{tilde,
	author = {Zhuang, Shengyao and Zuccon, Guido},
	title = {TILDE: Term Independent Likelihood MoDEl for Passage Re-Ranking},
	year = {2021},
	booktitle = {Proceedings of the 44th International ACM SIGIR Conference on Research and Development in Information Retrieval},
	pages = {1483–1492},
	numpages = {10},
	location = {Virtual Event, Canada},
}

@inproceedings{formal2022splade,
	author = {Formal, Thibault and Lassance, Carlos and Piwowarski, Benjamin and Clinchant, St\'{e}phane},
	title = {From Distillation to Hard Negative Sampling: Making Sparse Neural IR Models More Effective},
	year = {2022},
	booktitle = {Proceedings of the 45th International ACM SIGIR Conference on Research and Development in Information Retrieval},
	pages = {2353–2359},
	numpages = {7},
	location = {Madrid, Spain},
}

@inproceedings{thakur2021beir,
	title={{BEIR}: A Heterogeneous Benchmark for Zero-shot Evaluation of Information Retrieval Models},
	author={Nandan Thakur and Nils Reimers and Andreas R{\"u}ckl{\'e} and Abhishek Srivastava and Iryna Gurevych},
	booktitle={35th Conference on Neural Information Processing Systems Datasets and Benchmarks Track (Round 2)},
	year={2021},
}

@article{luan-etal-2021-sparse,
	title = "Sparse, Dense, and Attentional Representations for Text Retrieval",
	author = "Luan, Yi  and
	Eisenstein, Jacob  and
	Toutanova, Kristina  and
	Collins, Michael",
	journal = "Transactions of the Association for Computational Linguistics",
	volume = "9",
	year = "2021",
	pages = "329--345",
}

@inproceedings{wang2021bert,
	author = {Wang, Shuai and Zhuang, Shengyao and Zuccon, Guido},
	title = {BERT-Based Dense Retrievers Require Interpolation with BM25 for Effective Passage Retrieval},
	year = {2021},
	booktitle = {Proceedings of the 2021 ACM SIGIR International Conference on Theory of Information Retrieval},
	pages = {317–324},
	numpages = {8},
	location = {Virtual Event, Canada},
}

@inproceedings{chen2022ecir,
	author = {Chen, Tao and Zhang, Mingyang and Lu, Jing and Bendersky, Michael and Najork, Marc},
	title = {Out-of-Domain Semantics to the Rescue! Zero-Shot Hybrid Retrieval Models},
	year = {2022},
	booktitle = {Advances in Information Retrieval: 44th European Conference on IR Research, ECIR 2022, Stavanger, Norway, April 10–14, 2022, Proceedings, Part I},
	pages = {95–110},
	numpages = {16},
	location = {Stavanger, Norway}
}

@inproceedings{zhuang2021ictir,
	author = {Zhuang, Honglei and Qin, Zhen and Han, Shuguang and Wang, Xuanhui and Bendersky, Michael and Najork, Marc},
	title = {Ensemble Distillation for BERT-Based Ranking Models},
	year = {2021},
	booktitle = {Proceedings of the 2021 ACM SIGIR International Conference on Theory of Information Retrieval},
	pages = {131–136},
	numpages = {6},
	location = {Virtual Event, Canada},
}

@book{
	chuklin2015click,
	Author = {Chuklin, Aleksandr and Markov, Ilya and de Rijke, Maarten},
	Publisher = {Morgan \& Claypool},
	Title = {Click Models for Web Search},
	Year = {2015},
	Isbn = {9781627056489},
}

@inproceedings{oosterhuis-www2020-counterfactual,
	author = {Oosterhuis, Harrie and Jagerman, Rolf and de Rijke, Maarten},
	title = {Unbiased Learning to Rank: Counterfactual and Online Approaches},
	year = {2020},
	booktitle = {Companion Proceedings of the Web Conference 2020},
	pages = {299–300},
	numpages = {2},
	location = {Taipei, Taiwan},
}

@inproceedings{NEURIPS2021_b5200c61,
	author = {Swezey, Robin and Grover, Aditya and Charron, Bruno and Ermon, Stefano},
	booktitle = {Advances in Neural Information Processing Systems},
	editor = {M. Ranzato and A. Beygelzimer and Y. Dauphin and P.S. Liang and J. Wortman Vaughan},
	pages = {21644--21654},
	title = {PiRank: Scalable Learning To Rank via Differentiable Sorting},
	volume = {34},
	year = {2021}
}

@inproceedings{NEURIPS2020_ec24a54d,
	author = {Xie, Yujia and Dai, Hanjun and Chen, Minshuo and Dai, Bo and Zhao, Tuo and Zha, Hongyuan and Wei, Wei and Pfister, Tomas},
	booktitle = {Advances in Neural Information Processing Systems},
	pages = {20520--20531},
	title = {Differentiable Top-k with Optimal Transport},
	volume = {33},
	year = {2020}
}

@inproceedings{NEURIPS2019_d8c24ca8,
	author = {Cuturi, Marco and Teboul, Olivier and Vert, Jean-Philippe},
	booktitle = {Advances in Neural Information Processing Systems},
	pages = {},
	title = {Differentiable Ranking and Sorting using Optimal Transport},
	volume = {32},
	year = {2019}
}

@inproceedings{10.5555/3524938.3525027,
	author = {Blondel, Mathieu and Teboul, Olivier and Berthet, Quentin and Djolonga, Josip},
	title = {Fast Differentiable Sorting and Ranking},
	year = {2020},
	booktitle = {Proceedings of the 37th International Conference on Machine Learning},
	articleno = {89},
	numpages = {10},
}

@inproceedings{10.1145/3477495.3531849,
	author = {Jagerman, Rolf and Qin, Zhen and Wang, Xuanhui and Bendersky, Michael and Najork, Marc},
	title = {On Optimizing Top-K Metrics for Neural Ranking Models},
	year = {2022},
	booktitle = {Proceedings of the 45th International ACM SIGIR Conference on Research and Development in Information Retrieval},
	pages = {2303–2307},
	numpages = {5},
	location = {Madrid, Spain},
}

@inproceedings{10.1145/3289600.3290986,
	author = {Gallagher, Luke and Chen, Ruey-Cheng and Blanco, Roi and Culpepper, J. Shane},
	title = {Joint Optimization of Cascade Ranking Models},
	year = {2019},
	booktitle = {Proceedings of the 12th ACM International Conference on Web Search and Data Mining},
	pages = {15–23},
	numpages = {9},
	location = {Melbourne VIC, Australia},
}

@misc{bruch2022fusion,
	author = {Bruch, Sebastian and Gai, Siyu and Ingber, Amir},
	eprint = {2210.11934},
	archivePrefix={arXiv},
	primaryClass = {cs.IR},
	title = {An Analysis of Fusion Functions for Hybrid Retrieval},
	year = {2022}}

@article{santhanam2022moving,
	title={Moving Beyond Downstream Task Accuracy for Information Retrieval Benchmarking},
	author={Santhanam, Keshav and Saad-Falcon, Jon and Franz, Martin and Khattab, Omar and Sil, Avirup and Florian, Radu and Sultan, Md Arafat and Roukos, Salim and Zaharia, Matei and Potts, Christopher},
	eprint={2212.01340},
	archivePrefix={arXiv},
	primaryClass={cs.IR},
	year={2022}
}

@inproceedings{Ma2021DynaboardAE,
	title={Dynaboard: An Evaluation-As-A-Service Platform for Holistic Next-Generation Benchmarking},
	author={Zhiyi Ma and Kawin Ethayarajh and Tristan Thrush and Somya Jain and Ledell Yu Wu and Robin Jia and Christopher Potts and Adina Williams and Douwe Kiela},
	booktitle={Neural Information Processing Systems},
	year={2021}
}

@misc{doc2query-t5,
	author = {Nogueira, Rodrigo and Lin, Jimmy},
	title = {From doc2query to docTTTTTquery},
	year = {2019}}

@inproceedings{zhang-etal-2021-ltrmuppets,
	title = {Learning to Rank in the Age of Muppets: Effectiveness{--}Efficiency Tradeoffs in Multi-Stage Ranking},
	author = {Zhang, Yue  and
		   Hu, ChengCheng  and
		   Liu, Yuqi  and
		   Fang, Hui  and
		   Lin, Jimmy},
	booktitle = {Proceedings of the 2nd Workshop on Simple and Efficient Natural Language Processing},
	month = {11},
	year = {2021},
	pages = {64--73},
}

@inproceedings{liu-etal-2020-fastbert,
	title = "{F}ast{BERT}: a Self-distilling {BERT} with Adaptive Inference Time",
	author = "Liu, Weijie  and
	Zhou, Peng  and
	Wang, Zhiruo  and
	Zhao, Zhe  and
	Deng, Haotang  and
	Ju, Qi",
	booktitle = "Proceedings of the 58th Annual Meeting of the Association for Computational Linguistics",
	month = {7},
	year = "2020",
	pages = "6035--6044",
}

@inproceedings{schwartz-etal-2020-right,
	title = "The Right Tool for the Job: Matching Model and Instance Complexities",
	author = "Schwartz, Roy  and
	Stanovsky, Gabriel  and
	Swayamdipta, Swabha  and
	Dodge, Jesse  and
	Smith, Noah A.",
	booktitle = "Proceedings of the 58th Annual Meeting of the Association for Computational Linguistics",
	month = {7},
	year = "2020",
	pages = "6640--6651",
}

@misc{repbert2020dr,
	author = {Zhan, Jingtao and Mao, Jiaxin and Liu, Yiqun and Zhang, Min and Ma, Shaoping},
	title = {RepBERT: Contextualized Text Embeddings for First-Stage Retrieval},
	year = {2020},
	eprint = {2006.15498},
	archivePrefix={arXiv},
	primaryClass = {cs.IR},
}

@misc{ma2021replication-of-dpr,
	author = {Ma, Xueguang and Sun, Kai and Pradeep, Ronak and Lin, Jimmy},
	title = {A Replication Study of Dense Passage Retriever},
	year = {2021},
	eprint = {2104.05740},
	archivePrefix={arXiv},
	primaryClass = {cs.IR},
}

@inproceedings{qu-etal-2021-rocketqa,
	title = "{R}ocket{QA}: An Optimized Training Approach to Dense Passage Retrieval for Open-Domain Question Answering",
	author = "Qu, Yingqi  and
	Ding, Yuchen  and
	Liu, Jing  and
	Liu, Kai  and
	Ren, Ruiyang  and
	Zhao, Wayne Xin  and
	Dong, Daxiang  and
	Wu, Hua  and
	Wang, Haifeng",
	booktitle = "Proceedings of the 2021 Conference of the North American Chapter of the Association for Computational Linguistics: Human Language Technologies",
	month = {6},
	year = "2021",
	pages = "5835--5847",
}

@inproceedings{reconc2022wsdm,
	author = {Zhan, Jingtao and Mao, Jiaxin and Liu, Yiqun and Guo, Jiafeng and Zhang, Min and Ma, Shaoping},
	title = {Learning Discrete Representations via Constrained Clustering for Effective and Efficient Dense Retrieval},
	year = {2022},
	booktitle = {Proceedings of the 15th ACM International Conference on Web Search and Data Mining},
	pages = {1328–1336},
	numpages = {9},
	location = {Virtual Event, AZ, USA},
}

@inproceedings{zhan2021joint,
	author = {Zhan, Jingtao and Mao, Jiaxin and Liu, Yiqun and Guo, Jiafeng and Zhang, Min and Ma, Shaoping},
	title = {Jointly Optimizing Query Encoder and Product Quantization to Improve Retrieval Performance},
	year = {2021},
	booktitle = {Proceedings of the 30th ACM International Conference on Information and Knowledge Management},
	pages = {2487–2496},
	numpages = {10},
	location = {Virtual Event, Queensland, Australia},
}

@misc{zhao2022dense-retrieval,
	author = {Zhao, Wayne Xin and Liu, Jing and Ren, Ruiyang and Wen, Ji-Rong},
	title = {Dense Text Retrieval based on Pretrained Language Models: A Survey},
	year = {2022},
	eprint = {2211.14876},
	archivePrefix={arXiv},
	primaryClass = {cs.IR},
}

@inproceedings{tonellotto2021colbert,
	author = {Tonellotto, Nicola and Macdonald, Craig},
	title = {Query Embedding Pruning for Dense Retrieval},
	year = {2021},
	booktitle = {Proceedings of the 30th ACM International Conference on Information and Knowledge Management},
	pages = {3453–3457},
	numpages = {5},
	location = {Virtual Event, Queensland, Australia},
}

@inproceedings{quality-biased-ranking,
	author = {Bendersky, Michael and Croft, W. Bruce and Diao, Yanlei},
	title = {Quality-Biased Ranking of Web Documents},
	year = {2011},
	booktitle = {Proceedings of the 4th ACM International Conference on Web Search and Data Mining},
	pages = {95–104},
	numpages = {10},
	location = {Hong Kong, China},
}

@article{cormack2011efficient,
	title={Efficient and Effective Spam Filtering and Re-ranking for Large Web Datasets},
	author={Cormack, Gordon V and Smucker, Mark D and Clarke, Charles LA},
	journal={Information Retrieval},
	volume={14},
	pages={441--465},
	year={2011},
	publisher={Springer}
}

@inproceedings{macdonald2012usefulness,
	title={On the Usefulness of Query Features for Learning to Rank},
	author={Macdonald, Craig and Santos, Rodrygo LT and Ounis, Iadh},
	booktitle={Proceedings of the 21st ACM International Conference on Information and Knowledge Management},
	pages={2559--2562},
	year={2012}
}

@misc{xu2022survey,
	title={A Survey on Model Compression and Acceleration for Pretrained Language Models}, 
	author={Canwen Xu and Julian McAuley},
	year={2022},
	eprint={2202.07105},
	archivePrefix={arXiv},
	primaryClass={cs.CL}
}
\end{document}